# Osteology of a Near-Complete Skeleton of *Tenontosaurus tilletti* (Dinosauria: Ornithopoda) from the Cloverly Formation, Montana, USA

Jonathan Tennant
70406912
MEarthSci Geology (Hons.)

School of Earth, Atmospheric and Environmental Sciences
University of Manchester
Williamson Building
Oxford Road
Manchester
M13 9PL

# Contents

## Figures and Tables

# 1.0    Abstract

The character diagnosis of *Tenontosaurus tilletti* has been revised and redefined into a more robust and quantifiable state. Significant emphasis is placed on constructing phylogenetic definition in such a method, as it prevents occlusion of true character states by alleviating potential individual interpretational bias. Previous placement within the Iguanodontia is refuted based on the lack of character affinity with the defining synapomorphies of the clade. The clade Hypsilophodontidae (=Hypsilophodontia), along with Iguanodontia, however is deemed to be in critical need of refinement to account for recent discoveries and re-classifications of certain euornithopods. Several of the synapomorphies are out-dated and deemed redundant in favour of a more quantifiable approach. Re-definition of these clades is critical if the current state of basal euornithopodan relationships is to be resolved. Phylogenetic studies must be approached from a multidisciplinary perspective; integration of tectonostratigraphical, ontogenetic, palaeoecological, and biomechanical data with sets of well-defined primary homologies are essential in increasing phylogenetic resolution and generating stratigraphically feasible ancestor-descendant relationships. Material attributed to *Tenontosaurus tilletti* is in need of strict re-analysis; the significant quantity of specimens attributed to this species is potentially the result of poor stratigraphic constraints and the vast spatiotemporal span occupied. Future revision of this material is expected to reveal temporal variations on the species-level inherently linked to environmental evolution, as well as possibly provide clues to sexual dimorphism in contemporaneous, yet morphologically distinct tenontosaurs.

## 2.0 Introduction

*Tenontosaurus tilletti* is a moderately-sized graviportal ornithopod from the Lower Cretaceous (Upper Aptian – Lower Albian) Cloverly Formation of the Bighorn Basin region in northwest Wyoming and south-central Montana, USA, and is known from approximately 80 skeletons of various ontogenetic stages, taphonomic conditions and degrees of completion (cranial and postcranial elements and teeth). It is also found less abundantly in the Paluxy Formation, Texas, and has been reported from fragmentary and poorly represented material from Lower Cretaceous deposits in Idaho, Utah, Arizona, and Maryland (Forster, 1990). Cifelli *et al.* (1997) also describe *Tenontosaurus* remains in the Antlers Formation, Oklahoma, amidst massive accumulations of articulated and disarticulated material.

First described properly by Ostrom (1970), the postcranial skeleton of *T. tilletti* was revised by Forster (1990). Throughout this period and to the present, many cladistic and phylogenetic analyses including the Ornithopoda have been undertaken (e.g. Dodson, 1980; Sereno, 1986; Forster, 1990; Weishampel and Heinrich, 1992; Butler *et al.*, 2008; Barrett and Han, 2009), each with independent and various outcomes. Currently a second species is recognised: *Tenontosaurus dossi* from the Aptian Twin Mountains Formation, Texas (Winkler *et al.*, 1997).

## 2.1 Aims and Objectives

The aim of this study is to provide a full morphological description of what is possibly the best-preserved specimen of *Tenontosaurus* to date. Dr. John Nudds purchased LL.12275 on behalf of the Manchester Museum in 1999, and it became the centrepiece of the lottery-funded refurbished Fossil Gallery of the University of Manchester Museum (fig. 1), until its replacement by a cast specimen of the notorious *Tyrannosaurus rex*. Since this displacement, it has remained in storage with the exception of several minor studies, and its significance only touched upon. Throughout dismantling and subsequent storage, the individual skeletal elements have undergone various alterations; many are still affixed to the frame, several contain plaster additions and visual modifications ('conservation' procedures); some are still affixed to the original armature, but most have been dismantled and varnish and aesthetic colour-wash removed. Those incomplete elements that had been carefully restored by the original preparators using synthetic fillers have inexplicably had their restored portions removed, and many of the more slender elements have become broken and damaged beyond repair. Despite this, the entire skeleton (estimated to be 85-90% complete) retains an incredible degree of preservation, with most significant structures largely intact rendering it a critical specimen for study. The intention is to compare this specimen with the holotype (AMNH 3040; paratypes YPM-PU 16338 and YPM 3456) and other previously described specimens by Ostrom (1970) and Forster (1990), to detect specific variations to ascertain true identity and test the reliability of these previous studies in terms of the completeness of the material described. The variable degree of restoration is taken into account, as in some areas this largely occludes detail (e.g. in the skull). The condition of the vertebral column is also problematic as many elements have been largely fragmented, distorted and disordered, and the assignment to any particular genus or species thus challenging.

A significant feature of this specimen is that it represents a sub-adult stage (assuming the given classification to *Tenontosaurus tilletti* to be unequivocal), so variations to the adult holotype will only be subtle if present. The description of this specimen will allow for more robust phylogenetic analyses rather than using a combination of several different specimens that could potentially represent multiple stages of tenontosaur growth, or even entirely new species. A complete comparison with *Hypsilophodon foxii* (and minor additional material) is also undertaken to elucidate the nature of the relationship between *Tenontosaurus* and the clade 'Hypsilophodontidae'. A full phylogenetic analysis unfortunately is beyond the scope of this study, but hopefully this will provide a firm basis for future revision. Understanding ontogeny to reconstruct phylogeny is also intended to be highlighted, as currently character matrices used in phylogenetic studies provide no account for ontogenetic variations, thus potentially occluding true primary homology identification and leading to phylogenetic instability.

## 2.2 Institutional Abbreviations

University of Manchester Museum (UoMM); University of Manchester (UoM); University of Cambridge (UoC); Natural History Museum, London (NHM) – formerly BMNH (British Museum of Natural History); Ohio Museum of Natural History (OMNH); Peabody Museum, Yale University (YPM); Peabody Museum, Yale University (YPM-PU) (originally in the collections of Princeton University); American Museum of Natural History (AMNH).

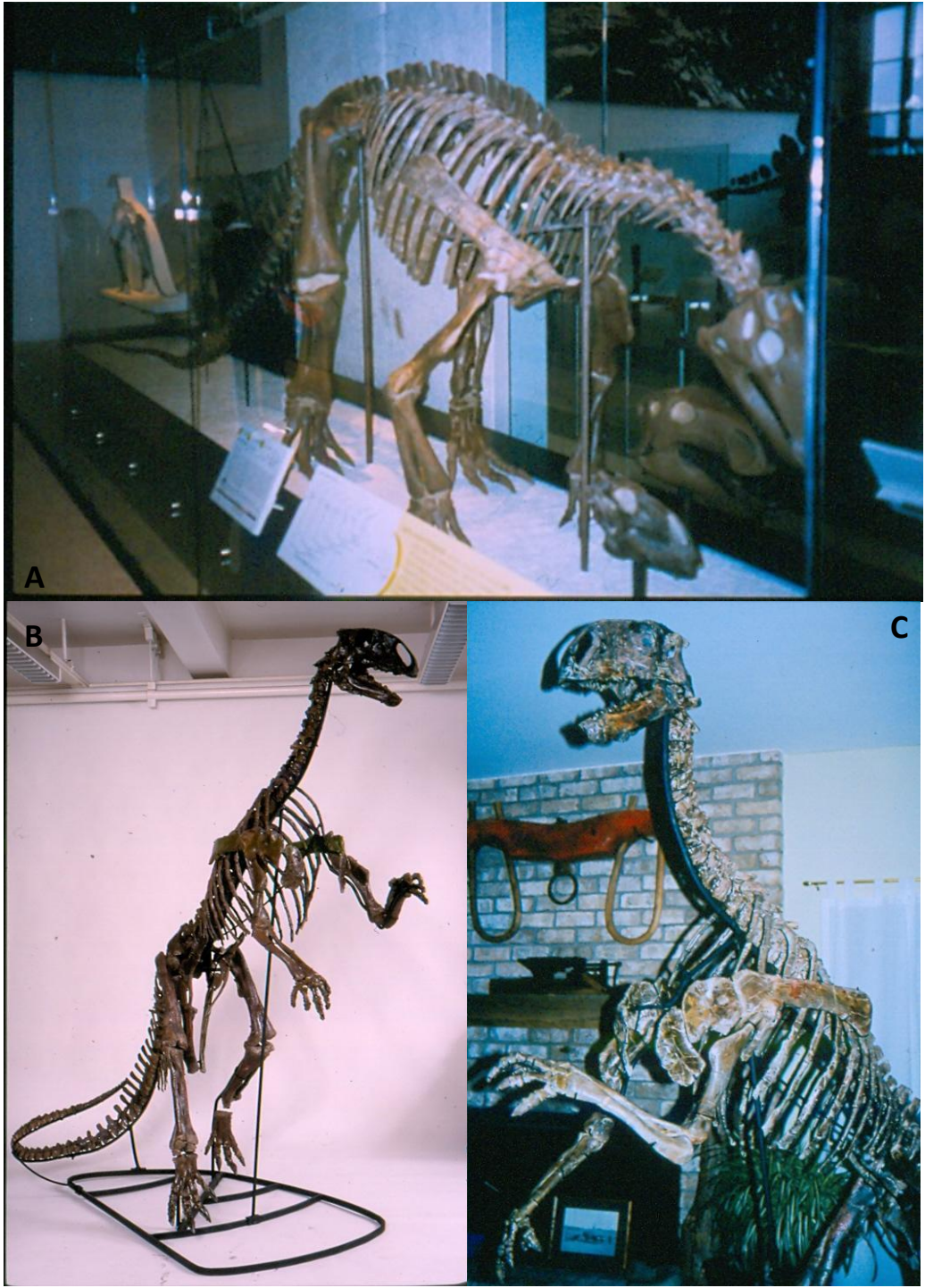

Figure 1 – **A:** *Tenontosaurus* specimen in quadrupedal stance **B:** LL.12275 fully articulated and mounted for display in bipedal stance **C:** LL.12275 close-up image of bipedal stance. Total length = 4.3m. Images courtesy of J. Nudds (UoM).

## 3.0 Previous Work

### 3.1 The Cloverly Formation

Fossiliferous units of this extensive formation are exposed along the eastern periphery of the Bighorn Basin in Montana and Wyoming, West of the Bighorn Mountains. The age of the Cloverly Formation is regarded to be Lower Cretaceous in age by a host of authors (e.g. Ostrom, 1970; Forster, 1984, 1990; Meyers *et al.*, 1992; Winkler *et al.*, 1997; Cifelli *et al.*, 1998; Nydon and Cifelli, 2002; Burton *et al.*, 2006), based on a variety of data including palynological, sedimentological and palaeontological analyses. Palaeomagnetic and zircon fission track data analysis indicates a Late Neocomian, Aptian and Early Albian age. Sparse microfossil data agrees with a Neocomian(?) to Albian age, although several authors question the validity of this data.

The most precise dating of the Cloverly Formation appears to be that of Burton *et al.* (2006), who calculate a single-crystal laser-fusion argon-argon age from an intraformational ashy horizon. The stratum occurs at approximately 75 metres above the contact with the underlying Upper Jurassic Morrison Formation (Kimmeridgian-Early Tithonian), and indicates an age of 108.5±0.2Ma implying deposition in the mid-Albian.

The middle fauna (of three distinct groups, dated at approximately 113-117Ma) of the Cedar Mountain Formation is suggested to be coeval with the Cloverly Formation due to the presence of *Sauropelta*, a common component of the Cloverly fauna. Other coeval deposits include part of the Lakota Formation of eastern Wyoming and western South Dakota, the upper half of the Gannett Group in the foredeep of westernmost Wyoming (Meyers *et al.*, 1992), and the Trinity Group of Northern Texas (Cifelli *et al.*, 1998; Burton *et al.*, 2006) based on analogous vertebrate assemblages.

Fauna include *T. tilletti*, ankylosaurs (e.g. *Sauropelta edwardsi*), sauropods, a small theropod, ornithomimids, a hypsilophodont (*Zephyrosaurus*), dromaeosaurids (e.g. *Deinonychus antirrhopus*), turtles, frogs, crocodiles, and triconodont mammals (e.g. Ostrom, 1969; Forster, 1984, 1990; Cifelli *et al.*, 1998). This represents a large, diverse parautochthonous taxa, the palaeoecology of which has been preliminarily studied by Oreska *et al.* (2007) based on vertebrate microfossil assemblages presumably from members of the above taxa, and is the first palaeoecological study of the Cloverly Formation since Forster (1984). Unfortunately, this is published only in abstract form, so a recent palaeoecological study is currently unavailable for integrative review.

## 3.2 Geological Setting

A sedimentological overview is fundamentally tied to palaeontology in its use to interpret palaeoecology; that is to say using palaeontology (e.g. functional morphology, morphometrics, and diet) to interpret behaviour is directly coupled with palaeoenvironmental data inferred from sedimentology to reconstruct the lifestyle and ecological behaviour of an organism - essentially the central aim for a palaeontologist. A stratigraphic summary is also critical in placing dinosaur-bearing units into an intraformational temporal and spatial context and for correlation with extraformational fossiliferous units. A synopsis is presented here for such a purpose.

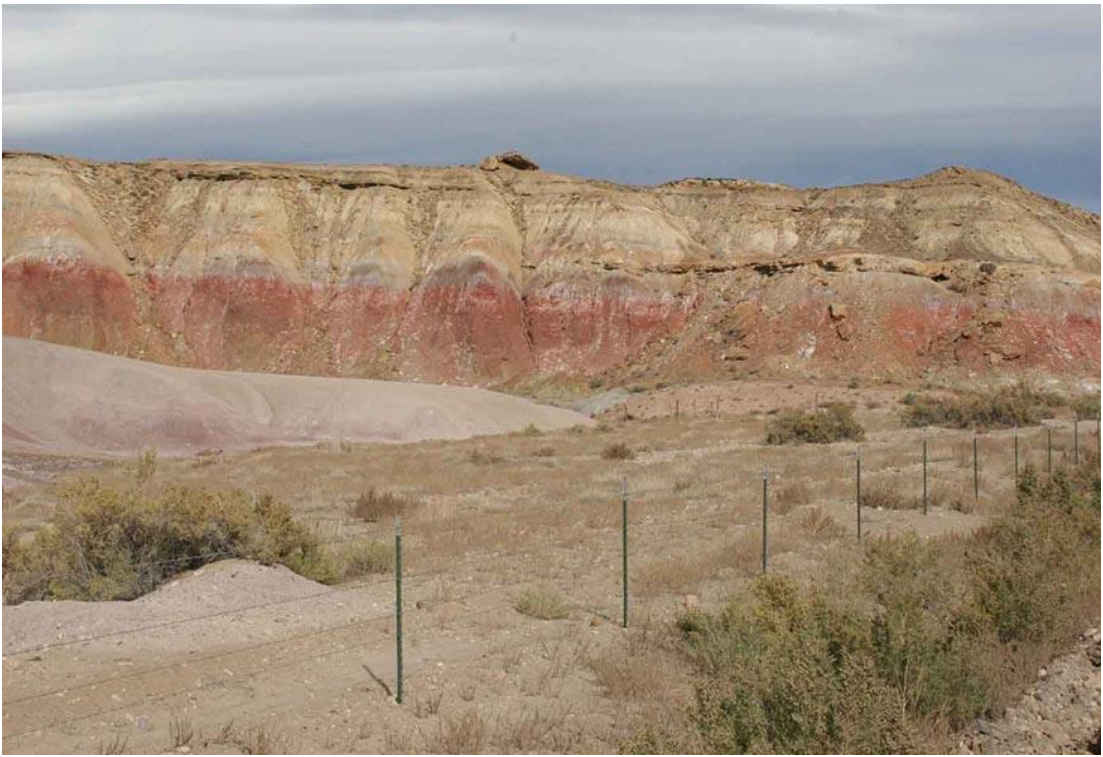

Figure 2 – Typical outcrop of the Cloverly Formation. *Wyoming State Geological Survey*

The type-section of the Cloverly Formation, as proposed by Meyers *et al.* (1992), is located on the eastern periphery of the Bighorn Basin and comprises a chert-pebble conglomerate overlain by chert arenite and variegated mudstone, lithic wacke and tan, cross-bedded quartz arenite (all non-marine strata – fig. 2). DeCelles and Burden (1992) divide the Cloverly Formation into two informal members: a lower mudstone deposited by muddy fluvio-lacustrine systems, and an upper chert-pebble conglomerate and sandstone deposited primarily by gravel-dominated braided rivers. Forster (1984) defines the Cloverly Formation as a 200m thick deposit characterised by variegated claystones with numerous sandstone and conglomeratic sandstone facies. Thus there is still no well-defined lithological consensus, and the boundaries between underlying and overlying formations are still disputed.

Ostrom (1970) provided an exhaustive revision of the Cloverly Formation, with a well-structured stratigraphic framework and summary of the vertebrate fauna within. The division and terminology has been generally accepted by palaeontologists to date. The author concludes that 8 stratigraphic members are present (units I-VIII) within the Upper Mesozoic of the Bighorn Basin, and of these, units IV-VII are defined as the Cloverly Formation, with V and VII being the principal fossiliferous units, but with scattered remains from unit VI. The absence of fossils in several lower units (e.g. the Pryor Conglomerate) may be partially responsible for the difficulty in identifying the Morrison-Cloverly faunal changeover. *Tenontosaurus* occurs in units V, VI and VII, with the most common accumulations in upper V and lower VII where specimens are arbitrarily located independent of lithology.

Elliott Jr. *et al*. (2007) take this analysis a step further, defining three successive depositional systems within the continental deposits: perennial to intermittent alluvial; intermittent to ephemeral alluvial; and playa; each of these is well-defined by distinct lithofacies. This facies evolution is attributed to the uplift of the Sevier Mountains in the Early Cretaceous leading to the development of a rain shadow and thus varying spatial distribution of depositional environments. However, it can also be created by varying accommodation leading to shifting depocentres or expansion of the tropics leading to climate-induced variations. The facies provide clues to the climatic regime, and suggest a change from humid and seasonal to wetlands and floodplain conditions in an arid to semi-arid environment. The exact evolutionary response of associated organisms to such climatic forcing is currently unknown, although one would expect Cloverly faunas (and flora) to be inherently distinct to those from the Morrison Formation, of which there was a distinctly drier climate with variable aridity.

The Cloverly Formation lies with unconformity on the Upper Jurassic Morrison Formation (e.g. Forster, 1984). The disparate palaeontological data (e.g. Ostrom, 1970) implies the presence of such an unconformable contact, as little-no evolutionary links are currently discernible between the two sequences. However, the lack of any typical features associated with unconformities (e.g. erosional scour at contact) suggests a generally conformable contact (see Meyers *et al.* (1992) for discussion). The authors also propose a 9Myr hiatus between the Morrison and Cloverly Formations based on preliminary palaeomagnetic data. Such a temporal hiatus, as well as the general lack of fossils in the lower Cloverly units, could be accountable for the apparently incongruous fauna in each setting.

The sedimentological information suggests that *Tenontosaurus* was primarily adapted to a semi-arid to arid environment, with seasonal climatic variations, analogous to modern mid-latitude regions. A complete floral overview would be expected to reflect this, and one would expect to see *Tenontosaurus* specimens to be specifically adapted to this designated ecological regime. Weishampel *et al.* (2010) propose that *Tenontosaurus* was an endemic species, with isolation induced by the Barremian breakup of Euamerica. Whether this is supported by stratigraphical analyses is currently unknown, but may prove to be useful in deciphering and understanding the suite of unusual characteristics *Tenontosaurus* exhibits.

## 3.3 Palaeoecology

*Tenontosaurus* is frequently cited as being found associated with remains of the dromaeosaur *Deinonychus antirrhopus*, such as the presence of shed teeth at multiple dig sites along with other more-complete skeletal material (Forster, 1984, 1990; Cifelli *et al.*, 1997; Ostrom, 1969, 1970; Maxwell and Ostrom, 1995). As it is rarely found with other prey taxa, it is apparent that the food of choice for *D. antirrhopus* was *T. tilletti* – it is unlikely that this association is due to a taphonomic or collecting bias. LL.12275 was also found with two *Deinonychus* teeth embedded in its neck (J. Nudds, pers. comm.)

Roach and Brinkman (2007) hypothesize that an average *D. antirrhopus* (70-100kg) should have been capable of solely combating and bringing down a half-grown, subadult *T. tilletti* (700-1000kg). This is analogous to modern day adult oras, with *D. antirrhopus* evidently being more agile and well equipped with features such as the extensively modified pedal digit II, long, clawed raptorial forelimbs and a rigid tail sheathed in elongate anterior-facing prezygapophyses (e.g. Ostrom, 1969; Roach and Brinkman, 2007).

Galton (1971b) outlines the affinities of the caudal series between *Deinonychus* and *Hypsilophodon*, *Tenontosaurus*, *Parksosaurus*, and *Thescelosaurus*, where the rigid tail acts as a dynamic counterbalance during locomotion. Given the size of *Tenontosaurus* however, it seems that it was not a rapid cursor similar to the others; thus, this feature becomes somewhat redundant and may represent a relict feature of the hypsilophodonts. This is coincident with Organ and Adams (2005) who, after observation of the osteohistology of ossified tendons in *Tenontosaurus*, conclude that the causes of intratendinous ossification are not related to the organisms body size, anatomical location or mechanical stresses. Thus it becomes probable that similar to *Deinonychus*, *Tenontosaurus* utilised its rigid tail to aid bipedal locomotion, but as more of a counter-balance than a rudder.

Maxwell and Ostrom (1995) and Li *et al*. (2008) report the findings of several specimens of *D. antirrhopus* based on teeth and fragmentary findings with a solitary *T. tilletti* specimen, and possible coeval deinonychosaur trackways respectively, concluding that *Deinonychus* engaged in pack-hunting and gregarious behaviour at least temporarily. However, particular ichnofossils such as this must be analysed with precaution, as 'coeval' in this sense could range from the tracks being made by multiple organisms during several seconds, or a matter of weeks (or more) in which case the apparent gregarious behaviour simply reflects the movement of several individuals over a longer period of time.

*D. antirrhopus* possessed a strongly recurved, hypertrophied and hyperextensible ungual claw on pedal digit II (e.g. Manning *et al*., 2005). The authors suggest that, based on mechanical models, the function of this claw was for traction during climbing, prey capture, and perhaps killing based on modern analogues from birds, reptiles and mammals (not the initial slashing and disembowelling suggested by Ostrom (1969)). However, this raises the question of the need for such a large and uniquely designed specialist feature, when a simple smaller claw would function in an equivalent manner (analogous to modern cats, and their hunting and occasional arboreal habits). This may relate to one being primarily quadrupedal, and the other an obligate biped, leading to distinctly different predation methods.

This hypothesis is countered by observation of other modern analogues, such as the ostrich (*Struthio camelus)* and cassowary (e.g. *Casuarius casuarius*) which are both fully capable of eviscerating adversaries such as large cats and even humans with powerful forward-directed thrust-kicks (Roach and Brinkman, 2007). This suggests that *D. antirrhopus* was more than capable of inflicting severe or mortal wounds to *Tenontosaurus*, probably regardless of size, although it is likely that predation was confined usually to smaller, sickly, or elderly tenontosaurs. In addition, it proposes the null-hypothesis to Maxwell and Ostrom (1995) and Li *et al.* (2008), that non-avian theropods were solitary hunters and at best formed loosely associated groups of scavengers or foragers.

*Tenontosaurus* was clearly herbivorous based primarily on tooth morphology. Norman and Weishampel (1985) studied ornithopod feeding mechanisms as during the latter half of the Mesozoic era, ornithopods diversified from an originally simple bauplan, becoming increasingly abundant with a range of body sizes, and thus dietary requirements. The authors conclude that a range of alternative modes of transverse food grinding were utilised by ornithopods achieved by a combination of an "isognathic" jaw frame and relatively simple adductor muscles with complex tooth batteries and either maxillary or mandibular rotation. Such motions are indicated by median-angle tooth wear on many ornithopods (Weishampel and Jianu, 2000). This ultimately led to a more efficient method of grinding plant fibres analogous to modern mammals, and may have been one of the critical factors contributing to the ascent and diversification of the Ornithopoda. Norman (1998) emphasizes similarly that functional improvements in cranial complexity are contiguous with increased proficiency in the gathering and processing of food. This is suggested to be related to niche partitioning of individual species (hence the substantial interspecific variation in cranial and tooth morphologies), or as a direct response to progressively arid ("xeric") adapted vegetation throughout the latter half of the Mesozoic.

Further evidence on the dietary habits of *Tenontosaurus* is provided in Stokes, (1987). The Cloverly Formation here is quoted as being part of a gastrolith-bearing sequence spread contiguously over 750,000 square miles of territory. This has significant implications, as it suggests that contemporaneous herbivores utilised gastroliths during feeding. Stokes, (1987) suggests that the dental morphology of *T. tilletti* (as well as Cloverly sauropods) was ineffective for chewing and grinding the flora available, and that mechanical assistance was provided through gastrolith consumption. LL.12275 was found with several such polished gastroliths in its stomach, together with two cycad seeds, supporting Stokes (1987) and also suggesting *Tenontosaurus'* preferred diet (J. Nudds, pers. comm.). This finding is also the first direct evidence of cycads within a cololite (fossilized gut contents), usually being located within coprolites (fossilized excrement) (Butler *et al.*, 2009), and could reveal a link between co-evolution of cycads and herbivorous dinosaurs upon future analysis.

Forster (1990b) described the possible aggregation of *T. tilletti* at a juvenile phase indicating that extended parental bonds, with juvenile congregation into groups, may have occurred. The implications are that at least within some herds, juveniles of *Tenontosaurus* remained within closely bound groups for significant periods of time after birth and initial maturation. Whether this trait is synapomorphic within other ornithopod groups has yet to be determined. Eggshells have also been described from within the Cloverly Formation along with neonate ornithopod remains by Maxwell and Horner, (1994), and although presently unclassifiable, could possibly provide future insights into parental behaviour and reproduction within Early Cretaceous ornithopods. Varrichio *et al.* (2007) highlight an exhibition of extensive parental care amongst hypsilophodonts, suggesting such traits were plesiomorphic for Ornithopoda.

Recent studies by Lee and Werning (2008) and Scheyer *et al.* (2009) describe *Tenontosaurus* in terms of sexual maturity; the presence of medullary bone (endosteally-derived bone tissue) from the mid-diaphyses of an associated fibula and tibia of the specimen OMNH 34784 indicates that reproductive maturity in tenontosaurs was achieved by the age of 8 years. This appearance in Ornithopoda as well as Theropoda (e.g. *Tyrannosaurus rex* and *Allosaurus fragilis*) suggests that this feature is plesiomorphic for the Dinosauria (Scheyer *et al*., 2009). The discovery of such tissue in other dinosaurs could prove critical in understanding their reproductive habits, as well as providing clues on sexual dimorphism.

Presently, it is near impossible to differentiate between intraspecific gender distinction and variations between different but closely related species. The result is that the erection of some species, or genera, may be erroneous in that they simply represent the alternative gender to another organism. Whether such an incongruity has ever been recorded is presently unknown; however determining such potential fundamental flaws would be extremely difficult to discern, and requires careful consideration in future phylogenetic studies.

## 3.4 Biomechanics

The structure and morphology of the appendicular and axial skeleton of *T. tilletti* indicate that it was a moderately sized, but robustly built obligate biped capable of limited quadrupedal locomotion (e.g. Forster, 1990).

Organ (2006) undertook a biomechanical analysis of the locomotor abilities of *Tenontosaurus* by focussing strictly on the function of the epaxial (dorsal to the transverse processes) and hypaxial (ventral to the transverse processes) ossified tendons that run along the sacral and caudal series of the vertebral column (fig. 3). This characteristic is considered plesiomorphic for *Tenontosaurus*, and is actually synapomorphic amongst nearly all ornithischians despite the huge diversity of forms present in the clade (e.g. Sereno, 1999; Organ and Adams, 2005).

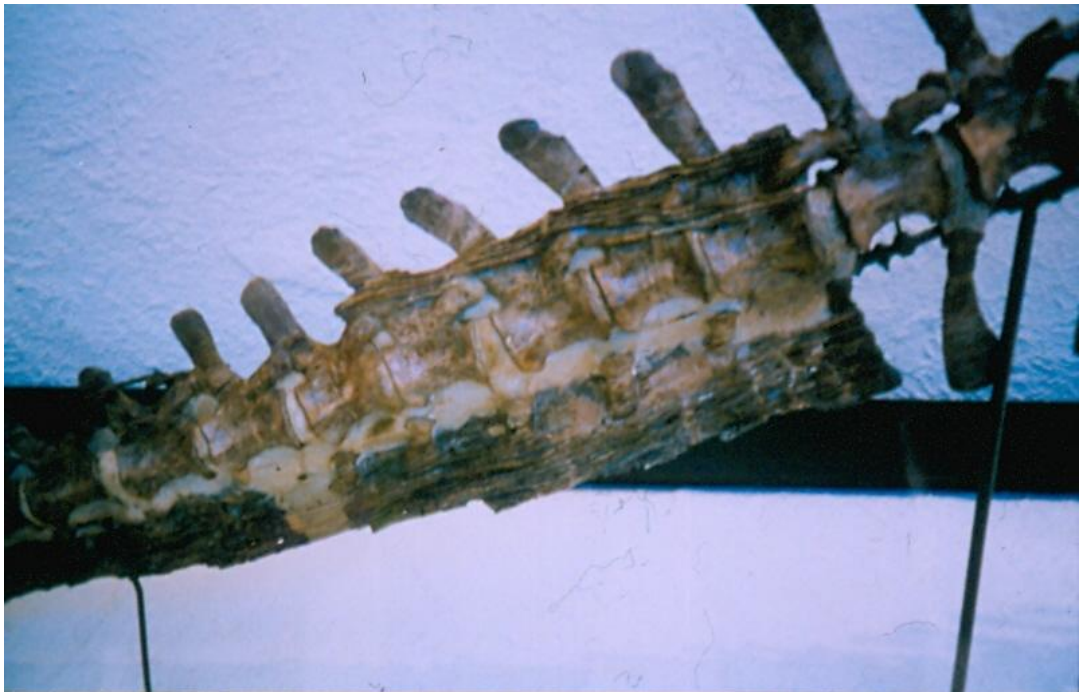

Figure 3 – Hypaxial and epaxial ossified tendons dorsal and ventral to the caudal series, occupying positions both laterally and distally to the neural spines and chevrons. Unfortunately, these elements are now absent. Field of view approximately 60cm. Image courtesy of J. Nudds (UoM).

Forster (1990) states that such structures are utilised to cantilever the body of *Tenontosaurus* over the acetabulum, with the inflexible tail acting as a counter-balance to the torso. Maxwell and Ostrom (1995), who specify that the presence of these ossified tendons in conjunction with near vertical to vertical articular facets on the pre- and postzygapophyses of the caudal series probably acted to restrict lateral motion in all but the most distal extremity of the tail, reinforce this interpretation. Thus, the conclusion is that it possessed a poor defensive function, and hence was possibly utilised in a role proposed by Forster (1990). However, modelling by Organ (2006) suggests that ossified tendons did not restrict mediolateral motion; tendon networks must be situated laterally and distal to the neural spines to have any effect on the tail, whereas in *Tenontosaurus* they are primarily confined to the parasagittal plane. This would suggest a preferred bipedal stance with the stiffened tail acting primarily to support the trunk during locomotion, browsing, and social activities. Reorganisation of the pelvic girdle allowed the mass of the gut to be positioned between the hindlimbs, which would also help contribute to a bipedal stance (Norman, 2004). However, it is not a simple case of mass balance and tail function that dictates stance; one must account for additional factors such as pedal morphology and ontogenetic variation (e.g. Moreno, 2007).

## 3.5 Palaeobiogeography

Basal iguanodontians (i.e. non-hadrosaurians) and 'hypsilophodontids' inhabit an extensive geographic distribution, occurring in Asia, Europe, North and South America, Africa and Australia, as well as a broad temporal distribution, occurring in Late Jurassic to latest Cretaceous sediments (table 1). Accordingly, ornithopods are found in a wide distribution of palaeoenvironmental settings. Non-hadrosaurian ornithopods are commonly affiliated with coastal depositional environments, and negatively associated with terrestrial deposits (Butler and Barrett, 2008); however they are found in inland settings (e.g. *Tenontosaurus*), indicating adaptation to a wide range of habitats. How this relates to environment-specific floral variations is currently unknown. The diversification of early iguanodontians can potentially be attributed to the Jurassic-Cretaceous convergence of the western European and North American continents, which upon cessation in the Hautevarian (130Ma) led to the segregation of North American fauna and their taxonomic and ecological divergence as endemic populations developed (Norman, 1998).

Their occurrence in South America is summarised by Coria and Salgado (1996) and Coria *et al.* (2007), who refer to fossil tracks in Chile, Brazil and Argentina, *Pisanosaurus mertii*, 3 Upper Cretaceous hadrosaur genera, the pre-Campanian *Anabisetia saldiviai,* and the Campanian *Gasparinisaura cincosaltensis* – both considered basal iguanodontians with similar ecological and morphological statuses to small, cursorial bipeds such as *Dryosaurus* and *Hypsilophodon*. The presence of these two taxa implies the existence of an isolated South American endemic basal iguanodontian lineage in the Late Cretaceous (Coria and Calvo, 2002), although their exact phylogenetic position is still disputed.

Within Central America they are poorly represented, known only from a single femur assigned to the Ornithopoda and of probable Campanian age (Horne, 1994).

| Taxon | Infraorder | Stratigraphic Distribution | Palaeogeographic Distribution |
|---|---|---|---|
| *Yandusaurus hongheensis* | Hypsilophodontidae | Bathonion (Middle Jurassic) | Sichuan, China |
| *Othnielia rex* | Hypsilophodontidae | Oxfordian-Tithonian (Late Jurassic) | Wyoming, Utah, Colarado, USA |
| *Hypsilophodon foxii* | Hypsilophodontidae | Barremian (Early Cretaceous) | Isle of Wight, UK; possibly Portugal and USA |
| *Zephyrosaurus schaffi* | Hypsilophodontidae | Aptian-Albian (Early Cretaceous) | Montana and Wyoming, USA |
| *Orodromeus makelai* | Hypsilophodontidae | Campanian (Late Cretaceous) | Montana, USA |
| *Parksosaurus warreni* | Hypsilophodontidae | Maastrichtian (Late Cretaceous) | Alberta, Canada |
| *Thescelosaurus neglectus* | Hypsilophodontidae | Campanian-Maastrichtian | North America; Canada |
| *Agilisaurus louderbacki* | Hypsilophodontidae | Bathonion-Callovian (Middle Jurassic) | Sichuan, China |
| *Oryctodromeus cubicularis* | Hypsilophodontidae | Cenomonian (Late Cretaceous) | Montana, USA |
| *Tenontosaurus tilletti* | Hypsilophodontidae /Iguanodontia | Aptian-Albian | Montana and Wyoming, USA |
| *Tenontosaurus dossi* | Hypsilophodontidae /Iguanodontia | Aptian-Albian | Texas, USA |
| *Rhabdodon sp.* | Hypsilophodontidae /Iguanodontia | Campanian | Europe |
| *Dryosaurus altus* | Iguanodontia | Kimmeridgian-Tithonian (Late Jurassic) | Wyoming, Utah, Colarado, USA |
| *Dryosaurus lettowvorbecki* | Iguanodontia | Kimmeridgian | Tendaguru, Tanzania |
| *Camptosaurus dispar* | Iguanodontia | Kimmeridgian-Tithonian | Wyoming, Utah, Colarado, Oklahoma USA |
| *Camptosaurus aphanoecetes* | Iguanodontia | Kimmeridgian-Tithonian | Utah, USA |
| *Camptosaurus prestwichii* | Iguanodontia | Kimmeridgian | UK |
| *Draconyx loureiroi* | Iguanodontia | Tithonian | Lourinhã, Portugal |
| *Gasparinisaura cincosaltensis* | Iguanodontia | Santonian (Late Cretaceous) | Patagonia, Argentina |
| *Anabisetia saldiviai* | Iguanodontia | Cenomonian | Patagonia, Argentina |
| *Iguanodon sp.* | Iguanodontia | Berriasian-Hautevarian (Early Cretaceous) | Europe; USA |
| *Valdosaurus canaliculatus* | Iguanodontia | Berriasian-Barremian | UK; possibly Romania |
| *Zalmoxes robustus* | Iguanodontia | Maastrichtian | Transylvania, Romania |
| *Zalmoxes shqiperorum* | Iguanodontia | Maastrichtian | Transylvania, Romania |

Table 1 – Stratigraphic summary and phylogenetic placement of well-known basal euornithopod taxa. Fragmentary and unidentified remains are not included due to difficulty in assignment. Summaries such as this should be carefully combined with tectonostratigraphic reconstructions to reproduce phylogenetic relationships that are consistent with spatial and temporal data.

Kobayashi and Azuma (2003) describe an ornithopod from the Kitadani Formation, Japan: *Fukuisaurus tetoriensis* classified as a non-hadrosaurian iguanodontian and a member of the Styracosterna, a monophyletic clade including *Probactrosaurus, Iguanodon, Ouranosaurus, Protohadros* and all hadrosaurs. Kim *et al*. (2009) illustrate trackways from Korea, with resultant assignment to an *Iguanodon*-like organism. Jun *et al*. (2008) provide a sufficient summary of Chinese iguanodontians

The recent re-analysis of the hypsilophodont *Jeholosaurus shangyuanensis* by Barrett and Han (2009) from the Early Cretaceous of China has served to fill a small gap in basal ornithopodan taxonomy. *Probactrosaurus mazongshanensis* was the first vertebrate fossil reported from the Xinminbao group, China (Tang *et al.,* 2001), a Late Barremian-Albian succession deposited in a fluvio-lacustrine setting within a semi-arid, subtropical climate (an apparently universal aspect of iguanodontian-bearing, Early Cretaceous deposits). It is noteworthy, that within this formation, *Siluosaurus zhangqiani*, a hypsilophodontid is present, which is somewhat coincident with the Cloverly Formation where *T. tilletti* and the hypsilophodontid *Zephyrosaurus* are found. This suggests the preferred ecological alliance (sympatric speciation) of these two families during this geological era. This grouping may be vital in unravelling hypsilophodontian and iguanodontian phylogeny; the same patterns of ancestor-descendant relationships are expected in both clades between stratigraphically associated organisms, if indeed Hypsilophodontidae is found to retain a valid cladistic status. This theory can possibly also be extended to other clades (e.g. within the Sauropodomorpha), if large-scale migration signals can be constrained.

This potential grouping may have even subtler implications: 'hypsilophodontians' may simply represent younger ontogenetic members of associated iguanodontians. For example, S. Maidment (pers. comm.) mentioned that *Hypsilophodon foxii* exhibits a remarkable similarity to *Iguanodon* when scaled up (excluding the skull). This would need to be tested rigorously with a wide sample range (destructive analysis) of all known hypsilophodonts; it may be that the clade *Hypsilophodontidae* is dissolved, as has been suggested previously but never fully enforced, with all taxa being assigned as younger ontogenetic stages of known iguanodontian genera. If this becomes the case, then the term "hypsilophodont" will not become redundant - instead of providing a specimen with a phylogenetic status, it will merely be a descriptive term for an ornithopod (possibly extendable to basal ornithischians) with a gracile bauplan, and a cursorial and bipedal mode of life. A possible taphonomic feature supporting this would be that the majority of known *H. foxii* specimens come from a single horizon (the infamous *Hypsilophodon* bed, Isle of Wight), which has never been studied in a taphonomic context (to the authors knowledge); it may perhaps be that the bed represents a single mass mortality event of a nesting site or juvenile congregation area. A point supporting such an event is that in the majority of specimens, the tibia has been snapped leaving the distal end articulated to the tarsus and pes. S. Maidment (pers. comm.) postulates that this could have resulted from many of the *Hypsilophodon* trying to escape as they became mired in a style of trap. Unfortunately, no explicit links between such events have currently been published, so presently such scenarios remain purely speculatative. Also, *Hypsilophodon* skulls appear well-fused, a feature exhibited in non-juvenile dinosaurs usually, although allometric growth rates within 'hypsilophodonts' have only been preliminarily studied with *Orodromeus makelai* (Horner *et al*. 2009).

Non-hadrosaurian ornithopods are found with a wide spatial distribution in Europe, especially in the Late Cretaceous of Spain, south France, Austria, Romania and Hungary (e.g. Sachs and Hornung, 2006). *Camptosaurus prestwichii* is found within the early Late Jurassic Lower Kimmeridge Clay, as well as contemporaneously on the other side of the Atlantic in the Morrison Formation (Galton and Powell, 1980). The family Camptosauridae has also been reported from the Late Jurassic Lourinhã Formation of Portugal (Mateus and Antunes, 2001; Galton, 2009), revealing yet another link within the fauna of Europe and North America during this period. *Callovosaurus leedsi* has been re-identified by Ruiz Omeñaca *et al*. (2007), confirming the presence of a dryosaurid in the Middle Jurassic (Callovian) of England, together with the genus *Valdosaurus*.

The currently monospecific *Hypsilophodon foxii* (Huxley, 1869) is a persistent specimen found within the Lower Cretaceous Wealden Marls (e.g. in the *Hypsilophodon* bed at the top of the Wessex Formation, of Barremian (132-125my) age) on the South-western shore of the Isle of Wight, England (Galton, 1971a; 1971b; Galton, 1974a; Butler and Galton, 2008), together with *Iguanodon*, which also occurs within the Lower Cretaceous of western North America (Galton and Jensen, 1975; Weishampel and Bjork, 1989). The assignment of material to this taxon however has recently been questioned and analysed by Brill and Carpenter (2007); the authors invalidate the American genus *Iguanodon lakotaensis*, stating that the material requires erection as a new genus, *Theiophytalia kerri*, with systematic placement between *Camptosaurus* and *Iguanodon*. Several Wealden specimens previously recognised as *Iguanodon* have been renamed as *Mantellisaurus atherfieldensis* by Paul (2007), with this genus representing a smaller, gracile iguanodont-like form. Particularly largely built iguanodonts have also recently been described from the Upper Barremian of France (Knoll, 2009). In addition, *Hypsilophodon* has been reported from both the Upper Jurassic of Portugal and Spain, and the early Cretaceous of North America, although some of the European material is poorly represented (Sanz *et al.*, 1983). Galton (2009) classifies the Spanish and Portuguese material as Euornithopoda indet. and the American *Hypsilophodon* is regarded *nomen dubium*. However, as Spanish and Portuguese are becoming increasingly similar to English fauna, it does hint at an ecological similarity and plausible contemporaneity too. Texan specimens however do appear to represent a grade of *Hypsilophodon*, although the Proctor Lake specimen is regarded by Galton (2009) as an unnamed ornithopod taxon, occupying a phylogenetic position between *Hypsilophodon* and *Tenontosaurus*. The description of this specimen may be crucial in unravelling the complex systematic placement of North American ornithopods, as currently a distinct morphological gap is present prior to *Tenontosaurus*.

Iguanodontians are found in the Upper Cretaceous (Upper Maastrichtian) of Transylvania, Romania. *Zalmoxes* is a member with two species currently acknowledged: *Z. robustus* and *Z. shqiperorum* (Weishampel *et al.*, 2003). The former is a medium-sized, rotund species, which despite its size displays a striking contrast to the more gracile *Dryosaurus*, *Hypsilophodon* and *Gasparinisaura*. The latter species is more comparable to larger ornithopods such as *Camptosaurus*. Both are associated with the clade Rhabdodontidae (also contains the taxon *Rhabdodon*), which incorporates the Euornithopoda and Iguanodontia (Sereno, 1996).

*Camptosaurus dispar* and *Camptosaurus aphanoecetes* represent Iguanodontia in the Upper Jurassic Morrison Formation. These ornithopods are both primarily quadrupedal and have a similar anatomy to *Tenontosaurus* (Norman, 2004; Carpenter and Wilson, 2008). Rare and isolated allochthonous remains including metatarsals have been located and assigned to the Ornithopoda from the Budden Canyon Formation, California (Hilton et al., 1997). North American hypsilophodont-grade dinosaurs include *Oryctodromeus cubicularis*, *Zephyrosaurus schaffi* and *Orodromeus makelai* (Varrichio *et al*., 2007).

Canadian iguanodonts are evident based upon assignment of bipedal trackways to small bipedal forms, with a gregarious nature being inferred from the presence of parallel and presumed coeval trackways. Similar tracks are also found in the Cretaceous of South Korea, and the Lower Cretaceous of England, Colorado and New Mexico (Lockley and Matsukawa, 1999). The authors suggest that 25cm maximum pes length can be assigned to a cursorial form rather than subcursorial or graviportal. This assignment based on track size is somewhat speculative, as prints observed preserved in rock are not always a simple function of pes size. The implications of such statements suggests however that as an organism develops, it will generally alter from a rapid cursor into a progressively graviportal form.

The hypsilophodontid *Laellynasaura* is found in the Aptian-Albian of Australia. Molnar and Galton (1986) report on both hypsilophodontids and iguanodontids from the Lower Cretaceous (Albian), with *Fulgotherium australe* and *Muttaburrasaurus* respectively. *Muttaburrasaurus* is postulated to represent an anomalously over-sized hypsilophodontian, analogous to *Tenontosaurus* in the northern hemisphere. Wiffen and Molnar (1989) describe a *Dryosaurus*-like ornithopod from the Upper Cretaceous of New Zealand, concluding that it is in fact a hypsilophodontian, extending the temporal range of this clade and implying that they pervaded into polar regions, and perhaps even into Antarctica. Given the endemic nature of Australian species, unravelling their relationships could provide clues to the evolution of endemic Early Cretaceous North American taxa.

Currently, few ornithopods from Africa are known: *Dryosaurus* from the well-known Tendaguru Formation of Tanzania, and an unidentified iguanodont from the Lower Cretaceous of Niger (Taquet and Russel, 1999). *Kangnasaurus coetzeei* (Cooper, 1985) has been tentatively confirmed as a dryosaurid from South Africa by Ruiz-Omeñaca *et al.* (2007), increasing the palaeogeographical range of the Dryosauridae.

Sizes range from small lightly built cursorial bipedal forms (2-3m long, e.g. *Dryosaurus*) to larger facultative quadrupeds (10-11m long, e.g. *Iguanodon*). The earliest recognisable presence of iguanodontians comes from the Early Kimmeridgian of England with *Camptosaurus prestwichii*. Other Kimmeridgian forms include *C. dispar* from the USA, and *Dryosaurus* from the North America and Africa. The latest currently know is *Zalmoxes* from the Latest Cretaceous of Europe.

The presence of basal ornithischians up to the early Jurassic (e.g. *Lesothosaurus*) implies that ornithopods may not have appeared prior to the Middle Jurassic (Butler *et al*. 2007).

Given the current prevailing cladistic positioning of *Tenontosaurus*, a significant ghost lineage is implied within the Iguanodontia at the origin of the clade from the Upper Jurassic to Aptian (e.g. Weishampel and Heinrich, 1992). However, the use of ghost lineages to resolve phylogeny is fundamentally flawed due to the dynamic nature of phylogenetic studies. Many authors generate these lineages (e.g. they are numerous within the Maniraptora) by extending the phylogenetic range of a known taxon or clade back to the first occurrence of the sister taxon or group, in spite of the lack of evidence from the fossil record. Although this does appear to be a logical approach to increasing phylogenetic resolution, it contravenes the very nature of science by generating a theoretical ancestry that can only be proved by either potential future acquisition of conspecific specimens, or the recoding of existing data matrices with the result that they become stratophylogenetically stable based on currently known taxa. One could alternatively view the implications of ghost lineage reconstruction in terms of weaknesses in the fossil record, and thus direct future research and exploration.

With regards to *Tenontosaurus*, a robust analysis is required to resolve the interactions between basal euornithopod taxa. An understanding of how, or if, ancestor-descendant relationships are coupled to tectonic processes is an essential prerequisite; that is to say that to draw a link between two taxa requires an 'event' (e.g. tectonics creating a dispersal or vicariant signal; co-evolution and adaptation;

palaeoclimatic variation) to force an evolutionary response in an organism or group of organisms. If no such event is recorded, then phylogenetic relationships become unfeasible. Therefore great care must be taken to ensure that when generating cladograms one forges taxa associations that are consistent with tectonostratigraphic records. Thus palaeontology becomes a multidisciplinary study, combining knowledge of not only biological processes, but also an understanding of large-scale and local geological interactions which are inherently, but often subtly, linked to a palaeoecosystem.

One point that requires emphasis may be something that has been broadly overlooked in the past: does a relatively complex morphology necessarily represent a 'derived' state within a clade? The question arises, as within a given ecosystem, the organisms will often be in steady-state equilibrium for the ecological duration, as each species will be specifically adapted to its individual partitioned niche within the system. If the ecosystem were to change, and suddenly there is a more 'plesiomorphic' successor present, then this may not represent a more primitive state within a clade, simply that the adaptations which have been acquired in the younger organism are specifically suited to the new environment and create the illusion of being primitive. Again, this relies on the intuition of individual authors, and the characterisation of what classifies as a 'derived' state, as well as a careful understanding of organism interactions within a complex system.

This question is raised based on the stratigraphic position of *Camptosaurus* and *Tenontosaurus* within North America, where the stratigraphically younger *Tenontosaurus* is consistently placed as the primitive ancestor to the clade

Camptosauridae. The two fauna occupy different palaeoecological niches, and it may be that the environment during the faunal transition became less harsh, and thus more derived states were made redundant in favour of a simpler but equally effective morphology. Dodson (1980) argues in favour of a *Camptosaurus-Tenontosaurus* ancestor-descendant relationship; however this has largely been undone by the work of multiple more recent phylogenetic studies.

### 3.6 Phylogeny

Ornithischians are defined by having a bird-hipped configuration of the pelvic girdle, which is to say the pubes have been rotated posteriorly to lie alongside the ischia (e.g. Sereno (1986, 1999a)). Ornithopoda Marsh 1981 has the phylogenetic definition: "all genasaurians more closely related to *Parasaurolophus walkeri* Parks, 1922 than to *Triceratops horridus* Marsh, 1189" (Butler *et al.*, 2008). The classification of *Tenontosaurus* within the Ornithopoda is deemed stringent and unequivocal.

Butler *et al.* (2008) propose that Ornithopoda is a polyphyletic clade including rhabdodontids, tenontosaurs, dryosaurids, and ankylopollexians (i.e. a paraphyletic assemblage of hypsilophodontids and iguanodontians, conforming to Sereno (1986, 1999a) and Norman (2004)) (fig. 4). One conclusion, albeit an unlikely one, is that the definition of Ornithopoda may be expanded to include heterodontosaurids, ceratopsians and pacycephalosaurs. Barrett and Han (2009) presented a revised form of Butler *et al*. (2008)'s extensive phylogenetic analysis (fig. 4). They too concluded that relationships within the Cerapoda required considerable further detailed investigation to resolve current disparities. The authors also significantly convey that this situation is somewhat paradoxical, due to the high degrees of material known for associated taxa (e.g. *Hypsilophodon, Tenontosaurus, and Jeholosaurus*).

Hypsilophodontids appear to represent a paraphyletic grade of neornithischian (Ornithopoda and Marginocephalia) and basal ornithopod taxa (e.g. Butler and Galton, 2008; Butler *et al*., 2008; Maidment and Porro, 2010). This is converse to Weishampel and Heinrich (1992), who define Hypsilophodontidae as the monophyletic sister clade to Iguanodontia, together comprising the Euornithopoda (fig. 5). Unfortunately, no strict consensus on an autapomorphy-based phylogenetic definition of Hypsilophodontidae currently appears to exist. Galton (2007) conveys that the members of 'Hypsilophodontidae' can be viewed as a suite of successive sister taxa to Iguanodontia (i.e. entails the dissolution of the clade), sensu Scheetz (1998, 1999), rather than a stand-alone clade as the sister group to Iguanodontia. Phylogenetic analyses resolved traditional hypsilophodonts as a broad grouping of polytomic relationships. Regrettably, one of these is published only in abstract form and the other is a currently unobtainable thesis, so analysis of the studies is presently impossible. Galton (2009) correspondingly disregards the clade name 'Hypsilophodontidae'; the bucket-term 'basal euornithopod' is instead favoured, which is perhaps the best option until conclusive systematic relationships of 'hypsilophodonts' can be elucidated with confidence. Iguanodontia represents a paraphyletic assemblage of increasingly derived taxa (e.g. Galton, 2009), yet appears to remain problematic despite the robust cladistic analysis of Butler *et al.* (2008); the phylogenetic definition appears to be balanced on the placement of *Thescelosaurus*, which is currently a point of deliberation. Butler *et al.* (2008) conclude with the remark that the "instability of basal ornithopod phylogeny" currently restricts any unambiguous phylogenetic definition regarding the Iguanodontia. Whereas this is somewhat challenging towards this study, it provides a direct target for future cladistic analyses to be aimed.

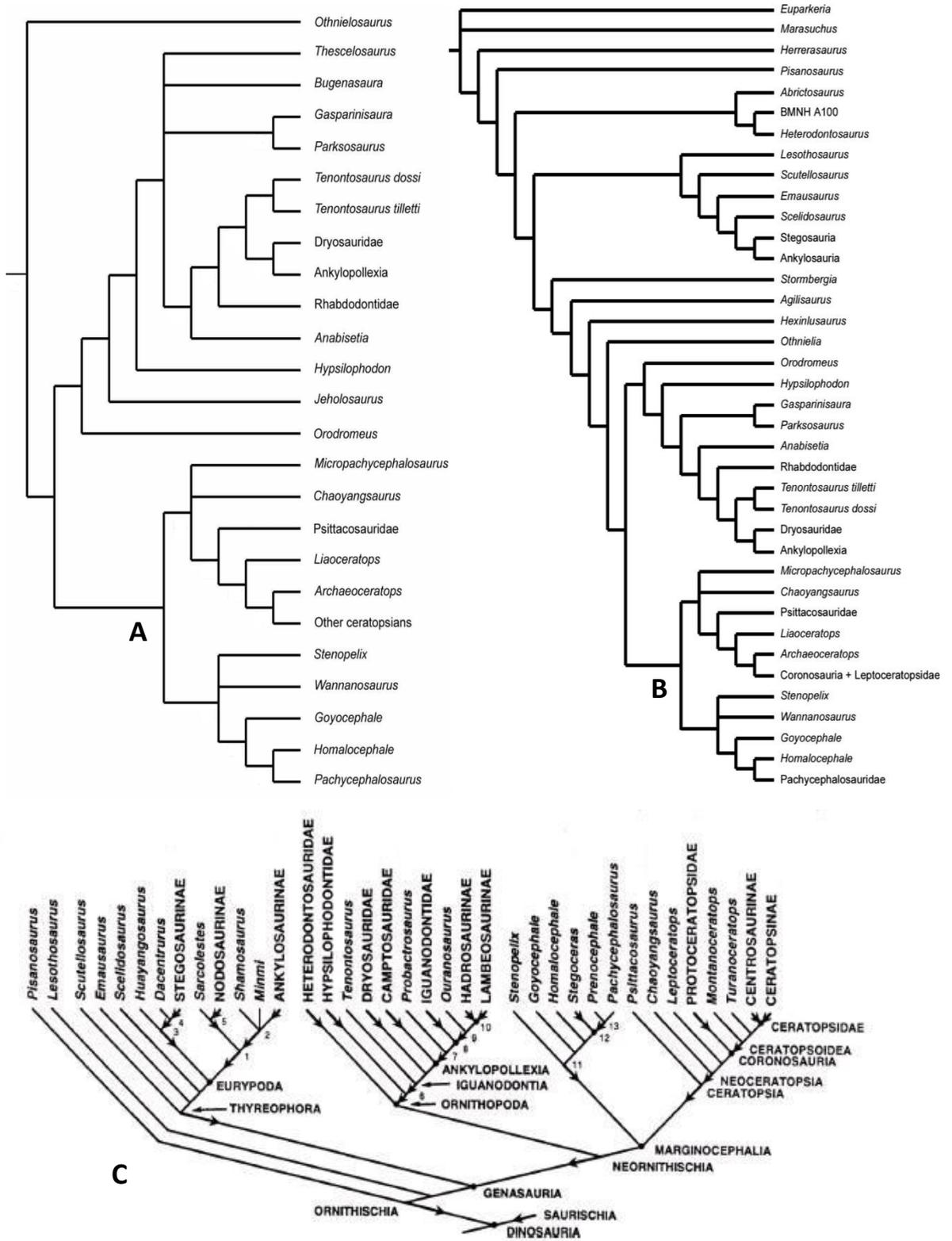

Figure 4 – cladograms representing the phylogeny of the Ornithischia; **A:** Barrett and Han, (2009) – edited from Butler *et al.*, (2008) **B:** Butler *et al.*, (2008) **C:** Sereno, (1999). Note exclusion of several basal euornithopodan taxa in **A** and **B,** and the association of *Tenontosaurus*.

Slightly different results are recovered in the analysis by Boyd *et al*. (2009) (fig. 6): North American basal neornithischians are resolved into a divergence between two distinct subclades with discrete morphologies. One comprises taxa proposed to be equipped to occupy a fossorial (digging) mode of life (e.g. *Oryctodromeus, Zephyrosaurus*), whilst the other is the morphologically larger clade including *Thescelosaurus* and *Parksosaurus*. This division is placed as the sister clade to *Hypsilophodon, Gasparinisaura, Tenontosaurus* and Iguanodontia. Considering the problematic placement of *Thescelosaurus* appears to be resolutely resolved, the phylogeny presented by Butler *et al*. (2008) requires revision to integrate this new data, which may lead to stabilising the positions of basal ornithopodan taxa.

Presently the taxonomic affinities of *Tenontosaurus* are disparate, being sited either within Iguanodontia as a basal taxon, or Hypsilophodontidae. The problem herein lies with the application of coherent and robust phylogenetic analyses, the use of inadequate material, and the absence of a rigorous testing process for primary homologies; this ultimately leads to incorrect relationship assumption, and the breakdown of the cladogram (see Sereno (2005) for an in-depth analysis of phylogenetic taxonomy). The position of several taxa including *Gasparinisaura*, and clades such as Rhabdodontidae also occlude relationships, as they remain disparate in spite of several recent analyses.

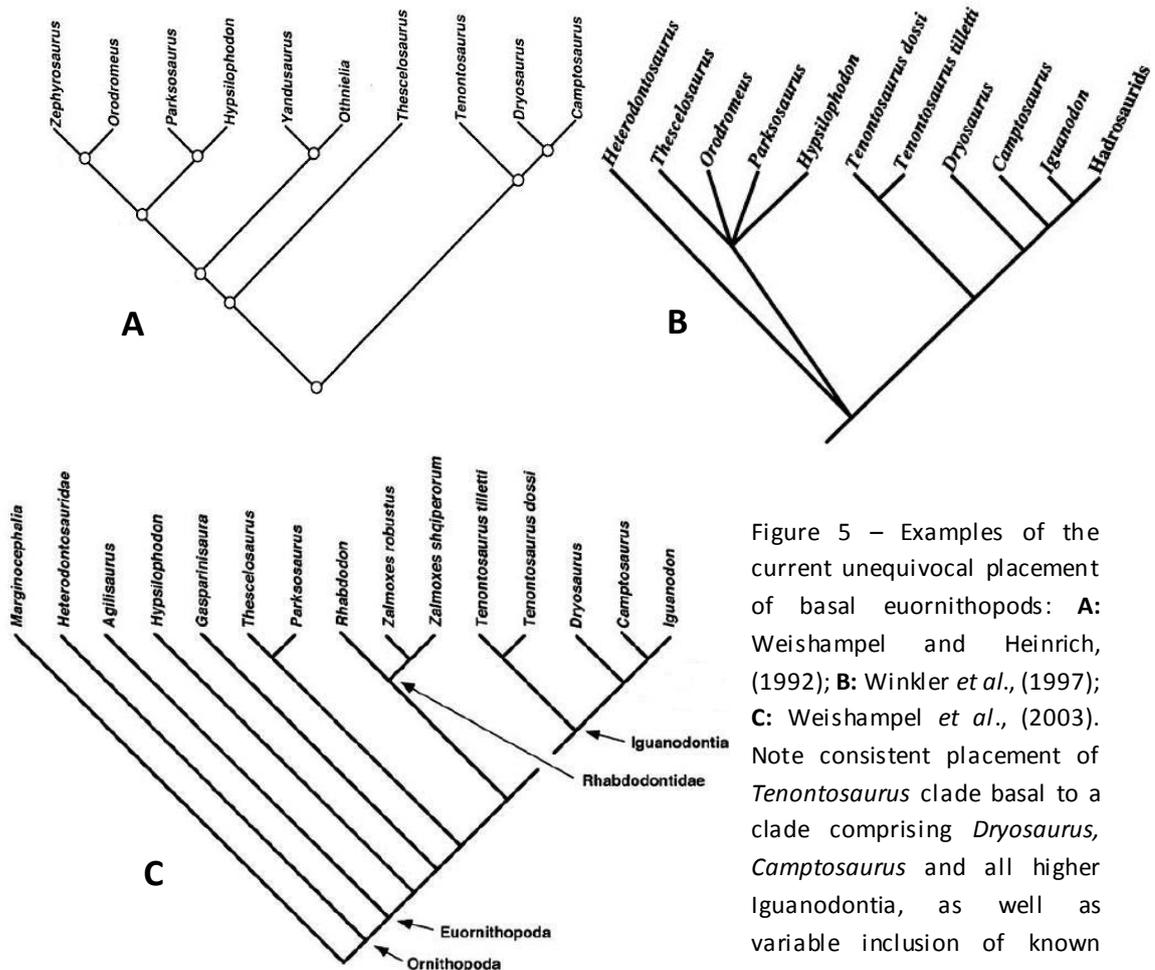

Figure 5 – Examples of the current unequivocal placement of basal euornithopods: **A:** Weishampel and Heinrich, (1992); **B:** Winkler *et al.*, (1997); **C:** Weishampel *et al.*, (2003). Note consistent placement of *Tenontosaurus* clade basal to a clade comprising *Dryosaurus, Camptosaurus* and all higher Iguanodontia, as well as variable inclusion of known taxa.

Initially, Ostrom (1970) assigned *Tenontosaurus* to the family Iguanodontia based on its resemblance to *Camptosaurus* and *Iguanodon*; similar assignments include the classification of the now redundant taxon *Vectisaurus valdensis* to Iguanodontidae by Galton (1976) based on its larger graviportal form compared to the generally accepted smaller, cursorial Hypsilophodontidae. However, modern phylogenetic studies are strict and more reliable, being based on a series of established synapomorphies (homologies) rather than a simple assessment of total character resemblance. Sereno (1984, 1986, 1999a, b) places *Tenontosaurus* as the primitive member of Iguanodontia with *Muttaburrasaurus* (Galton, 2009), conforming to Forster (1990) and being reinforced by Weishampel and Heinrich (1992), Coria and Salgado (1996) and Winkler *et al.* (1997).

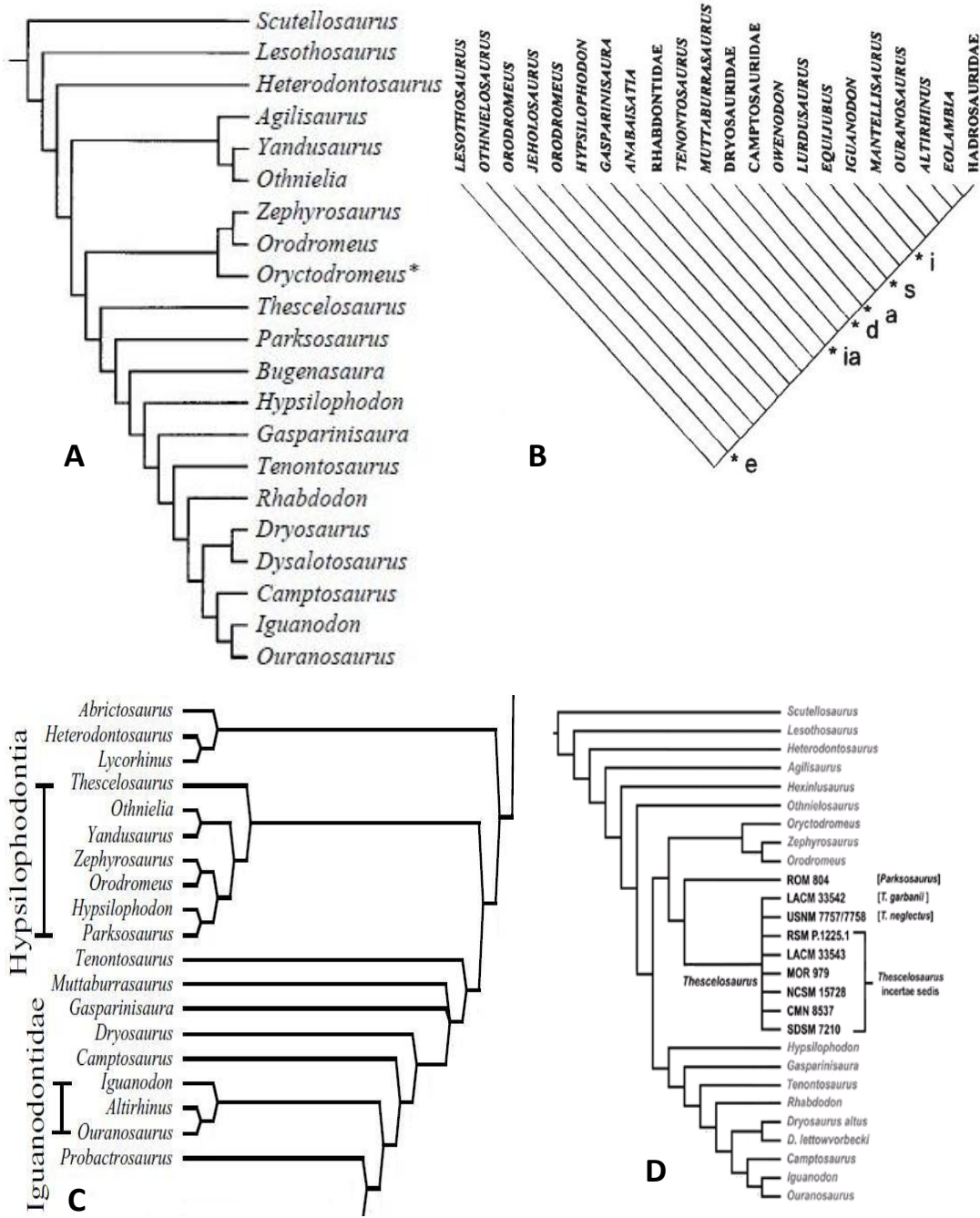

Figure 6 – More recent phylogenetic studies incorporating basal euornithopods and basal ornithischian taxa (usually as outgroups). **A:** edited from Pisani *et al.* (2002) **B:** Galton, (2009) e – Euornithopoda, ia – Iguanodontia, d – Dryomorpha, a – Ankylopollexia, s – Styracosterna, I - Iguanodontidae **C:** Varrichio *et al.*, (2007) **D:** edited from Boyd *et al.* (2009). Note the equivocal placement of *Tenontosaurus* either as the sister taxon to Iguanodontia, or as the basal member of the clade with *Muttaburrasaurus*. Galton (2009) (**B**) includes *Orodromeus* twice for an unknown reason.

Authors classifying *Tenontosaurus* amongst the Hypsilophodontidae include Dodson (1980), Norman (1990, 1998), and Coria and Salgado (1996); tenontosaurs are considered to be anatomically convergent with later iguanodontians. Dodson (1980) states *Tenontosaurus* cannot be a member of a monophyletic Iguanodontia due to the lack of dental specialisations characterising it apart from members such as *Iguanodon*, *Camptosaurus* and *Ouranosaurus*. The positioning as a hypsilophodont still implies *Tenontosaurus* is the basal sister taxon to all higher euornithopods (or iguanodontians), simply that it is not contained within the actual clade Iguanodontia. Thus the conclusion is somewhat supportive of Forster (1990) for example, but the exact definition of the clade Hypsilophodontidae remains problematic in that *Tenontosaurus* appears to retain several primitive characters consistent with this more basal clade (i.e. a 'hypsilophodontid' appearance paralleled with unequivocal derived characters). Norman (1998) similarly states that *Tenontosaurus* is a "morphologically oversized, derived hypsilophodontid".

Galton (1974a) defines Hypsilophodontidae based upon several characteristics: head small with short snout, large orbits, and no canine teeth; cursorial with distal part of hind limb elongate; and that they persistently represent the most basal and primitive members of the Ornithopoda. Unfortunately, such a simplistic scheme is now deemed obsolete in favour of the synapomorphic method of phylogenetic systematics; therefore although *Tenontosaurus* does not exhibit the above suite of characters, this does not necessarily mean that by modern standards it can be classified within the Iguanodontia.

Within the phylogenetic analysis undertaken by Weishampel and Heinrich (1992), missing data comprises 28% of character matrix, and character absence is variably distributed. The use of incomplete data sets is therefore potentially the cause of disparity within incongruent analyses (and the source of polytomies); however, at any given time of analysis, the most-complete data sets are compiled based on what is currently known, and therefore subject to change as gaps are filled over time. Nonetheless, caution is emphasized when utilising partial data sets, as the relationships drawn will often appear conclusive when in fact are occluded to an undeterminable degree. This problematic approach is also discussed by Sereno (1999c), who outlines several problems with the autapomorphy-based approach to phylogenetics, such as variations in character coding and homoplasy. The author similarly emphasizes the problem with numerous missing centres in phylogenetic analyses, in that they serve only to decrease phylogenetic resolution by generating multiple equally parsimonious cladograms. The effect of such aspects masking cladistic relationships is resplendent if one observes "the suite of homoplastic features" exhibited within *Tenontosaurus*, *Ouranosaurus* and *Altirinhus*, as discussed in Norman (1998).

Currently, two species of tenontosaur are recognised: *T. tilletti* and *T. dossi*. Paul (2008) provides a rationale for the classification of an organism at the species level: "a fundamental requirement for including more than one species within a genus is a reasonably consistent standard of skeletal material variation within a given genus". Based upon this, and assuming precision in the descriptions of Forster (1990) and Winkler *et al.* (1997), the dissection of the *Tenontosaurus* genus into two separate species appears authentic. Given the vast amount of material assigned to *T. tilletti* from the Cloverly Formation, and the associated large temporal range, it would not be unexpected that if upon careful examination of all specimens that more than one species would be present.

Forster (1990) provides the following diagnosis for *T. tilletti*:

1. Vertebral count of 12-16-5-60(+)
2. Deep tail comprising two thirds of the total length of the animal
3. Ossified tendons run axially along either side of the neural spines in dorsal, sacral and caudal vertebrae, and along the caudal centra and chevrons
4. Scapula with straight caudal margin
5. Coracoid with strong sternal process and coracoid foramen completely closed off from articulation
6. Forelimb relatively long and robust
7. Humerus dominated by strong extensive deltopectoral crest
8. Carpus comprised of intermedium, radiale and ulnare
9. Manus short and broad with phalangeal formula of 2-3-3-1?-1?
10. Ilium with long decurved preacetabular process
11. Ilium with dorsally expanded and rugose caudal margin
12. Ilium with very narrow brevis shelf
13. Pubis with short, straight pubic rod
14. Obturator foramen closed off from articulation
15. Prepubic blade laterally compressed, moderately deep and unexpanded at tip
16. Shaft of ischium straight and laterally compressed with tab-like obturator process one-third down the shaft
17. Femur with a finger-like lesser trochanter and pendant-like fourth trochanter
18. Femur with shallow extensor groove and deep flexor groove
19. Pes with phalangeal formula of 2-3-4-5-0, with vestigial 5[th] metatarsal

This revision of Ostrom (1970) excludes all initial cranial defining characters. Although a direct analysis with LL. 12275 and various other ornithopodan skulls has been unavailable (excluding with *H. foxii*), it does appear that many of these characters are either too general or synapomorphic to be defining autapomorphies. Thus, for the purposes of this study, only a revision of the diagnosis of Forster (1990) is undertaken. Future revision of cranial characteristics (of numerous specimens) is expected to reveal true distinguishing features of *Tenontosaurus*.

*T. dossi* appears primitive to *T. tilletti* based on the presence of several characteristics (Winkler *et al.*, (1997)):

1. Presence of premaxillary teeth
2. A long postpubis
3. A larger metatarsal V
4. Lacks a brevis shelf on the ventral border of the ilium
5. Possibly having fewer cheek teeth
6. Lesser eversion of the premaxilla
7. Less denticulation of the predentary

However, it does appear more derived based on the presence of a relatively longer humerus, where humeral length equal to or exceeding the scapular length is presumed synapomorphic of hypsilophodontids; thus this character may need revision in its use in defining this clade (Winkler *et al.*, 1997). A problem arises within this classification: several of these characters are often quoted as being primitive, and yet are often derived states (e.g. wide brevis shelf) are observed within hypsilophodonts (e.g. *Hypsilophodon foxii*); despite this, these organisms have been consistently placed as more primitive members of a clade (Hypsilophodontidae) basal to Iguanodontia, of which *Tenontosaurus* is often cited as the basal-most genera. It appears that homologies used to define clades require strict assessment in terms of their 'derived' statuses, using a strict nomenclature to avoid confusion, and including quantifiable character states where possible.

Winkler *et* al. (1997) position *Tenontosaurus* as the sister group of all remaining higher iguanodontians (i.e. placed as the most basal member), concordant with Sereno, (1986, 1999b), Forster, (1990) and Weishampel and Heinrich, (1992). The placing of *Tenontosaurus* between 'primitive' genera (e.g. *Hypsilophodon*) and more 'derived' genera (e.g. *Iguanodon*, and all higher iguanodonts) is based on several distinguishing characteristics (Winkler *et al.*, 1997):

| Higher than more primitive genera | Lower than more derived genera |
|---|---|
| Ventrally everted premaxilla | Less everted premaxilla |
| Smaller and/or few/no premaxillary teeth | A single anterior maxillary process |
| At least 3 denticles on the predentary | Small or absent brevis shelf on the ilium |

The clade Hypsilophodontidae has previously been defined by the following characters (from Forster 1990, sensu Sereno 1986):

1. Ossified hypaxial tendons in the tail
2. Rod-shaped pre-pubic process
3. Partially ossified sternal segments of the cranial dorsal ribs with corresponding flattening of the caudal edges of the sternal plates
4. Length of humerus either greater to or equal the length of the scapula

Compared to the well-defined Iguanodontia, this set of synapomorphies poorly represents the Hypsilophodontidae clade, and if it genuinely still exists requires thorough re-analysis to determine a broad suite of defining characters. Many of the above require revision: the first is widely synapomorphic amongst ornithischians; the third may relate to ontogeny; and the fourth may relate more to functional variability (e.g. changing stance through ontogenetic growth) than taxonomic status.

Sereno (1986) defines the following synapomorphies of the Iguanodontia clade:

1. Premaxillary teeth absent
2. Ventral premaxillary margin everted
3. Maxilla with paired ventral processes: primitive rostromedial process and new rostroventral process which laps the premaxilla palate ventrally
4. Dentary with parallel dorsal and ventral borders
5. Denticulate predentary bill margin
6. Leaf-shaped denticles on teeth
7. External opening of antorbital fossa small or absent
8. Manus digit III reduced to three phalanges
9. Femur with shallow cranial intercondylar groove and deep caudal intercondylar groove
10. Premaxilla contacts the lachrymal
11. Rostral edge of quadrate notched into a "quadrate foramen"
12. Space separating the ventral margin of the quadratojugal from the jaw articulation
13. Quadratojugal reduced in size relative to the quadrate
14. Diamond-shaped maxillary and dentary crowns
15. Enamel absent from the medial side of maxillary teeth and lateral side of dentary teeth
16. Maxillary crowns narrower axially than opposing dentary crowns
17. Primary ridge on maxillary teeth stronger than primary ridge on dentary teeth
18. Low postacetabular blade and well-developed brevis shelf
19. Proximally placed obturator process on ischium
20. Ischial shaft round in cross-section, decurved and footed

Comparison with this character suite is essential in defining the position of *Tenontosaurus* relative to the Iguanodontia clade. One would assume that a majority of character states must be present for inclusion within a clade.

# 4.0    Systematic Palaeontology

Ornithischia (Seeley, 1887)

Genasauria (Sereno, 1999c) [Ornithopoda and Thyreophera]

Neornithischia (Sereno, 1999c) [Ornithopoda and Marginocephalia]

Ornithopoda (Marsh, 1881)

Euornithopoda (Sereno, 1999a)

*Tenontosaurus tilletti* (Ostrom, 1970)

## 4.1 Revised Diagnosis

1. Vertebral count 12-15-6-61(+)
2. Tail comprising at least 57% total animal length
3. Scapula with straight but with slight proximally and distally expanded caudal margin to approximately 110-120% depth
4. Coracoid with strong, subtriangular sternal process, and coracoid foramen completely closed off from articulation (40% coracoidal width)
5. Forelimb 68% hindlimb length, with equally robust humerus and ulna
6. Humerus dominated by strong, rounded, proximally confined deltopectoral crest; expansion approximately equal to humeral shaft thickness
7. Carpus comprising unfused intermedium, radiale and ulnare
8. Manus width approximately 150% length, with phalangeal formula 2-3-3-2?-2?
9. Ilium with gently ventrally dipping preacetabular process; approximately 60% iliac body length
10. Ilium with dorsoventrally expanded postacetabular blade and rugose caudal margin, and concave ventral border
11. Ilium brevis shelf less than 1.5 times mediolateral width of postacetabular ventral margin
12. Pubis with straight postpubic rod comprising 1.5x pubic body length, and projecting at 100° to prepubic blade
13. Obturator foramen closed off from articulation
14. Prepubic blade laterally compressed, depth non-linear and approximately 75% maximum ischium depth, with distal tip variably expanded
15. Shaft of ischium non-linear, proximal half dominated by lateral ridge, with crescentic obturator process one-third length of shaft distally

16. Femur with hemicylindrical lesser trochanter extending one-fifth femoral length longitudinally, one-third greater trochanter width and separated by thin fissure
17. Pes with phalangeal formula 2-3-4-5-(0)

## 4.2 Material Studied

LL.12275: semi-complete skull; complete vertebral column (atlas, axis, cervical, dorsal, sacral and caudal series) minus most caudal neural spines; pectoral girdle comprising paired scapulae, coracoids and sternal plates; both forearms (humeri, ulnae and radii) complete; manus near-complete with carpus; pelvic girdle comprising both ischia, ilia and pubes; both hindlimbs (femora, fibulae and tibiae), tarsus and pes complete; absence of any ossified tendons.

The degree of preservation and restoration is highly variable: several bones are entirely restored, and are not described here; others are partially restored, and only characters that are fully visible are described, with inferences to concealed morphology. Unfortunately, most vertebrae are in poor condition, variably warped and most do not articulate; whether this status is the result of taphonomy and preservation or due to restoration is unknown, and are thus largely omitted from this study. The significance of this is low, as it is unlikely that any new autapomorphies or synapomorphies would be found and study of such damaged bones is not deemed profitable. The ribs and chevrons are also dismissed from study, partially due to their simple structures that are described elsewhere and also that on dismounting the specimen by the museum in 2004, apparently it was deemed necessary to break them into multiple disassociated pieces.

A full morphological description can be found in appendix I; only the most significant variations from previous studies are provided here. Other slight variations are encountered; however, these subtleties are likely attributed to minor intraspecific or individual distinction or ontogeny as oppose to true interspecific variation and are outlined in appendix 1. It is worth noting that there is currently no distinction between the masculine and feminine tenontosaur states.

Material from the NHM was also examined in the hope of discerning morphological variations between *Tenontosaurus* and other species. Material studied includes *Hypsilophodon foxii*, *Thescelosaurus neglectus*, *Valdosaurus canaliculatus* and *Lesothosaurus diagnosticus.* Descriptions of the various elements studied for comparative material can be found in appendices II-V.

The most striking variation between *Hypsilophodon* and *Tenontosaurus* is of course, the size variation. The specimen R2477 contains a magnificently preserved cranium, and although mostly disarticulated now, was fully articulated and restored during the production of the *Hypsilophodon* monograph by Galton (1974a). The skull formed an extremely brittle assemblage of complex structures; however, most of the internal structure of *Tenontosaurus* is missing or impossible to view in the specimen analysed, thus they are largely disregarded here too. Sutural relationships between several elements are difficult to ascertain, and thus reference to Galton (1974a) is strongly recommended for an accurate reconstruction.

| | LL. 12275 | | YPM 5436 | | YPM 5459 | | OU 11 | | PU 16338 | | AMNH 3040 | | BB1 | |
|---|---|---|---|---|---|---|---|---|---|---|---|---|---|---|
| | Left | Right | Left | Right | Left | Right | Left | Right | Left | Right | Left | Right | Left | Right |
| **Cranium** | | | | | | | | | | | | | | |
| Skull length (premaxilla to occipital condyle) | 335 | | 460 | | 270a | | | | | | | | | |
| Skull width (maximum) | 161 | | | | | | | | | | | | | |
| Skull height (quadrate articular to parietal) | 152 | | 225 | | 120a | | | | | | | | | |
| Maxillary tooth row length | 140 | 133 | 200 | 185 | | | | | | | | | | |
| Orbit: length | 59 | 59 | 80 | 79 | 43 | | | | | | | | | |
| height | 60 | 65 | 62 | 68 | 61 | | | | | | | | | |
| Lateral fenestra: height | 51 | 54 | 58 | 54 | | | | | | | | | | |
| length | 38 | 32 | | 59 | | | | | | | | | | |
| Mandible length | >267 | 316 | 480a | 450 | 260a | | | | | | | | | |
| Mandibular tooth row length | 133 | 144 | 135 | 135 | | | | | | | | | | |
| Maxillary tooth positions | 12 | 12 | 11 | 11 | | | | | | | | | | |
| Dentary tooth positions | 11 | 11 | 12 | 13 | | | | | | | | | | |
| **Vertebral Dimensions** | | | | | | | | | | | | | | |
| Length of odontoid and atlas intercentrum | 26 | | 36+30 | | | | 35+40 | | | | | | 24+25 | |
| Axis, greatest length of centrum | 37 | | 55 | | | | 47 | | | | | | 36 | |
| Axis, greatest height of centrum | 35 | | | | | | 35 | | | | | | 34 | |
| Axis, greatest width of centrum | 32 | | 55 | | | | 46 | | | | | | 32 | |
| 5th cervical centrum: length | 42 | | 53 | | | | 60 | | | | | | 33 | |
| width | 26 | | 53 | | | | 43 | | | | | | 34 | |
| height | 38 | | 55 | | | | 58 | | | | | | 38 | |
| 10th cervical centrum: length | 46 | | | | | | 56 | | | | | | 42 | |
| width | 44 | | | | | | 50 | | | | | | 41 | |
| height | 42 | | | | | | 62 | | | | | | 45 | |
| 5th dorsal centrum: length | 48 | | | | | | | | 37 | | 45 | | | |
| width | 42 | | | | | | 60 | | 35 | | 50.0 | | | |
| height | 42 | | | | | | 50a | | 33 | | 47 | | | |
| neural spine height | | | | | | | 75 | | 65 | | 95 | | | |
| 10th dorsal centrum: length | 48 | | ?63 | | | | 60 | | 39 | | 50 | | | |
| width | 46 | | ?63 | | | | 65 | | 42 | | 55 | | | |
| height | 44 | | ?75 | | | | 63 | | 42 | | 50 | | | |
| neural spine height | >40 | | ?12.5 | | 50a | | 125 | | 74 | | 95 | | | |
| 15th dorsal centrum: length | 51 | | | | 45 | | 55 | | 40 | | 50 | | | |
| width | 51 | | | | 30a | | 80 | | 40 | | 60 | | | |
| height | 43 | | | | | | 85 | | 47 | | 60 | | | |
| Neural spine height | | | | | | | 155 | | >80 | | 115 | | | |
| 1st sacral centrum: length | | | | | | | 60 | | 40 | | 55 | | | |
| width | | | | | | | 65 | | | | 60 | | | |
| height | | | | | | | 90 | | 58 | | 66 | | | |
| 5th sacral centrum: length | 52 | | | | | | 60 | | 38 | | 50 | | 40 | |
| width | 46 | | | | | | 60 | | 40 | | 57 | | 43 | |
| height | 45 | | | | | | 70 | | 45 | | 70 | | 50 | |
| 1st caudal centrum: length | 43 | | | | | | | | 35 | | 48 | | 40 | |
| width | 51 | | | | | | | | 45 | | 62 | | 50.0 | |
| height | 62 | | | | | | | | 48 | | 65 | | 48 | |
| neural spine height | 73 | | | | | | | | 115 | | 140 | | >80 | |
| 5th caudal centrum: length | 42 | | | | | | | | 40 | | 50 | | 45 | |
| width | 51 | | | | | | | | 45 | | 58 | | 56 | |
| height | 52 | | | | | | | | 41 | | 52 | | 48 | |
| Neural spine height | 134 | | | | | | | | 135 | | 172 | | 145 | |
| 10th caudal centrum: length | 42 | | | | | | | | 48 | | 65 | | 57 | |
| width | 50.0 | | | | | | | | 42 | | 52 | | 48 | |
| height | 48 | | | | | | | | 37 | | 48 | | 47 | |
| neural spine height | | | | | | | | | 145 | | 205 | | 140 | |
| 15th caudal centrum: length | 41 | | | | | | | | 55 | | 72 | | 58 | |
| width | 35 | | | | | | | | 41 | | 50 | | 50.0 | |
| height | 40.0 | | | | | | | | 37 | | 48 | | 44 | |
| neural spine height | | | | | | | | | 100 | | 140 | | | |
| 20th caudal centrum: length | 54 | | | | | | | | 52 | | 72 | | | |
| width | 37 | | | | | | | | 38 | | 47 | | | |
| height | 45 | | | | | | | | 35 | | 47 | | | |
| neural spine height | | | | | | | | | 80a | | 90 | | | |
| 25th caudal centrum: length | 58 | | | | | | | | 49 | | 68 | | | |
| width | 33 | | | | | | | | 32 | | 43 | | | |
| height | 40.0 | | | | | | | | 32 | | 45 | | | |
| Neural spine height | | | | | | | | | | | 55 | | | |
| 30th caudal centrum: length | 57 | | | | | | ?55 | | 46 | | 65 | | | |
| width | 33 | | | | | | ?32 | | 25 | | 34 | | | |
| height | 36 | | | | | | ?33 | | 28 | | 38 | | | |
| 35th caudal centrum: length | 50.0 | | | | | | ?50 | | 42 | | 57 | | | |
| width | 29 | | | | | | ?26 | | 20 | | 27 | | | |
| height | 32 | | | | | | ?31 | | 25 | | 35 | | | |
| 40th caudal centrum: length | | | | | | | | | 38 | | | | | |
| width | 29 | | | | | | | | 17 | | | | | |
| height | | | | | | | | | 21 | | | | | |
| 45th caudal centrum: length | 44 | | | | | | | | 33 | | 47 | | | |
| width | 24 | | | | | | | | 13 | | 21 | | | |
| height | 26 | | | | | | | | 18 | | 27 | | | |
| 50th caudal centrum: length | 21 | | | | | | | | 32 | | | | | |
| width | 9 | | | | | | | | 12 | | | | | |
| height | 10.0 | | | | | | | | 16 | | | | | |
| 55th caudal centrum: length | 16 | | | | | | | | 25 | | | | | |
| width | 4 | | | | | | | | 8 | | | | | |
| height | 7 | | | | | | | | 12 | | | | | |
| 58th caudal centrum: length | 8 | | | | | | | | 25 | | | | | |
| width | 5 | | | | | | | | 7 | | | | | |
| height | 9 | | | | | | | | 11 | | | | | |

| Pectoral Girdle and Forelimb Dimensions | | | | | | | | | | | | | | |
|---|---|---|---|---|---|---|---|---|---|---|---|---|---|---|
| Scapula: length | 347 | 372 | 525 | 545 | 330 | 345 | 450 | 460 | | | | | | |
| maximum width | 107 | 111 | 180 | 182 | 64 | | 175 | 170 | | | | | | 80 |
| Coracoid: length | 114 | 103 | 200 | 135 | | | 220 | 220 | 125 | 125 | | | | 125 |
| height | 115 | 125 | 145 | 140 | | | 150 | 170 | 105 | 102 | | | | 85 |
| Sternal length | >155 | >157 | 180 | | | | 235 | 220 | 130 | 131 | | | | |
| Humerus: length | 307 | 318 | 435 | 465 | | | 442 | 440 | | | | | | 272 |
| proximal width | 100 | 93 | | 135 | | | 140 | 155 | | | | | | 92 |
| distal width | 79 | 78 | 120 | 115 | 35 | 85 | 113 | 130 | | | | | | 75 |
| deltopectoral crest length | 131 | 124 | 235 | 265 | | | 250 | 250 | | | | | | 135 |
| shaft circumference | 136 | 135 | | | | | 158 | 160 | | | | | | 100 |
| Ulna: length | 253 | 255 | 340 | 300a | 225 | | 340 | 340 | | | | | | 215 |
| proximal width | 75 | 77 | | 90 | 85 | | 110 | 100 | | | | | | 70 |
| distal width | 45 | 44 | | | | | 65 | 65 | | | | | | 41 |
| Radius: length | 239 | 242 | 320 | 260a | 210 | | 330 | 330 | | | | | | 205 |
| proximal width | 51 | 51 | 57 | 75 | 40 | | 70 | 73 | | | | | | 45 |
| distal width | 46 | 47 | 63 | | | | 80 | 70 | | | | | | 45 |
| Length: metacarpal I | 41 | 27 | | | 35 | | 67 | 62 | | | | | | |
| metacarpal II | 63 | >32 | | | 56 | | 92 | 93 | | | | | 62 | |
| metacarpal III | 67 | 64 | | | 72 | | 94 | | | | | | 66 | |
| metacarpal IV | 49 | 47 | | | 74 | | 71 | 72 | | | | | 48 | |
| metacarpal V | 23 | 37 | | | 42 | | 53 | | | | | | 34 | |
| digit I | >83 | >75 | | | | | 80 | 80 | | | | | | |
| digit II | >95 | 107 | | | | | | 90 | | | | | 60a | |
| digit III | 128 | 107 | | | | | | | | | | | 56 | |
| digit IV | >49 | ?64 | | | | | 38 | | | | | | 20 | |
| digit V | >44 | ?62 | | | | | 30a | 30 | | | | | | |

| Pelvic Girdle and Hindlimb Dimensions | | | | | | | | | | | | | | |
|---|---|---|---|---|---|---|---|---|---|---|---|---|---|---|
| Ilium: length | 232 | 219 | | | | | 510 | 280 | 285 | 420.0 | 420 | | | |
| height at acetabulum | 84 | 84 | | | | 32a | | 110 | 70 | 65 | 87 | 90 | 72 | |
| Ischium: length | 435 | 438 | | | | | 565 | 555 | 340 | 355 | | 460 | 295 | |
| distal width | 48 | 51 | | | | | 65 | 70 | 50 | 50 | | 70 | | |
| Prepubic blade: length | >120 | 164 | | | | | 170 | 180 | 132 | 143 | | 232 | | |
| height | 34 | 34 | | | | | 45 | 47 | 26 | 27 | | 52.0 | | |
| Postpubic rod length behind obturator foramen | - | ?116 | | | | | | >165 | | | | >270 | | |
| Femur: length | 451 | 447 | | | 415 | | 570 | 580 | 355 | 355 | | 480 | | |
| proximal width | 107 | 102 | | | | | | 120 | 95 | 95 | | 135 | | |
| distal width | 117 | 115 | | | 125 | | 165 | 150 | 107 | 110 | | >140 | 105 | |
| distance from femur head to 4th trochanter | 200 | 197 | | | | | >230 | 307 | 182 | 186 | | 265 | | |
| minimum circumference | 201 | 197 | | | | | 230 | 225 | 140 | 140 | | 202 | 145a | |
| Tibia: length | 379 | 395 | | | 500 | | | 535 | 350 | 355 | | | 350 | |
| proximal width | 118 | 123 | | | | | | | 105 | 100 | | | 100 | |
| distal width | 114 | 118 | | | 105 | | | | 103 | 110 | | | 108 | |
| Fibula: length | 369 | 384 | | | 432 | | | 510a | 317 | 328 | | | 325 | |
| proximal width | 70 | 53 | | | 50a | | | | 68 | 62 | | | 70 | |
| Astragalus: length | 70 | 70 | | | | | | | 33 | 32 | | | 25 | |
| width | 47 | 46 | | | | | | | 59 | 61 | | | 55 | |
| Calcaneum width | 35 | 35 | | | | | | | 44 | 42 | | | | |
| Length: metatarsal I | 33 | 36 | | | | | 140 | | 86 | 85 | | | 90.0 | |
| metatarsal II | 127 | 139 | | | | | 185 | | 122 | 125 | | | 130 | |
| metatarsal III | 151 | 155 | | | | | 215 | | 137 | 140 | | | 145 | |
| metatarsal IV | 127 | 129 | | | | | 180 | | 113 | 115 | | | 118 | |
| Length: digit I | 111 | >98 | | | | | >140 | | 95 | 95a | | | 105 | |
| digit II | 138 | 138 | | | | | 180 | | 125 | 135 | | | 140 | |
| digit III | 162 | 163 | | | 195 | | 215 | | 145 | 148a | | | 170 | |
| digit III | 139 | 132 | | | 130 | | 135 | | 150 | 150 | | | 150 | |

Table 2 – Comparative measurements between LL.12275 and those described by Ostrom (1970) (including both holotype and paratype). Measurements indicate that YPM 5459 is also of sub-adult stage, although variations within the appendicular skeleton are significant, perhaps suggesting intraspecific variation either in size or allometric growth, or perhaps interspecific variation (hence, incorrect diagnosis). Measurements in millimetres.

| | | |
|---|---|---|
| Forelimb length (excluding carpus and manus) | 566 | 573 |
| Hindlimb length (excluding tarsus and pes) | 830 | 842 |
| Humerus/ulna | 1.19 | 1.25 |
| Femur/tibia | 1.19 | 1.13 |
| 3rd metatarsal+tibia/femur | 1.18 | 1.23 |
| Forelimb/hindlimb | 0.68 | 0.68 |

Table 3 – Several ratios of appendicular aspects of LL.12275; data would appear to suggest a bipedal mode of life, although this is largely speculatative in favour of numerical biomechanical modelling. Measurements in millimetres.

**4.31 Skull**

Currently the only skull of *T. tilletti* described is that of the paratype, YPM 5456, by Ostrom (1970). The following is a comparison of a mostly-complete cranium with this specimen, taking into account ontogenetic variability and also the fact that several of the bones in the upper skull are missing, represented purely by restorative material.

Skull length approximately 1.5 times height, revealing broad form in lateral view somewhat similar to *Hypsilophodon foxii* and *Iguanodon* (fig. 7); posterior margin relatively straight and vertical; dorsal margin post-nares posteriorly horizontal; ventral margin straight between quadrate and premaxilla; sub-rectangular to sub-circular orbit dominates lateral view (figs. 9a, 10a, b); exact two-dimensional profile variably obscured by restoration of pre-orbital 'cheek' elements (maxilla, lachrymal, prefrontal and supraorbital all largely absent or reconstructed).

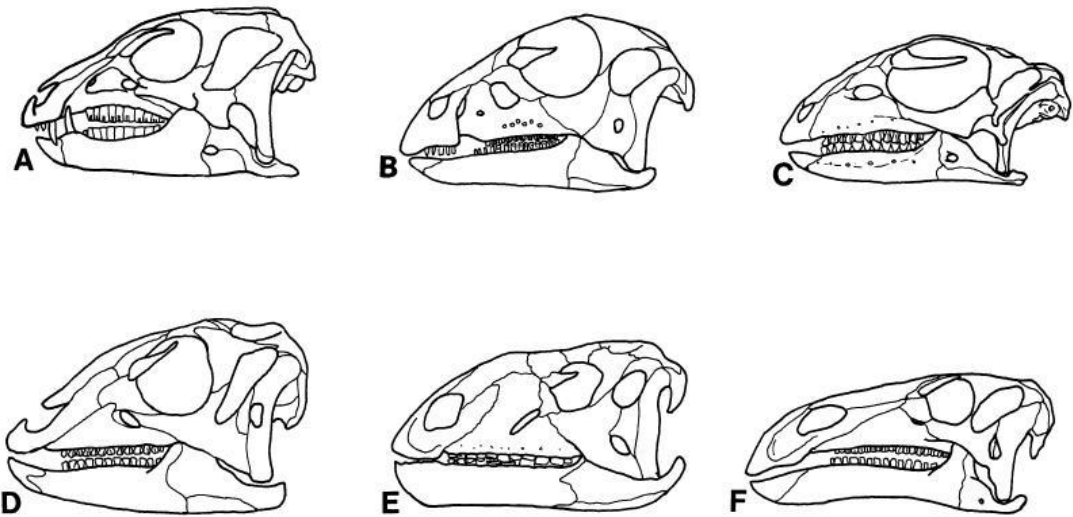

Figure 7 – Various ornithischian skulls; **A:** *Heterodontosaurus tucki* **B:** *Hypsilophodon foxii* **C:** *Gasparinisaura cincosaltensis* **D:** *Dryosaurus* **E:** *Tenontosaurus tilletti* **F:** *Iguanodon* (Coria and Salgado, 1996).

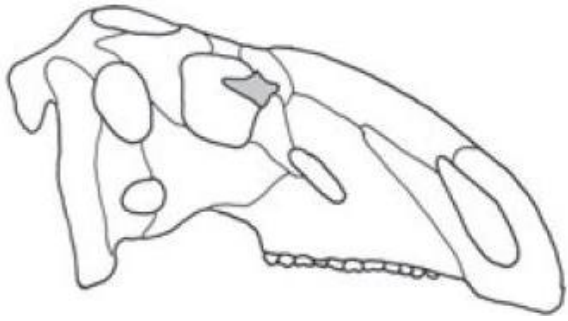

Figure 8 – Reconstructed sketch of *Tenontosaurus tilletti* cranium; shaded element is the palpebral projecting into the orbit, Maidment and Porro, (2010). Note differences to figure 7, e.g. extent of external nares and form of quadratojugal.

Lateral temporal fenestra varies: left side (sinistral) comprises one single fenestra due to absent quadratojugal and partial restoration; on right side (dextral), lateral temporal fenestra single well-developed oval structure elongated dorsoventrally - no second auxiliary fenestra on either flank as described by Ostrom (1970); can be explained by restoration on sinistral side obscuring all detail, but dextrally definitely no development - quadratojugal instead forms continuous suture with jugal.

Upper border of external nares (narial opening) runs parallel to dorsal margin of premaxilla, - principally elongate oval form; character much more prominent than described by Ostrom (1970), but may partially or entirely result from reconstruction. However, upper projection of premaxilla appears to support elongate form - no apparent suture visible with absent nares, resulting in form similar to fig. 8.

Antorbital fenestra oriented obliquely with long axis directed posterodorsally-anteroventrally paralleling external nares; small, oval and slit-like opening, postulated to occur entirely within maxilla (sinistral fenestra is entirely restored, and dextral with restored posterior margin).

Skull wedge-shaped in dorsal aspect (figs. 9c, 10d) expanding posteriorly producing goblet-like form; dorsal premaxilla narrow leading rostrally into well-rounded 'beak', becoming increasingly curved medially; posterior margin broadly convex (excluding basioccipital); nares, prefrontals and dorsal-most fraction of premaxilla absent. Form of supratemporal fenestrae largely dictated by parietals, developing into transversely widened ovals, distinctly different to those depicted by Ostrom (1970). Lower jaw near-complete (distal tips absent); multiple dentary teeth well preserved – both as described in Ostrom (1970) – see appendix 1.

Dextral side of skull near complete, with everything posterior to orbit well preserved; sinistral side much poorer condition post-nares: quadratojugal absent; partial jugal, postorbital, frontal, quadrate and maxilla; complete paraoccipital process, squamosal and parietal. No supraorbitals observed, conversely to Ostrom (1970) and Maidment and Porro (2010).

*Premaxilla*: Ventral border smoothly rounded (figs. 9b, 10c) - no identifiable suture with anteriorly protruding nares described by Ostrom (1970). Resultantly premaxillary projections appear to entirely envelop and create border for external nares, excluding perhaps most posterodorsal edge.

*Quadratojugal*: primary difference to holotype absence of "small auxiliary temporal fenestra" of Ostrom (1970). Instead extends completely from ventral margin of lateral temporal fenestra forming cranial ventral border.

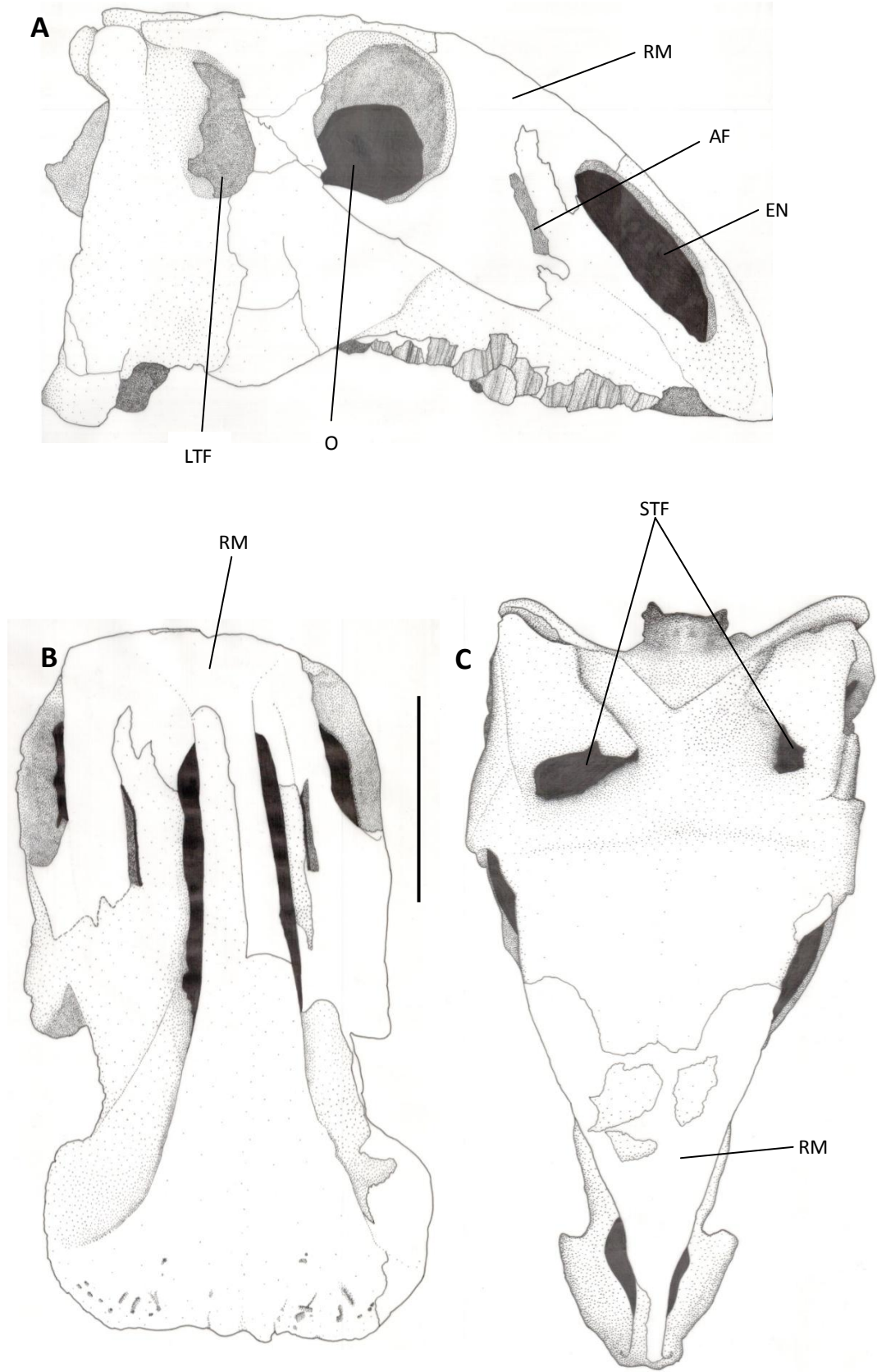

Figure 9 – Cranium in **A:** Lateral aspect **B:** Rostral aspect **C:** Dorsal aspect. Scale = 10cm. Unshaded areas represent restorative material. Note variations to figures 7 and 8 (e.g. size and shape of external nares, lack of auxiliary temporal fenestra). AF – antorbital fenestra, EN – external nares, LTF – lateral temporal fenestra, O – orbit, RM – restorative material, STF – supratemporal fenestra.

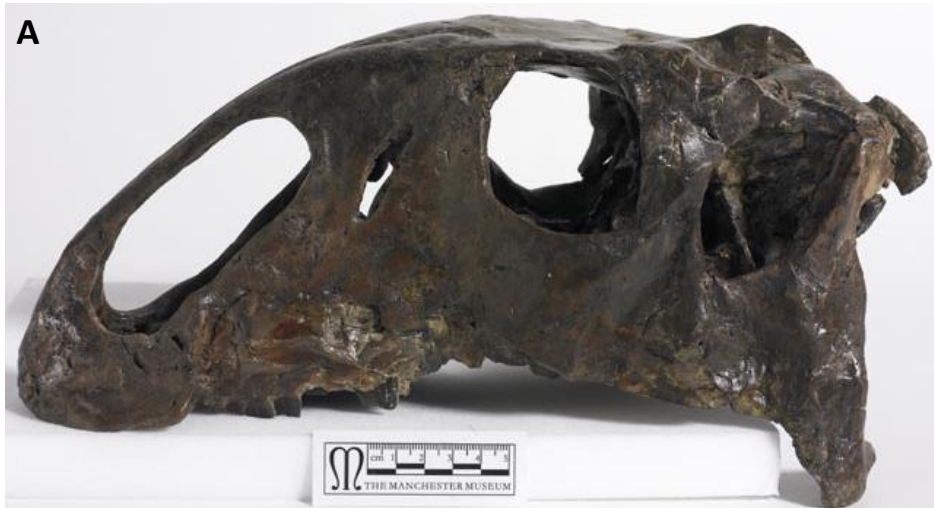

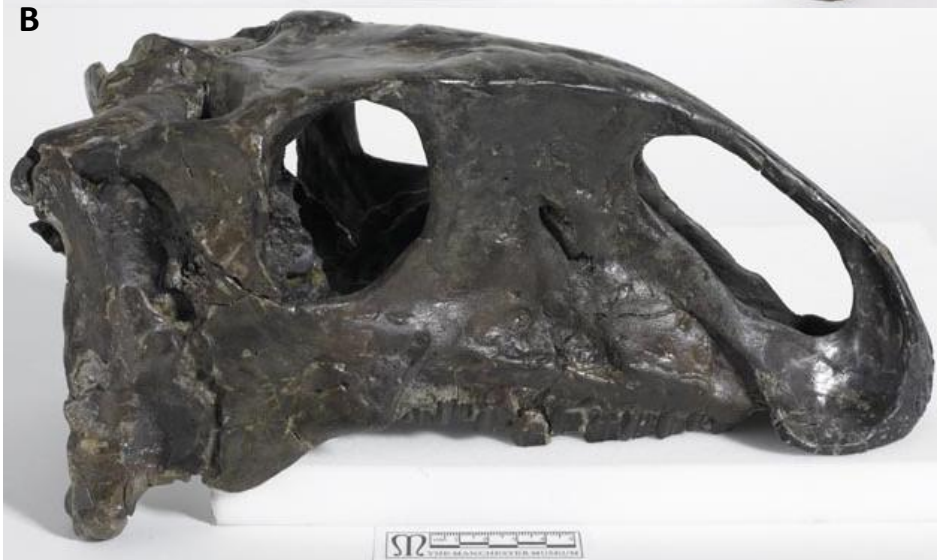

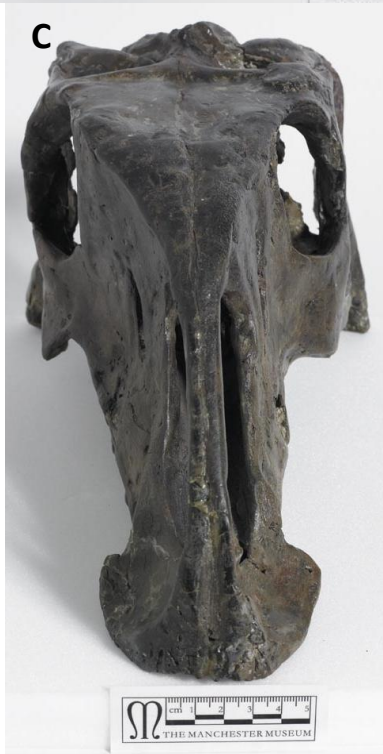

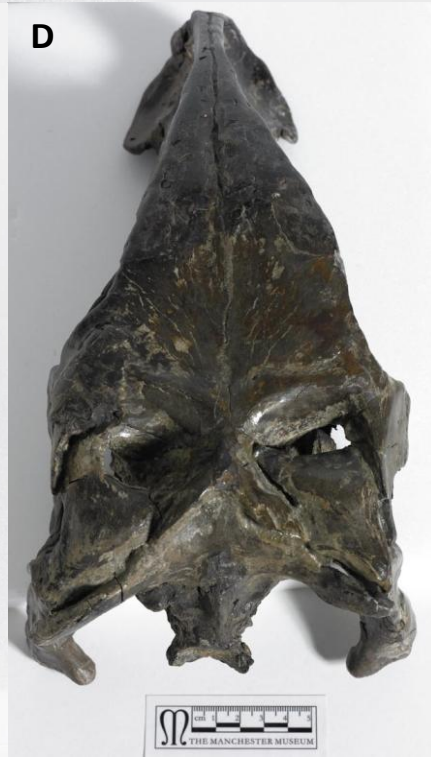

Figure 10 – LL.12275 cranium in **A:** left lateral **B:** right lateral **C:** rostral **D:** dorsal aspects. Scale in centimetres. Note *in situ* preservation of maxillary teeth, and variations in supratemporal fenestra morphology. Photographs courtesy of the UoM.

**A**

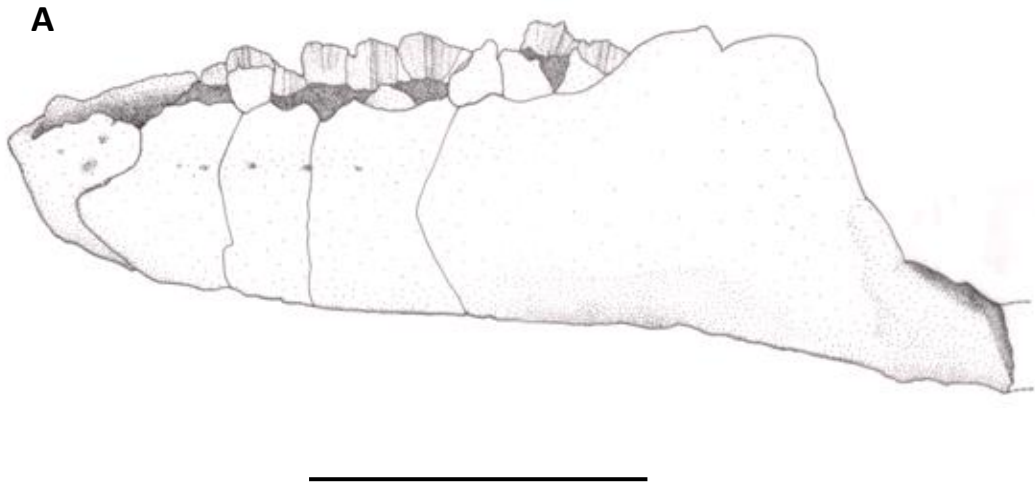

**B**

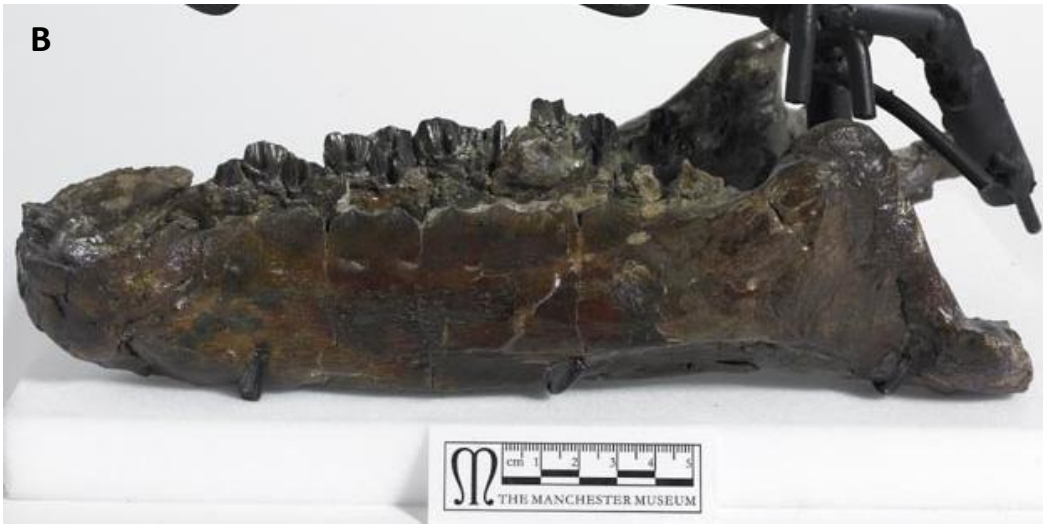

Figure 11 – **A:** sketch of lower jaw (left lateral aspect) **B:** Mostly complete lower jaw. Note poorly preserved labial tooth surfaces, and well-preserved medial surfaces (dextral side). Scale bar = 10cm. Photograph courtesy of the UoM.

*Lower* jaw: No visible surangular foramen as in Ostrom (1970); splenial not visible on medial dentary surface. "Glenoid" of Ostrom (1970) possibly refers to unfused foramen on cranium posterior at surangular-quadrate junction, possibly representing region of articulation. Articular absent, as is the "retroarticular surface" (possibly homologous to the "coronoid process"); posterior (distal) tips absent on both sides of jaw – appear transversely expanded and sub-oval in cross-section, composed mostly of surangular.

*Teeth*: enamelled only on medial surface ('iguanodont'-type) – lateral sides poorly preserved and weathered; mostly large, elongated-bowl forms (highly variable, e.g. irregular pentagonal); increase in size posteriorly; fit into small convex grooves imprinted into dentary and maxilla (alveoli); no premaxillary teeth; single large median ridge dominates, with several (4-5) smaller, faint, subsidiary parallel ridges on either side, forming denticulate crown margin. Oval form (dorsal aspect); possible wear facets on back teeth inclined laterally; medial surfaces mildly convex away from dominant ridge. No cingulum as in *Hypsilophodon*; crowns not diamond-shaped as in higher ornithopods; no visible root structures, or evidence of tooth replacement.

### 4.32 Vertebral Column

The vertebral column appears 100% complete, with every bone present (table 4; fig. 14). Unfortunately, the status of these bones is substandard; multiple elements are missing on nearly every vertebra, and the associated chevrons and ribs are completely disassembled and disassociated. There is no trace of hypaxial or epaxial ossified tendons, which essentially originally defined this genus. Many of the vertebrae are so incomplete and disassembled, that even a generic assignment is problematic, and it is suspected that many are sourced from different specimens.

The vertebral formula here is defined as 12-15-6-61(+), which is slightly different to 12-16-5-60(+) of Forster (1990) and 12-16-5-59(+) of Ostrom (1970). This difference can possibly be attributed to the interpretation here of a dorsosacral or caudosacral as a full sacral vertebra, in which case the identity would conform to previous studies. However, the sacrals here have not been seen *in situ* with the pelvic girdle, and the questionable condition makes configuration impossible, so a conclusive statement on the identity is only tentatively proposed.

Forster (1990) and Ostrom (1970) provide a comprehensive account of the vertebral column, and after examination of the better-preserved constituents, no significant distinctions were observed. As mentioned previously, the vertebral morphology does not provide detail of defining characters, and therefore a full analysis simply reiterating previous studies is deemed valueless. General points are mentioned simply to define columnar variation.

The cervical assemblage (including the atlas and axis (fig. 12B)) is in variable condition; most processes (pre- and postzygapophyses, parapophyses) are absent or poorly preserved, and often the 3-dimensional structure is variably warped or distorted. Commonly, elements have been displaced then adhered back improperly, or simply left separate, so a generalised structure or longitudinal variation analysis would be difficult, if required.

The dorsal series are more robust than the cervicals (fig. 12A), and exhibit variably better preservation than the cervicals as a result. The neural spines are greatly reduced compared to the sacral-most cervicals, often less than half the height. The distal tips of the spines are rarely preserved, but appear sub-rectangular and distally expanded compared to the more fin-like tips of the cervical neural spines. As with the cervicals however, too many elements are absent or displaced, and no variation with previous descriptions is noted. One point of particular interest though is that several of the vertebrae exhibit extremely strong fusion with each other at the centra, with no articulation possible.

Ostrom (1970) defines the first sacral as "the first vertebra bearing a lateral process (transverse process) that is not borne entirely on the neural arch", and the final sacral as the "most posterior segment bearing distally expanded lateral projections (sacral ribs) for extensive contact with the ilium". Unfortunately, the designated six sacrals are so poorly preserved, observation of any of these features is insurmountable, and the exact number of sacrals therefore can only be crudely speculated. Galton (1974a) stresses the 7[th] point of his diagnosis that the sacral count can vary intraspecifically within the Ornithopoda, which may account for the possible additional sacral observed here in *Tenontosaurus* (fig 13).

Of the caudal series, all but the most proximal three are missing the neural spines, although on the most distal third it is likely that they only occur as largely inconspicuous expansions or are morphologically absent. Where preserved, the neural spines are strongly curved with parallel sides. They are very narrow transversely and anteroposteriorly expanded.

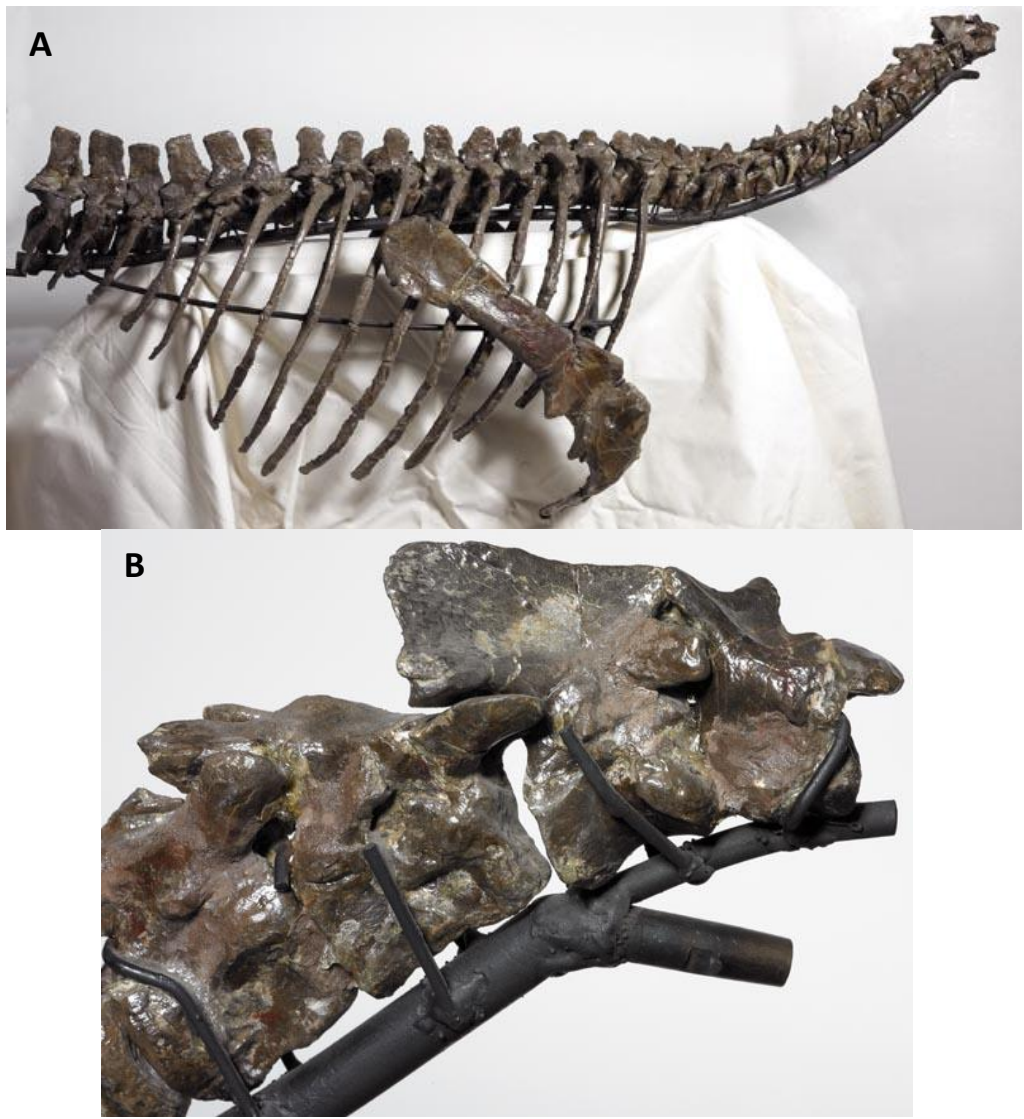

Figure 12 – **A:** cervical and dorsal series (plus pectoral girdle and ribs) mounted and articulated. Condition visible here exponentially better than state post-storage; condition implies all elements are from a single specimen. Field of view approximately 150cm. **B:** atlas and axis mounted (see Forster (1990) for description). Field of view approximately 15cm. Photographs courtesy of the UoM.

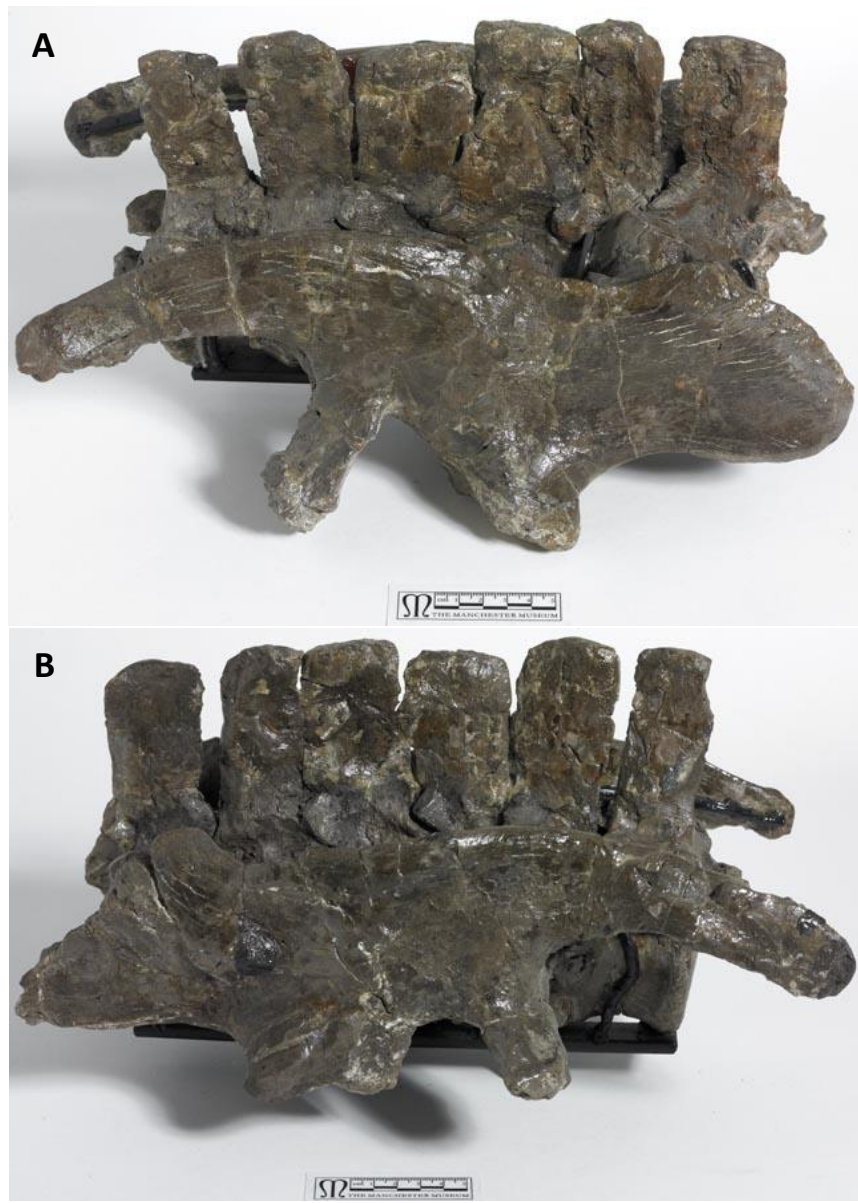

Figure 13 – Restored sacrum, with 6 sacral vertebrae and both complete ilia. **A:** left lateral aspect **B:** right lateral aspect. Scale in centimetres. Note variations in caudal margin of postacetabular blade. Photographs courtesy of the UoM.

| Series | Vertebra | Length (mm) | Width (mm) | Height (mm) |
|---|---|---|---|---|
| Atlas | 1 | 26 | 31 | 21 |
| Axis | 2 | 37 | 35 | 32 |
| Cervical | 3 | 45 | 21 | 38 |
| | 4 | 43 | 39 | 46 |
| | 5 | 42 | 26 | 38 |
| | 6 | 44 | 38 | 37 |
| | 7 | 45 | 36 | 37 |
| | 8 | 50 | 31 | 38 |
| | 9 | 44 | 36 | 42 |
| | 10 | 46 | 44 | 42 |
| | 11 | 45 | 30 | 50 |
| | 12 | 47 | 31 | 49 |
| Dorsal | 13 | 44 | 28 | 45 |
| | 14 | 55 | 34 | 47 |
| | 15 | 45 | 42 | 52 |
| | 16 | 48 | 40 | 47 |
| | 17 | 48 | 42 | 42 |
| | 18 | 48 | 39 | 44 |
| | 19 | 42 | 43 | 43 |
| | 20 | 44 | 35 | 43 |
| | 21 | 45 | 30 | 46 |
| | 22 | 48 | 46 | 44 |
| | 23 | 49 | 48 | 49 |
| | 24 | 50 | 46 | 45 |
| | 25 | 52 | 56 | 57 |
| | 26 | 50 | 51 | 47 |
| | 27 | 48 | 54 | 45 |
| | 28 | 55 | 49 | 54 |
| Sacral | 29 | | | |
| | 30 | 49 | 51 | 51 |
| | 31 | 44 | 37 | 51 |
| | 32 | 52 | 46 | 45 |
| Sacral | 33 | 48 | 43 | 46 |
| | 34 | 43 | 51 | 62 |
| | 35 | 46 | 61 | 57 |
| Caudal | 36 | 43 | 54 | 61 |
| | 37 | 43 | 50 | 54 |
| | 38 | 43 | 51 | 52 |
| | 39 | 43 | 47 | 54 |
| | 40 | | | - |
| | 41 | 49 | 43 | 49 |
| | 42 | 48 | 46 | 51 |
| | 43 | 42 | 50 | 48 |
| | 44 | 44 | 48 | 40 |
| | 45 | 46 | 53 | 41 |
| | 46 | 40 | 53 | 35 |
| | 47 | 39 | 49 | 32 |
| | 48 | 41 | 35 | 40 |
| | 49 | 45 | 42 | 53 |
| | 50 | 45 | 44 | 44 |
| | 51 | 39 | 42 | 45 |
| | 52 | 49 | 29 | 40 |
| | 53 | 54 | 37 | 45 |
| | 54 | 56 | 36 | 42 |
| | 55 | 58 | 36 | 38 |
| | 56 | 58 | 36 | 39 |
| | 57 | 60 | 30 | 37 |
| | 58 | 58 | 33 | 40 |
| | 59 | 57 | 31 | 39 |
| | 60 | 58 | 30 | 34 |
| | 61 | 59 | 33 | 37 |
| | 62 | 55 | 35 | 41 |
| | 63 | 57 | 32 | 36 |
| | 64 | 58 | 33 | 36 |
| Caudal | 65 | 52 | 34 | 34 |
| | 66 | 50 | 30 | 31 |
| | 67 | 49 | 30 | 30 |
| | 68 | 50 | 29 | 32 |
| | 69 | 50 | 25 | 32 |
| | 70 | 50 | 30 | 25 |
| | 71 | 47 | 26 | 30 |
| | 72 | 48 | 21 | 24 |
| | 73 | 46 | 29 | 31 |
| | 74 | | 28 | 24 |
| | 75 | | | |
| | 76 | | | |
| | 77 | 43 | 24 | 26 |
| | 78 | 44 | 24 | 26 |
| | 79 | | 28 | 28 |
| | 80 | | | |
| | 81 | | | |
| | 82 | | | |
| | 83 | 21 | 9 | 10 |
| | 84 | | | |
| | 85 | 18 | 13 | 11 |
| | 86 | | | |
| | 87 | | | |
| | 88 | 16 | 5 | 9 |
| | 89 | 11 | 5 | 7 |
| | 90 | 11 | 6 | 7 |
| | 91 | 10 | 6 | 7 |
| | 92 | 9 | 5 | 7 |
| | 93 | 8 | 5 | 7 |
| | 94 | 6 | 4 | 6 |
| | 95 | 6 | 3 | 5 |

Table 4 – Measurements of the vertebral column of LL.12275: length, width and height of individual centra to ascertain any possible pattern within the series. See associated graphs below for summary.

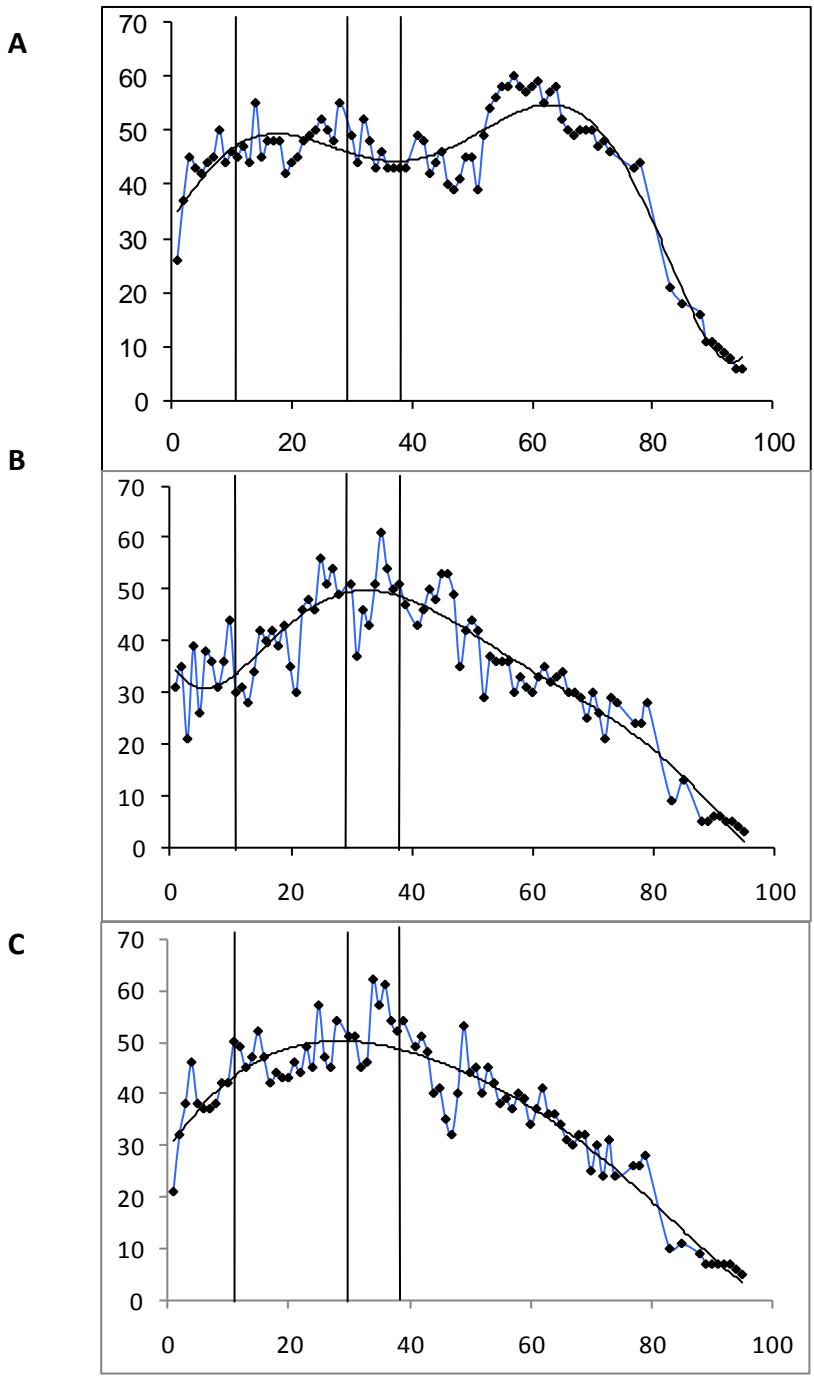

Figure 14 – Graphs illustrating variety in vertebral centra dimensions (y-axis). **A:** length (mm) **B:** width (mm) **C:** height (mm). Centra number along x-axis (0 = atlas, 95 = terminal caudal vertebra). Vertical lines represent individual series (see Table 4 for summary). Revealed is the lack of smooth congruity between adjacent vertebrae, either as the result of taphonomy, mishandling, or genuine variation.

The assignment of these vertebrae to a particular organism is non-specific, as Winkler *et al*. (1997) note that in the entire vertebral series, there are no variations between *T. tilletti* and *T. dossi*; therefore classification based on vertebral morphology is not attempted here.

As has been mentioned previously, the chevrons and ribs do not warrant description, and such a task would be futile considering the status of current preservation. Ossified tendons are not currently present, and even if they were it would be highly unlikely they would yield any significant features than those described before (Ostrom, 1970; Forster, 1990; Winkler *et* al. 1997; Organ and Adams, 2005; Organ, 2006). Their initial presence is only conjecturally insinuated by the uncommon presence of longitudinal grooves on the centra of several vertebrae; this could however represent a taphonomic, preservation or mortality process.

**4.33 Pectoral Girdle**

The pectoral girdle constitutes two scapulae, two coracoids and two sternal plates; the pairs exhibit near-perfect symmetry, with minor differences occurring as a direct result of either morphological contrast or preservation.

*Sternum*: not as broad or thick as those imaged in figure 9 of Forster (1990) (fig. 15) - inconsistency may be function of ontogeny; it is logical that these elements become progressively structurally solid (ossified) and integrated with increasing age and growth.

Maximum length achieved approximately 2.5-3 times maximum height (coincident with *Camptosaurus* (Dodson and Madsen Jr., 1981); minor proportional variation to that described by Forster (1990); therefore one can speculate that lateral broadening is a function of ontogeny. With exception to the discrepancies outlined above, the sternal plates are otherwise identical to those of Forster (1990). Generally resembles sternum of *Heterodontosaurus*, *Hypsilophodon*, *Thescelosaurus*, *Othnielia*, *Parksosaurus* and *Dryosaurus* (Dodson and Madsen Jr., 1981; Galton, 1981).

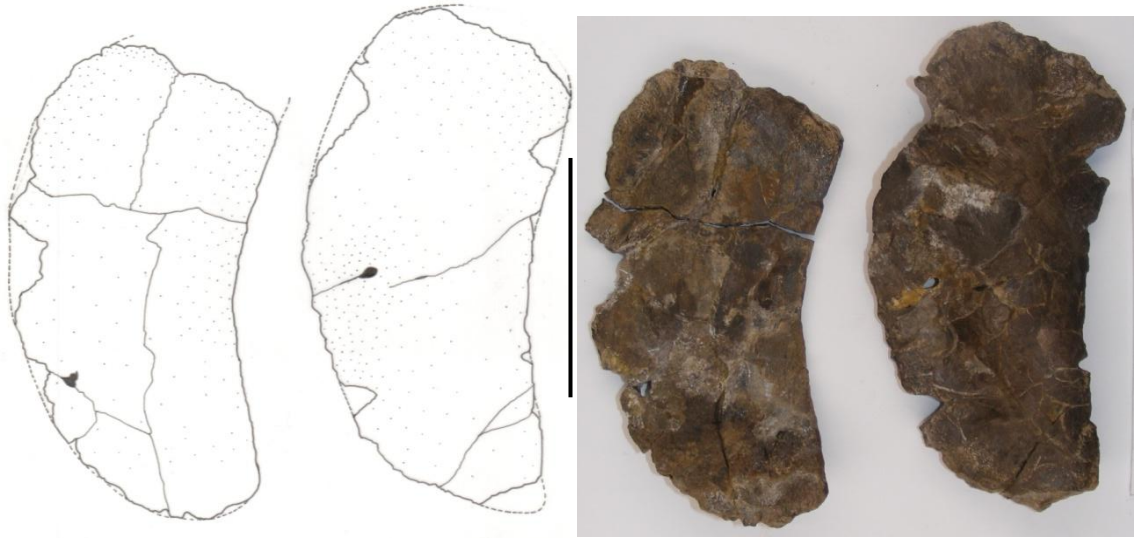

Figure 15 – Paired sternal plates; both are moderately distorted, and defining the left and right elements is problematic, but not of critical importance. Scale = 10cm.

*Scapula*: "constricted neck" of Forster (1990) requires explanation, as dorsoventral thickness appears uniform, unless it refers to entire section distal to respectively expanded glenoid. Cranial edge not straight as Forster (1990) – actually moderately concave in lateral view due to unsymmetrical divergence at either extremity (figs 16a, b). Otherwise, cranial and caudal borders of distal two-thirds parallel. "Acute caudodorsal angle" of dorsal border described by Forster (1990) not observed - slightly obtuse or at least orthogonal. Based on observations from Forster (1990) however, placement of morphology between adult and juvenile confirmed (i.e. adolescent or subadult form), due to non-symmetrical dorsal borders in both scapulae. "Slight flaring" of Forster (1990) not observed - thickness uniform except for mild cranial convergence, conforming to slight cranial thinning at ventral-most tip of cranial border. Glenoid surface does not appear rugose as Forster (1990) describes - instead rather smooth or variably 'pock-marked'.

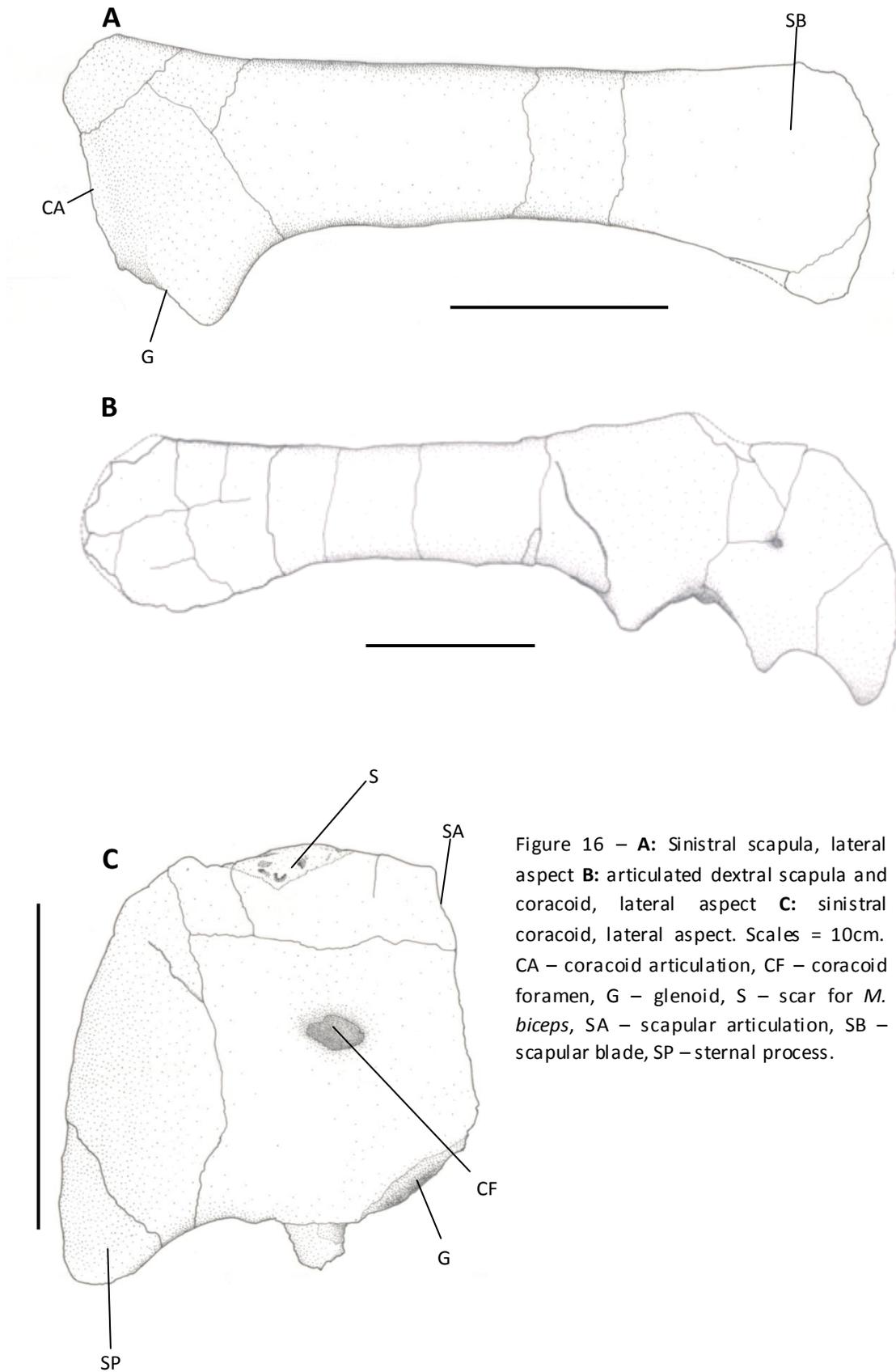

Figure 16 – **A:** Sinistral scapula, lateral aspect **B:** articulated dextral scapula and coracoid, lateral aspect **C:** sinistral coracoid, lateral aspect. Scales = 10cm. CA – coracoid articulation, CF – coracoid foramen, G – glenoid, S – scar for *M. biceps*, SA – scapular articulation, SB – scapular blade, SP – sternal process.

*Coracoid*: several variations are apparent to the description by Forster (1990) (figs. 16b, c): glenoid fossa equally distributed between scapula and coracoid, and not observed being "laterally canted"; sloping more medial with resultant profile appearing as a 180° rotation of associated scapular surface of fossa; cranial and ventral borders slightly rugose as Forster (1990) mentions although not thickened as author describes. Forster (1990) also states "the body of the coracoid is bent perpendicular to the long axis of the scapula" (not observed); statement also somewhat ambiguous, as perpendicular angle could be anywhere within 360° of a one-dimensional axis. "Deep medial concavity" also not observed although slight concavity perhaps present but largely obscured by previously mentioned ridge.

Statement "the [coracoid] foramen courses caudolaterally as it passes medially through the coracoid" of Forster (1990) perhaps slightly misleading as it implies a continuous opening whereas foramina are completely enclosed and not connected; morphology on either surface is arbitrary, possibly reflecting distinct individual functions opposing foramina play.

**4.44 Forelimb**

*Humerus*: no noticeable variation to Forster (1990) (figs. 17, 18, 19).

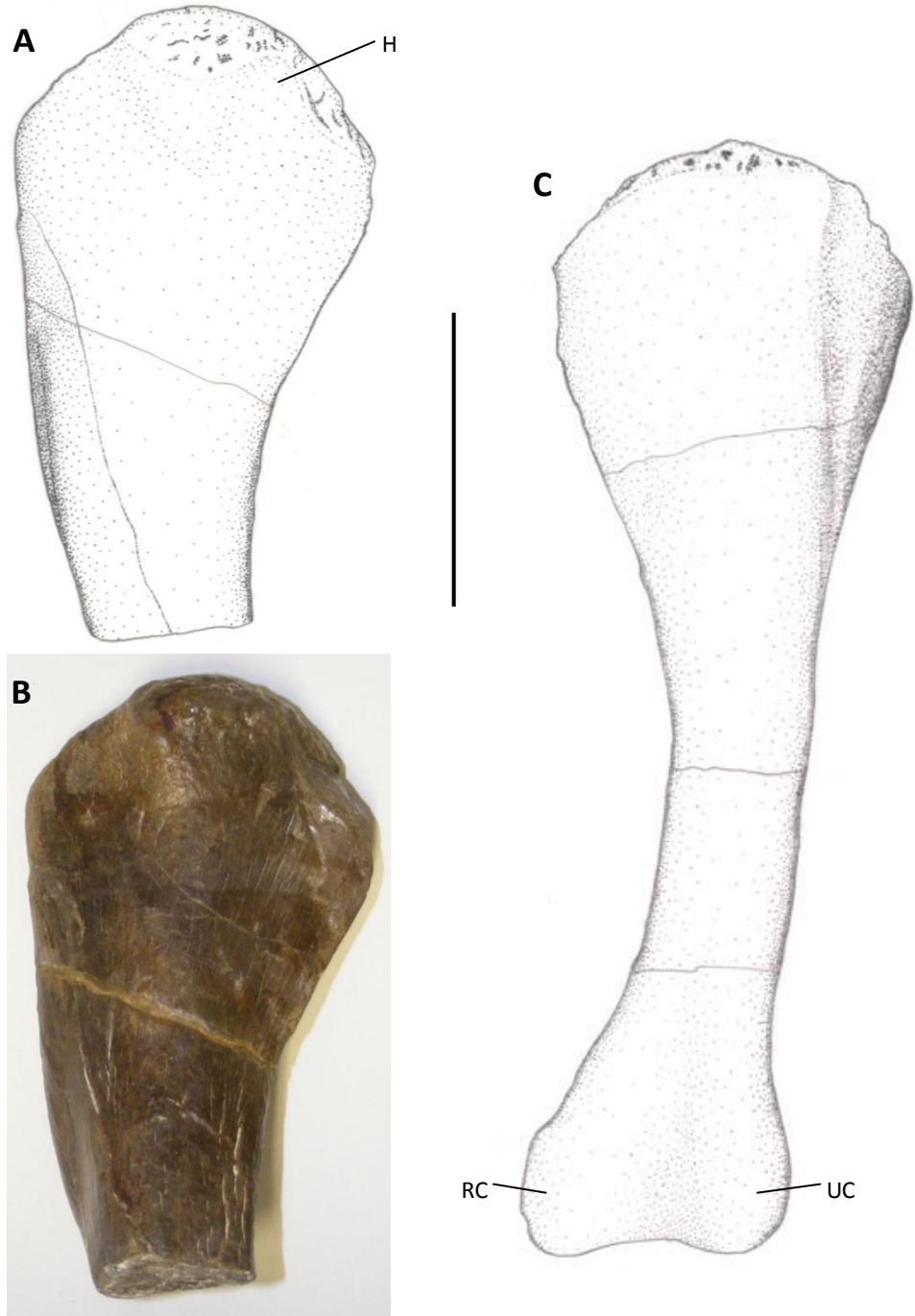

Figure 17 – **A:** Sinistral humerus, proximal end, caudal aspect **B:** as A **C:** dextral humerus, caudal aspect. Scale = 10cm. H – humeral head, RC – radial condyle, UC – ulnar condyle.

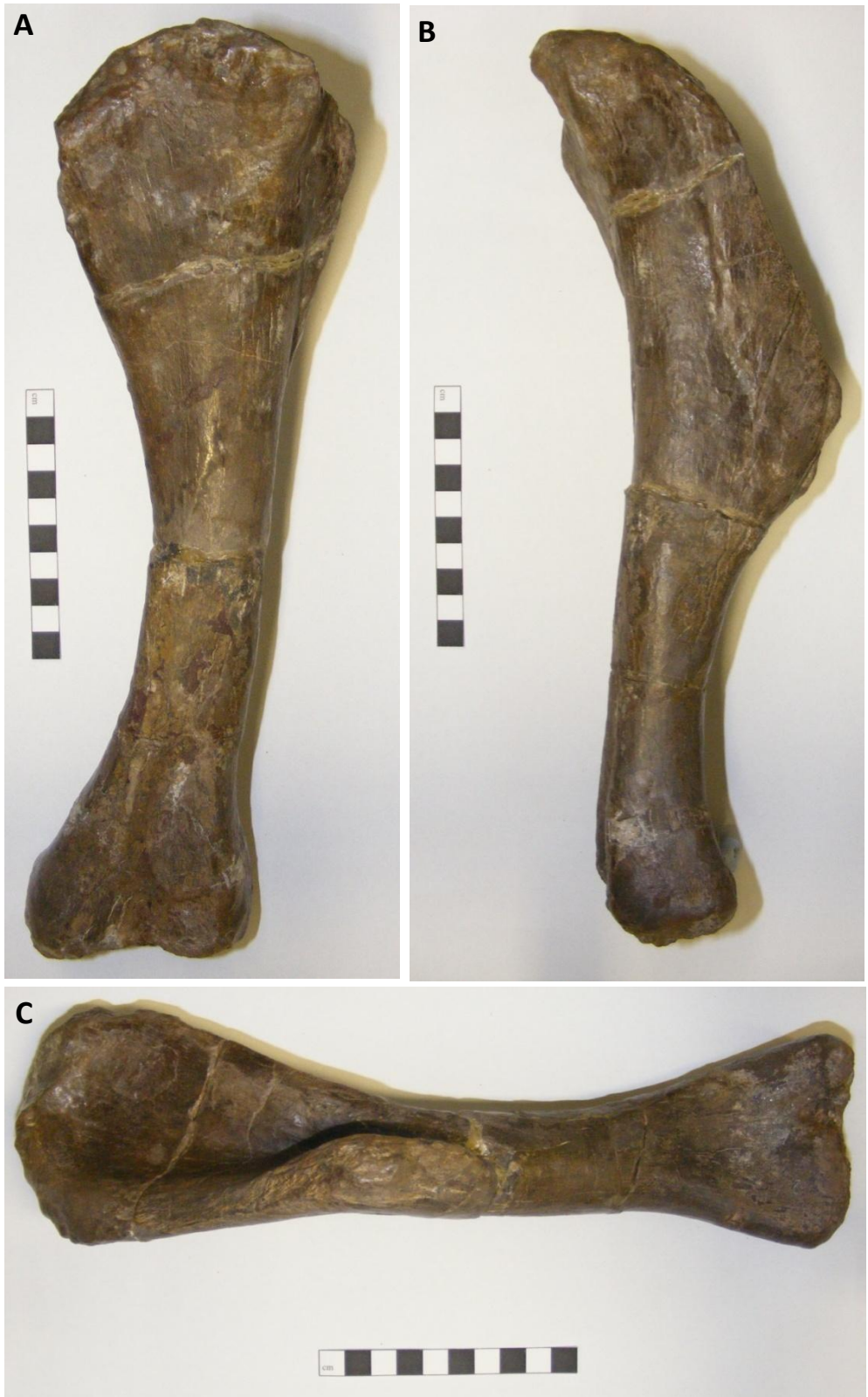

Figure 18 – Dextral humerus in **A:** caudal aspect **B:** lateral aspect and **C:** cranial aspect. Scales = 10cm. Note slight overlap of proximally placed and dominating deltopectoral crest.

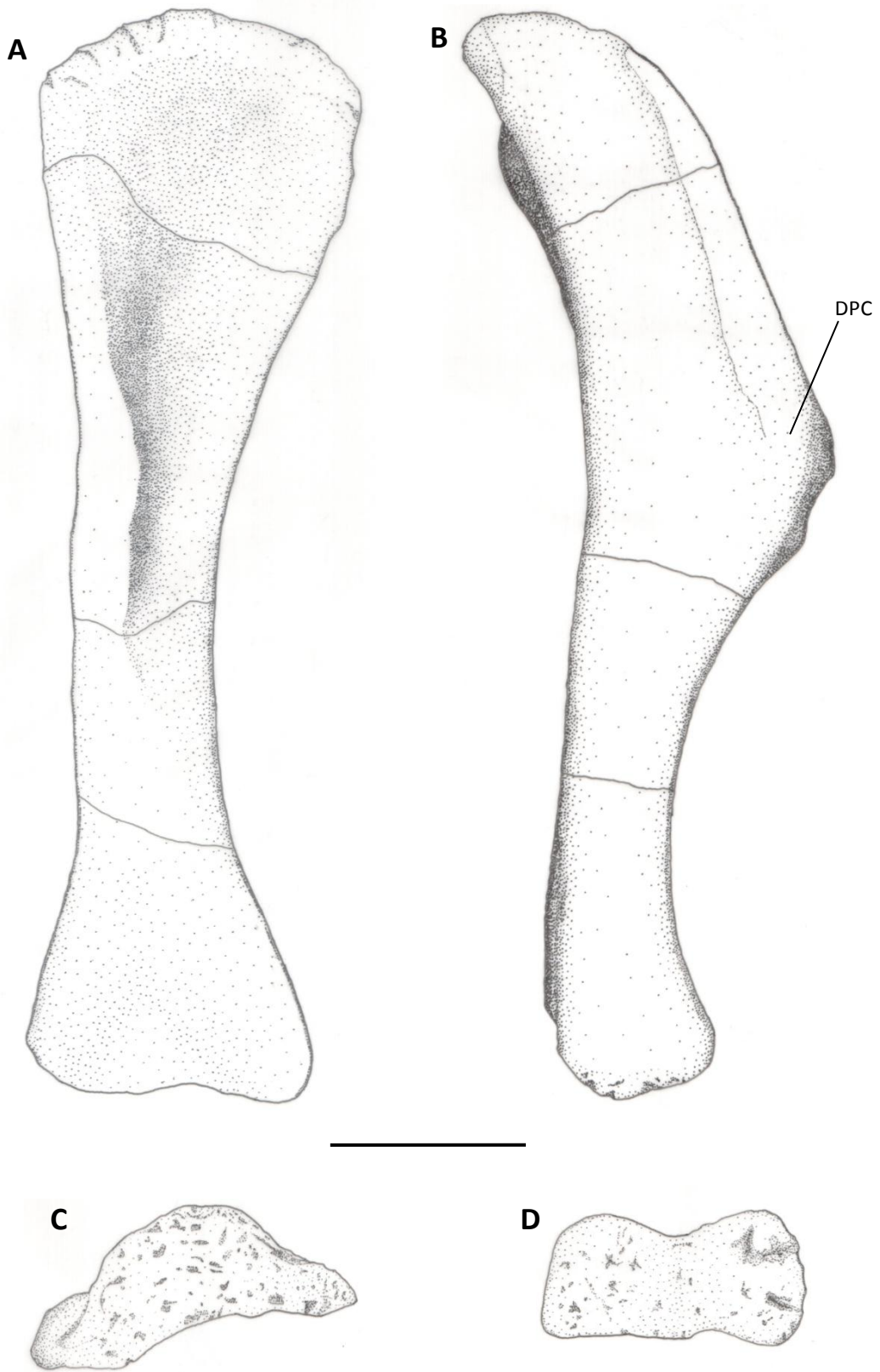

Figure 19 - Dextral humerus in **A:** cranial and **B:** lateral aspect **C:** proximal aspect **D:** distal aspect. Scale = 10cm. DPC – deltopectoral crest.

*Ulna*: lateral surface formed by twisting of shaft so craniolateral distal surface diverges to conform to complex proximal expansions (figs. 20, 21). Forster (1990) states that ulnar shaft is straight in juveniles, and in adults has a slight s-shape in lateral and cranial views. In this specimen, 's-shape' is well developed in cranial view, but in lateral view has a simple arcuate form with terminal expansions.

The twisting of the shaft is very complex. It may relate to the origin of quadrupedalism from bipedalism (S. Maidment, pers. comm.), where rotation of the distal forelimb leads to reorganisation of the carpus and thus the rotation of the manus to conform to a planar substrate. Chinsamy (1995) suggests that a transition may be recorded in the ontogenetic variations of *Dryosaurus*, where juveniles were facultative quadrupeds, and adults were exclusively bipedal. Such an ontogenetic coupled sequence may also be apparent within *Tenontosaurus*; it would be interesting to see if future biomechanical studies could reveal if such an ontogenetic pattern was reflected in stance style (presently, only briefly mentioned in Dodson (1980)). However, this study is converse to a more recent analysis by Moreno (2007), who suggests the acquisition of quadrupedalism is unified to pedal modifications, and that *Dryosaurus* represents an intermediate stage in the transition to bipedalism. The author also states that the bone morphology varies through an ontogenetic series, primarily dependant on the biomechanical role the bones must play; this is known as "Wolff's law of bone transformation", and somewhat guides interpretation of form in terms of function at a given growth stage.

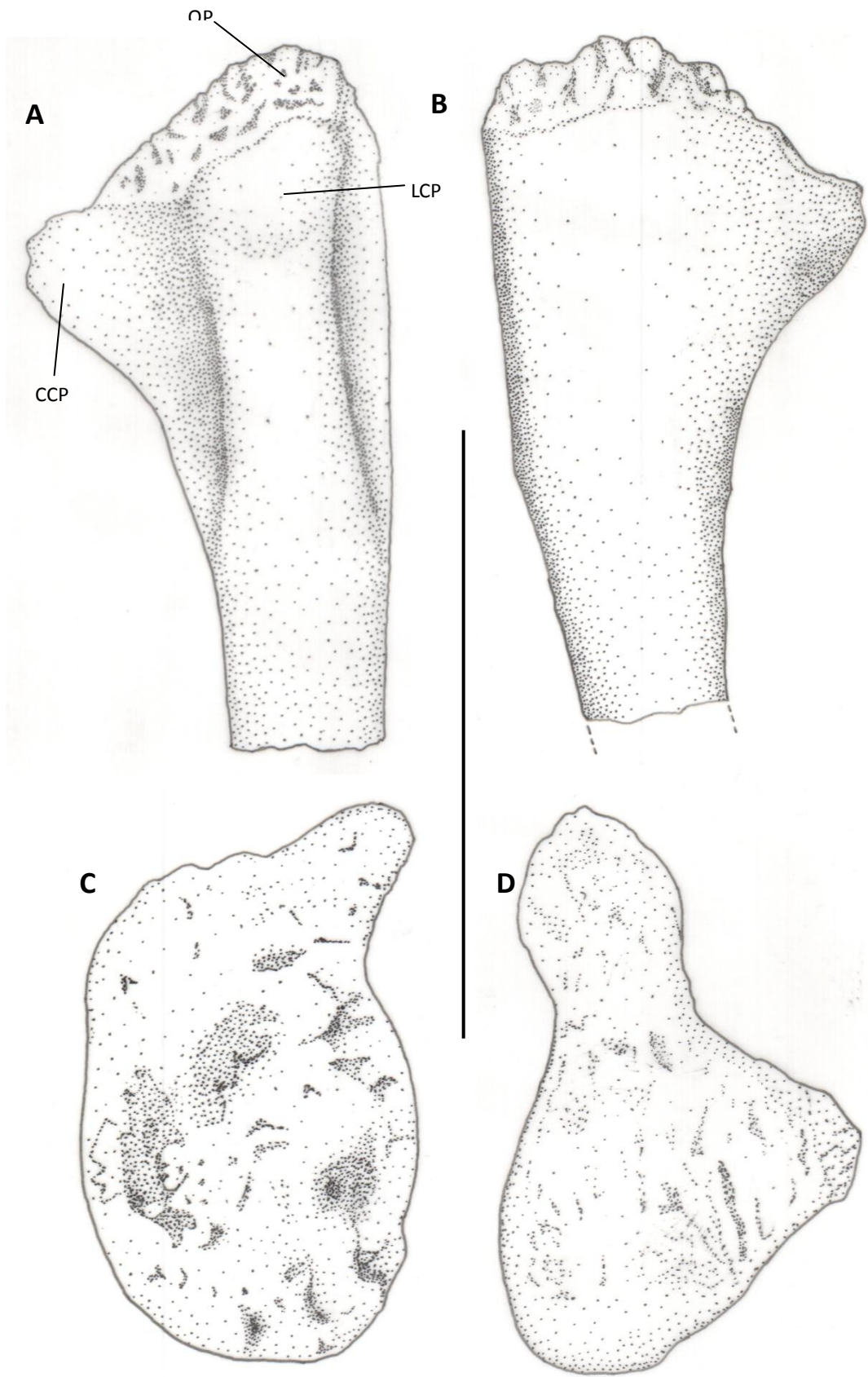

Figure 20 – Dextral ulna in **A:** cranial aspect **B:** caudal aspect **C:** distal aspect **D:** proximal aspect. Scale = 10cm. CCP – cranial coronoid process, LCP – lateral coronoid process, OP – olecranon process

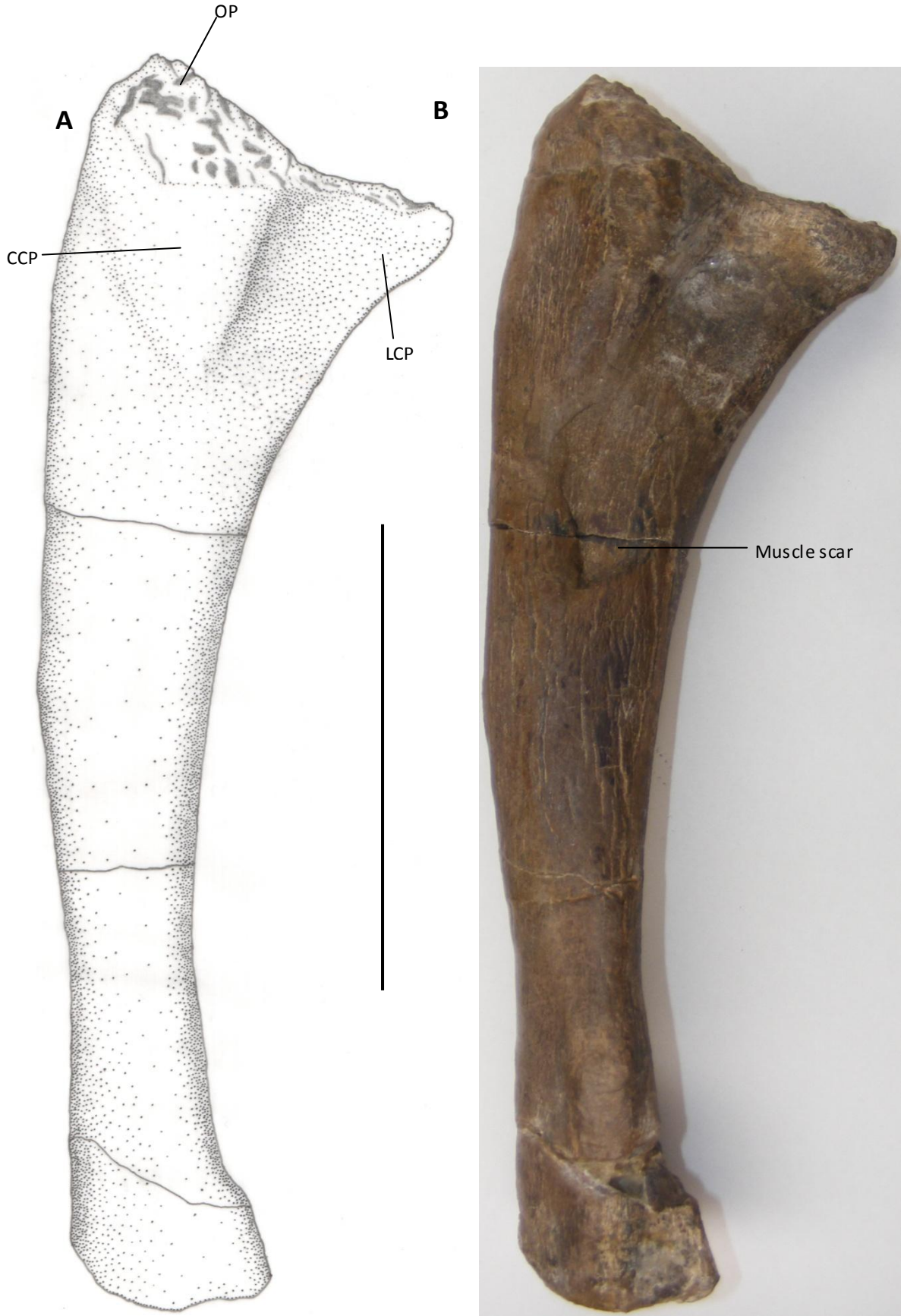

Figure 21 – Dextral ulna, lateral view. Scale = 10cm. Abbreviations as fig. 20.

*Radius*: no noticeable variation to Forster (1990) (fig. 22).

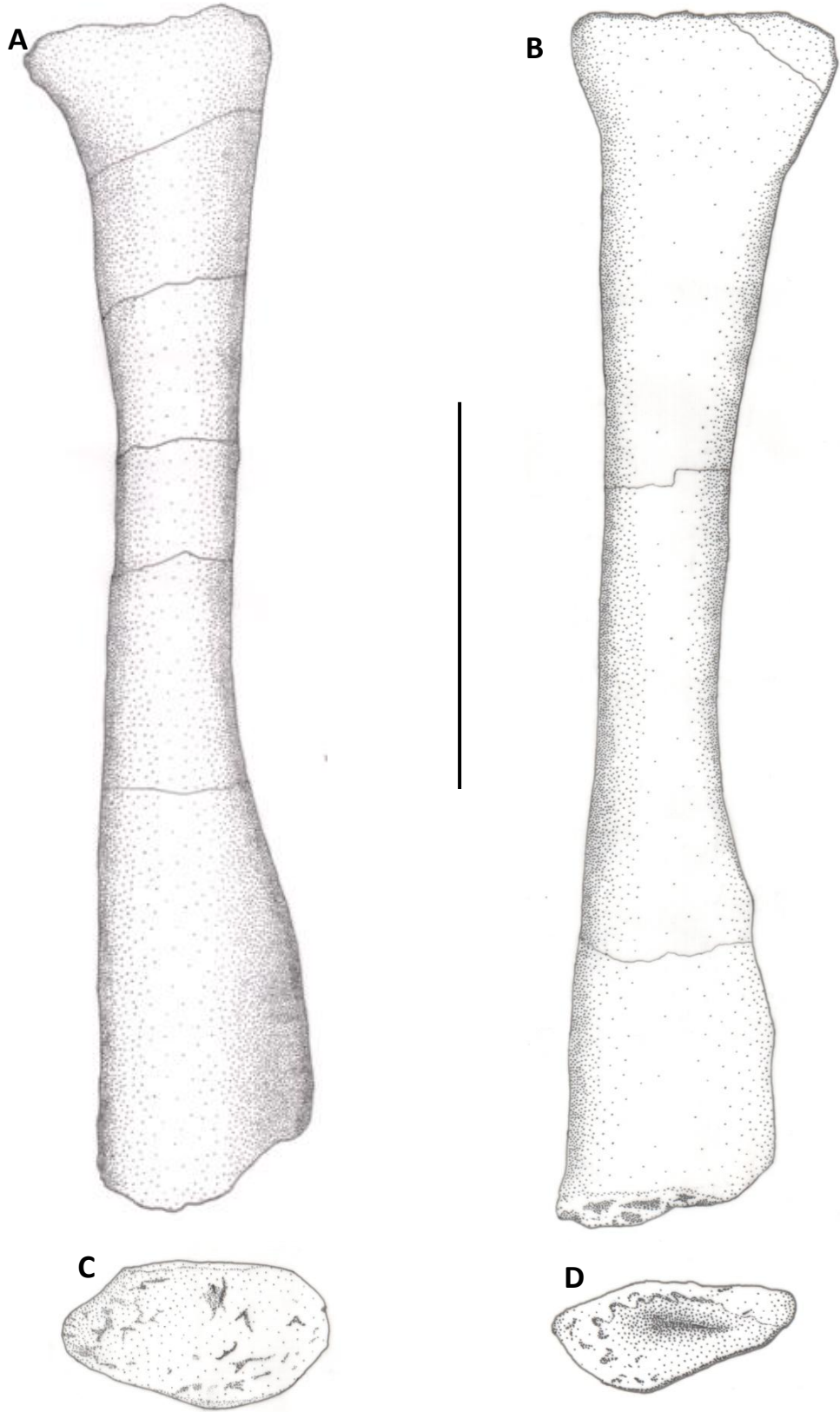

Figure 22 – **A:** sinistral radius, lateral aspect **B:** dextral radius, lateral aspect **C, D:** sinistral radius in proximal and distal aspects. Scale = 10cm.

*Carpus*: comprises three elements: presumably the unfused radiale, intermedium and ulnare based on Ostrom (1970) and Forster (1990). Unfortunately, they are all badly distorted and preserved, so any comparative analysis would be inadequate in true detail. The additional "distal carpals" of Forster (1990) are not found; however this does not mean they were not initially present. The unfused status of the carpal elements is considered plesiomorphic, as in *Orodromeus*, *Dryosaurus*, and *Hypsilophodon* (Galton, 2009), but more probably reflects a bipedal stance.

*Manus*: comprises five digits, with a general width of approximately 1.5 times the length (fig. 23) – dissimilar to more slender *Dryosaurus* and *Hypsilophodon*. Left manus: digit I complete; digit II missing phalanx II; digit III complete; digit IV missing distal phalanges; digit V missing terminal ungual phalanx; all 5 metacarpals are present. Although partially incomplete, this phalangeal formula does differ to that proposed by Forster (1990), (2,3,3,1?,1?), and reverts back to that suggested by Ostrom (1970): 2,3,3,2?,2?. This supports the image of this specimen's sinistral manus depicted in Schachner and Manning (2008), and is coincident with the phalangeal formula for *Camptosaurus* (Dodson, 1980). This alteration however is highly tentative, as several of the phalanges are damaged and possibly originally assigned incorrectly to their positions. Note that reduction of digit III to 3 phalanges is considered the derived iguanodontian state.

Right manus slightly less complete: digit I complete; digit II distal phalanges missing; digit III ungual phalanx absent; digit IV distal phalanges absent; digit V terminal phalanx absent. On the medial phalanx (phalanx I) of digit V, there is a clear distal surface for articulation with the ungual phalanx, which strongly supports the proposed phalangeal formula.

Proximal dorsal surfaces of phalanges articulate smoothly with associated distal surfaces of adjacent metacarpals, extending onto dorsal surfaces as far as one-quarter metacarpal length. This feature is the "hyperextension" mentioned by Forster (1990), and is possible with all elements in at least the medial digits. Hyperflexion does not appear to have been possible. This function possibly served to aid grasping during feeding, social activities, or perhaps even provided aid as a defence mechanism, giving the ability to clasp predators. Another possibility is that it could enhance locomotion over irregular terrain compared to a planar surface. It could serve to reduce tension during quadrupedal locomotion, by providing a more 'springy step' rather than a 'gallop' by allowing manual flexure.

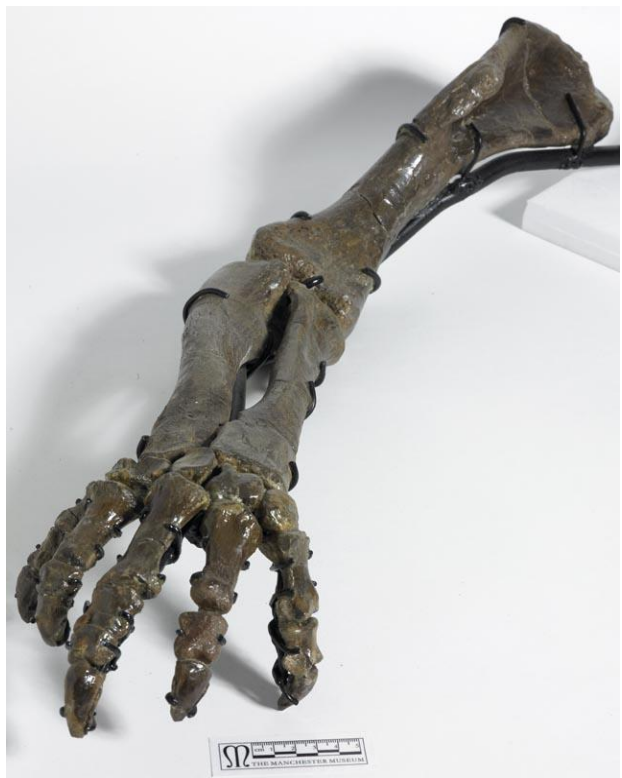

Figure 23 – Artificially articulated and mounted complete dextral forelimb. Additional fourth carpal element visible here not found. Also note phalangeal formula (specifically for digits IV and V) Scale in centimetres. Photograph courtesy of the UoM.

### 4.35 Pelvic Girdle

*Ischium*: shaft slightly bowed longitudinally – not straight in dorsal view as Forster (1990) (fig. 24). "Square to round obturator process" of Forster (1990) not observed - process crescentic in section. Lateral twisting of distal shaft described in Forster (1990) not witnessed - surface remains uniplanar with a linear longitudinal axis. Actually occurs proximal to obturator process (i.e. cranially, not caudally) so both proximal ischial processes are variably oblique to the long-axis of the shaft.

Sinistral ischium with gently sinuous long-axis (to higher degree than linear dextral ischium) – this characteristic possibly the "pronounced ventral bend" of Forster (1990), although probably not on the magnitude the author described. No twisting of shaft observed either on sinistral element, although slight illusion generated by termination of longitudinal ridge, which occurs slightly offset from median plane distally. Distal tip in both not flared as Forster (1990) also declares; a slight dorsal deflection does however occur, although not reflected in ventral border, and not what should be described as "flaring", compared to more recognisable occurrences of this feature. Both ischia are distally rugose; the "distal scarring" mentioned by Forster (1990) is not observed; however this may be due to failure of recognition of this feature, partial obscuration, or absence of distally fused ischia in specimens of *T. tilletti* that are not fully matured.

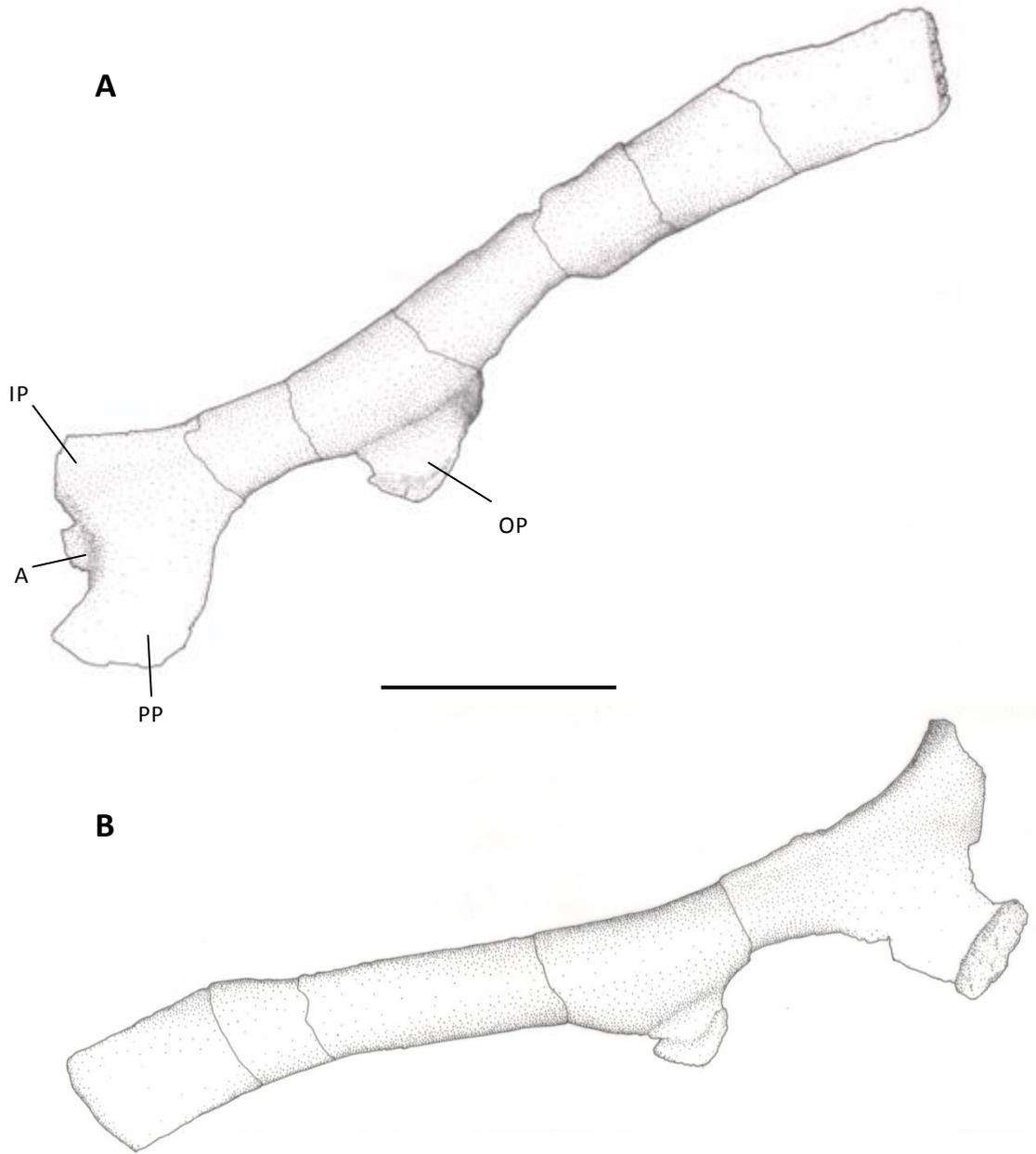

Figure 24 – **A:** sinistral ischium, lateral aspect **B:** dextral ischium, lateral aspect. Scale = 10cm. Note non-linear shaft, and proximal placement of the obturator process. A – acetabulum, IP – iliac process, OP – obturator process, PP – pubic process.

*Ilium* (figs. 25, 26, 27): brevis shelf develops as strongly concave and broad depression bordered by steep medial edge and well-developed ventral ridge projecting approximately equidistant to iliac width providing strong attachment area for the *caudo-femoralis brevis* muscles to fourth trochanter (S. Maidment, pers. comm.). Relative to iliac thickness brevis shelf well-developed: defined as surface area being greater than iliac mediolateral thickness; therefore the description of Forster (1990) is true compared to most more-advanced ornithopods, but still a well-developed structure even at sub-adult stage; no supracetabular shelf/rim stated by Forster (1990); possible presence suggested by gentle discontinuous rim forming on dorsal margins of ischiac and pubic peduncles (not as prominent as *Lesothosaurus*).

Caudal border not "laterally thickened" as Forster (1990) describes – instead retains uniformly thick postacetabular blade; illusion created by an extremely gentle dorsoventral concavity in lateral surface of blade - thickness retained by an equally gentle convexity on medial surface.

Forster (1990) states that there is a "near absence of a brevis shelf", which is considered to be slightly inaccurate as an autapomorphy. Although not as well developed as *Hypsilophodon foxii*, it still forms a prominent muscle attachment region on the postacetabular ventromedial margin. Further ambiguity is raised as Galton and Powell (1980) state that *Camptosaurus prestwichii* possesses a "narrow brevis shelf"; Galton and Taquet (1982) depict a 'narrow' brevis shelf in both *Iguanodon* and *Hypsilophodon*, although personal observation of the latter indicates it is considerably more developed than *Tenontosaurus*, although not to the same degree as members of the Dryosauridae when scaled (Galton and Taquet, 1982; Butler *et al*. 2008). Dodson (1980) describes the brevis shelf of the genus *Camptosaurus* as a "moderately developed medial reflection of the postacetabular margin", which matches the character observed in this specimen. It appears that this feature is too unclear to distinguish genera between camptosaurs and tenontosaurs, therefore it is suggested that it be removed as a defining character of the genus *Tenontosaurus*, or redefined in a quantifiable manner: brevis shelf less than 1.5 times mediolateral width of postacetabular ventral margin (fig. 26C). The same problem appears to be a recurrent theme in specific/generic diagnosis; use of ambiguous phrases such as "relatively long", "short" and "robust" are insubstantial, detracting from analytical value if not placed into a quantifiable context. It becomes dependant on the author's judgment of an imprecise character, which may significantly differ from another's, developing from interpretational bias. It is thus stressed that characters expressed in such a style be redefined in a quantifiable manner so an unequivocal set of autapomorphies can be elucidated.

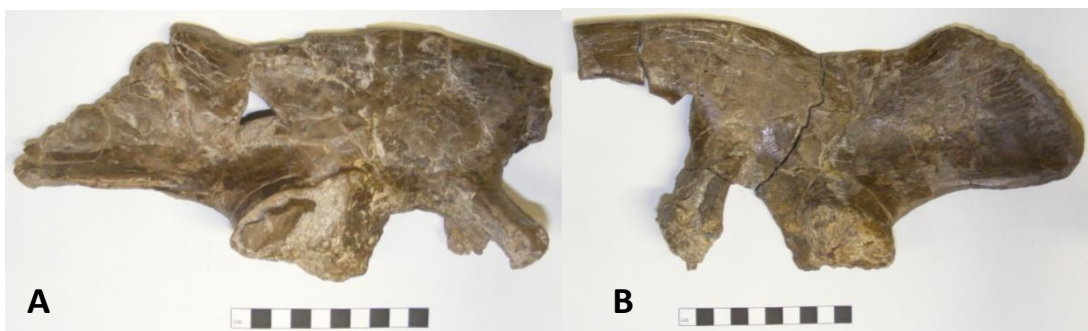

Figure 25 – **A:** sinistral ilium, lateral aspect; note apparent divergence of pubic peduncle **B:** dextral ilium, lateral aspect; note contrasting caudal border to A – morphology otherwise similar, therefore may relate to element being derived from a different specimen and exhibiting intraspecific variation. Scale in centimetres.

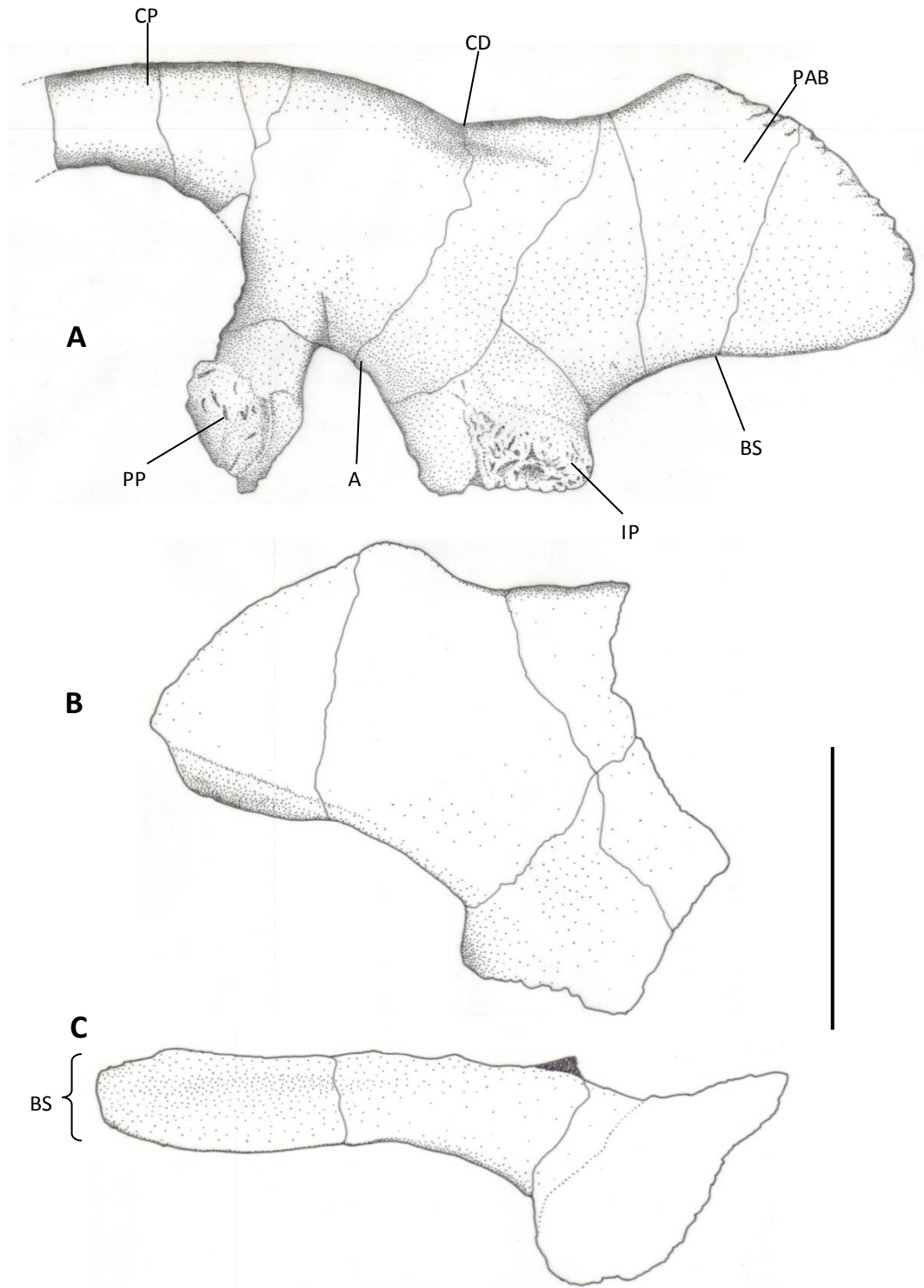

Figure 26 – Dextral ilium **A:** lateral aspect **B:** distal section, medial aspect **C:** postacetabular ventral aspect, exhibiting form of brevis shelf. Scale = 10cm. A – acetabulum, BS – brevis shelf (ventromedial deflection), CD – concavity of dorsal margin, CP – cranial process, IP – ischiac process, PAB – postacetabular blade, PP – pubic process.

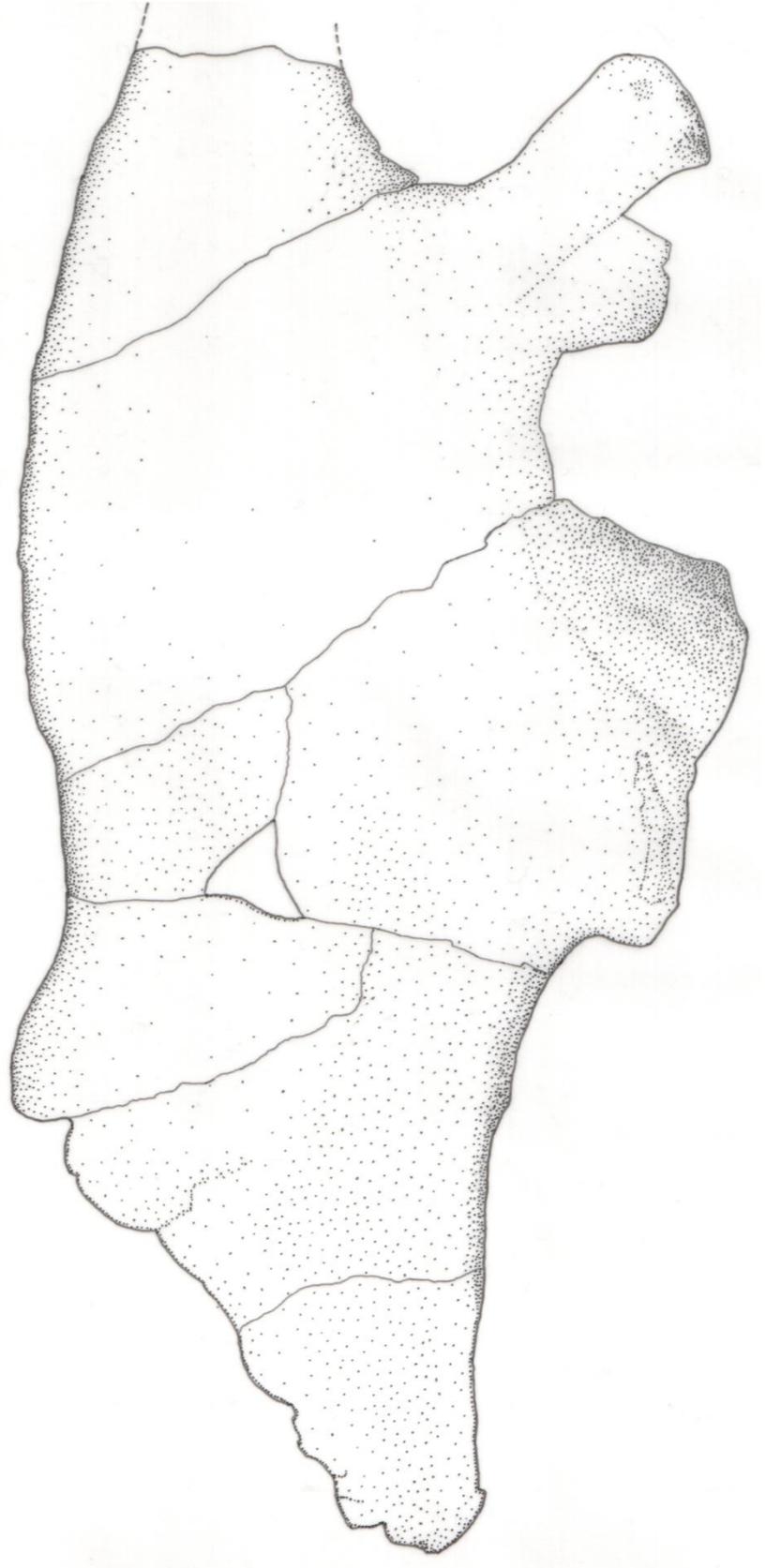

Figure 27 – Sinistral ilium, lateral aspect. Scale = 10cm.

*Pubis*: ventral border relatively straight, with slight ventral deflection approximately half blade length (fig. 28) – provides gentle proximal convexity and sub-parallel convergence of dorsal and ventral borders distally. This feature is not as prominent as Forster (1990) describes in specimen YPM 5460, and may relate to the "kink" in the holotype, AMNH 3040, although it does not appear rugose as described. "Tubercles" of Forster (1990) unidentified, or not present – nothing observed fitting description.

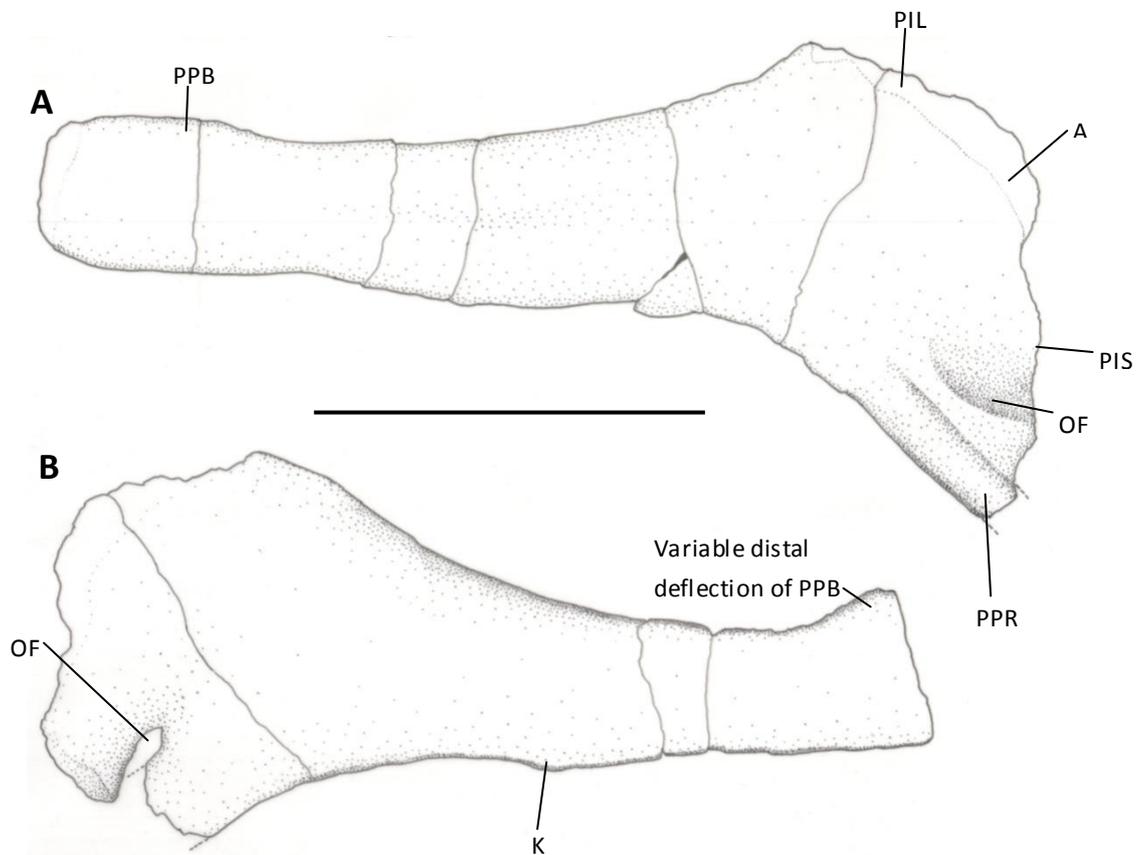

Figure 28 – **A:** sinistral pubis, lateral aspect **B:** dextral pubis lateral aspect. Scale = 10cm. Note variable degree of enclosure and perforation of obturator foramen, and inconsistent expansion of distal tip of preacetabular blade. A – acetabulum, K – kink in ventral border, OF – obturator foramen, PIL – puboiliac articulate, PIS – puboischial articulate, PPB – prepubic blade, PPR – postpubic rod.

**4.36 Hindlimb**

*Femur*: tip of fourth trochanter does not appear to be pointed, scarred or caudoventrally hooked as Forster (1990) expresses. Medial condyle also not "more transversely expanded" than the lateral condyle (fig. 29).

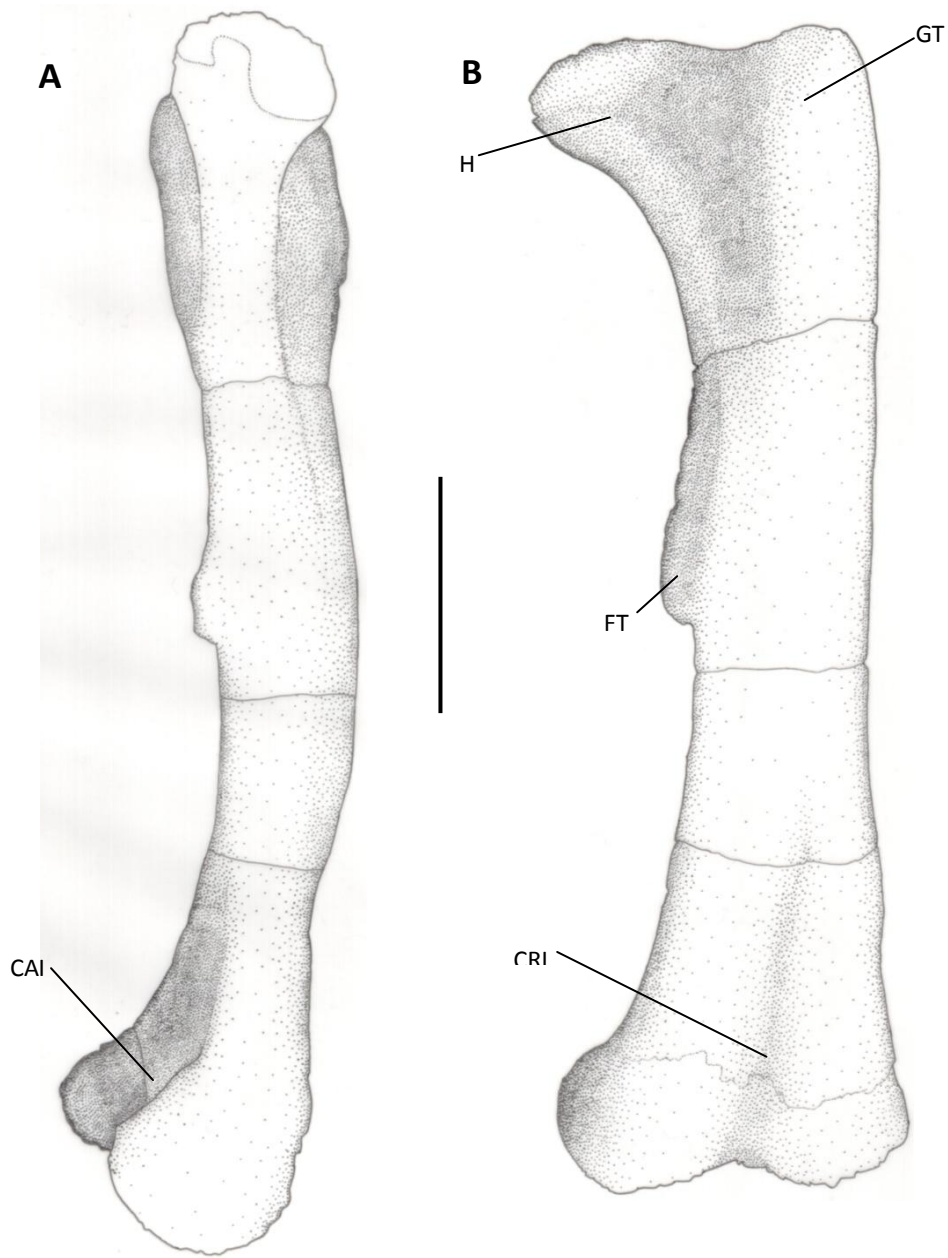

Figure 29 – sinistral femur in **A:** medial aspect, **B:** cranial aspect. Scale = 10cm. Note shallow cranial intercondylar groove and relatively deeper caudal intercondylar groove. CAI – caudal intercondylar groove, CRI – cranial intercondylar groove, FT – fourth trochanter, GT –greater trochanter, H – femoral head.

*Tibia*: cnemial crest not as "narrow" as Forster (1990) – approximately twice the size depicted in Fig. 20B (fig. 30). Proximal gentle cranial concavity propagates into small scar mark – succeeded by deeper perforation on sinistral tibia (not mentioned in Forster (1990)). Caudolateral surface reflects this at same distance from distal end (same arcuate, listric scar form) on both tibiae; shaft here resumes transverse expansion, developing broad, sub-triangular distal end. Long axis oblique to proximal end (slightly less than 60° as quoted by Forster (1990)) – degree of twisting either variable within individuals or function of ontogeny.

*Fibula* (fig. 30): slight cranial expansion one-fifth length of shaft (distally) – manifests as small, rugose, discontinuous ridge (not identified in Forster (1990)).

Tarsus: one slight difference occurs to Forster (1990): in Fig. 21B, the author depicts the astragalus completely capping the tibia; here apparently it does not cover the lateral malleolus, possibly leaving it free to articulate with metatarsal V.

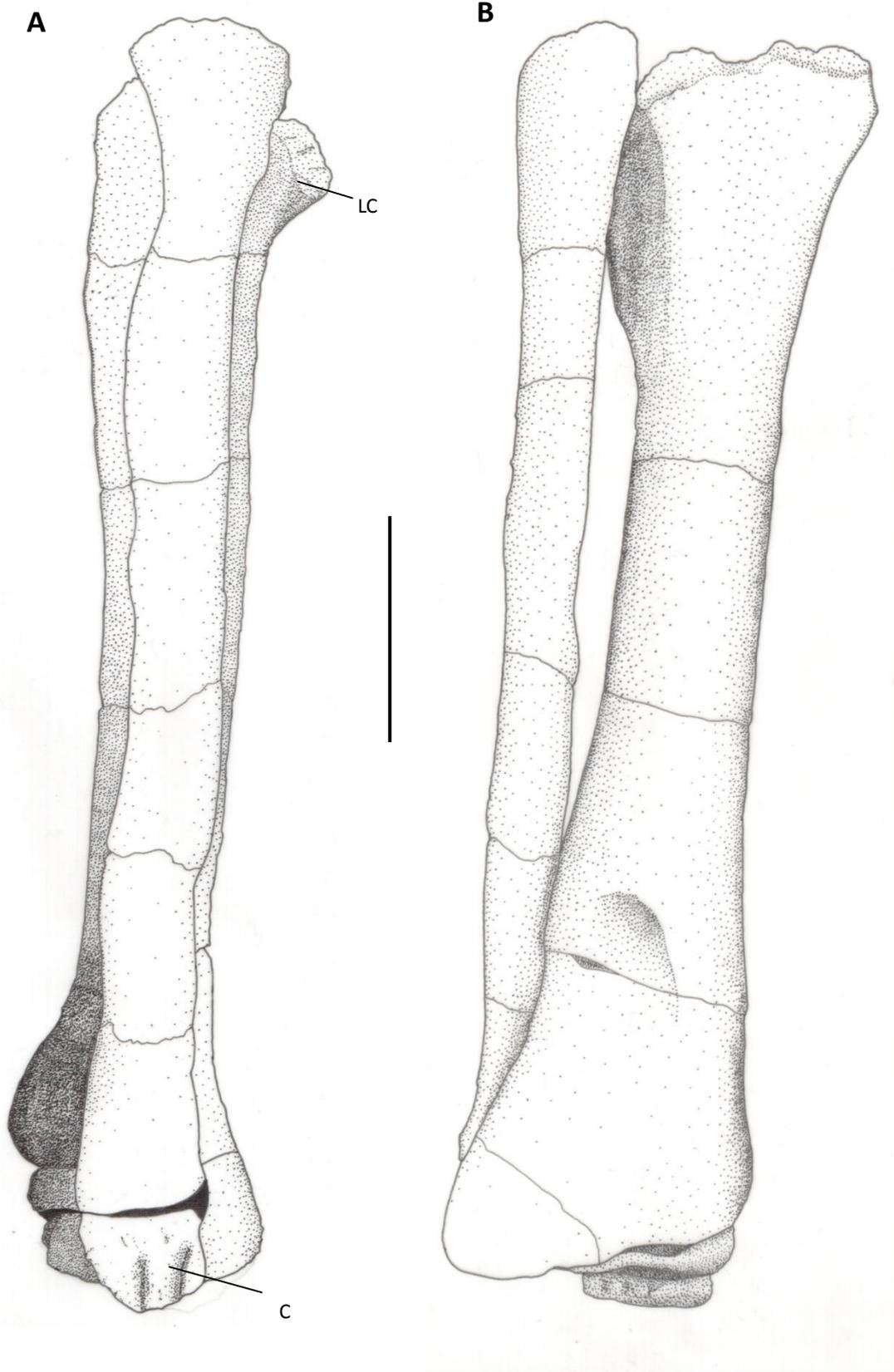

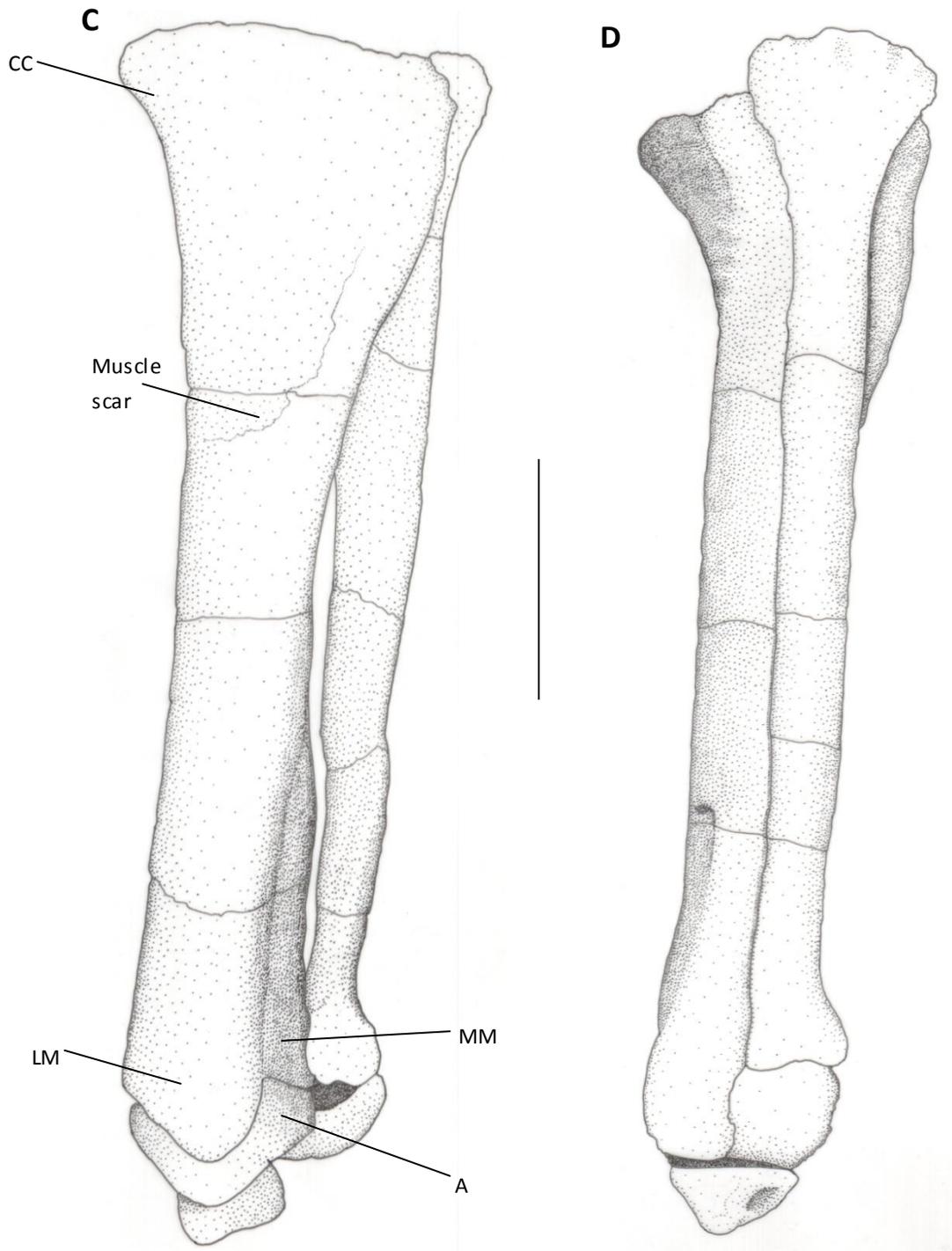

Figure 30 – dextral tibia, fibula and tarsus artificially assembled in **A:** caudal aspect **B:** medial aspect **C:** cranial aspect **D:** sinistral tibia, fibula and calcaneum, lateral aspect. Scale = 10cm. Note prominent muscle scar in B. A – astragalus, C – calcaneum, CC – cnemial crest, LC – lateral condyle, LM – lateral malleolus, MM – medial malleolus.

*Pes*: four metatarsals forming tightly bound component ("compact metapodial unit" of Forster (1990)) inducing a digitigrade stance (e.g. Moreno, (2007)); phalangeal formula 2-3-4-5(-0), as Forster (1990) and *Camptosaurus* (Dodson, 1980), without vestigial fifth metatarsal (fig. 31). Forster (1990) states: "digit I, while well-developed, is short and probably did not reach the ground" – this largely depends on the positioning of metatarsal I - lateral rotation/articulation is possible, potentially placing the digit in a functional manner, rather than having a fully developed digit practically redundant. Pedal elements all artificially attached – articular nature largely obscured. Each digit bears well-developed distal ungual phalanx.

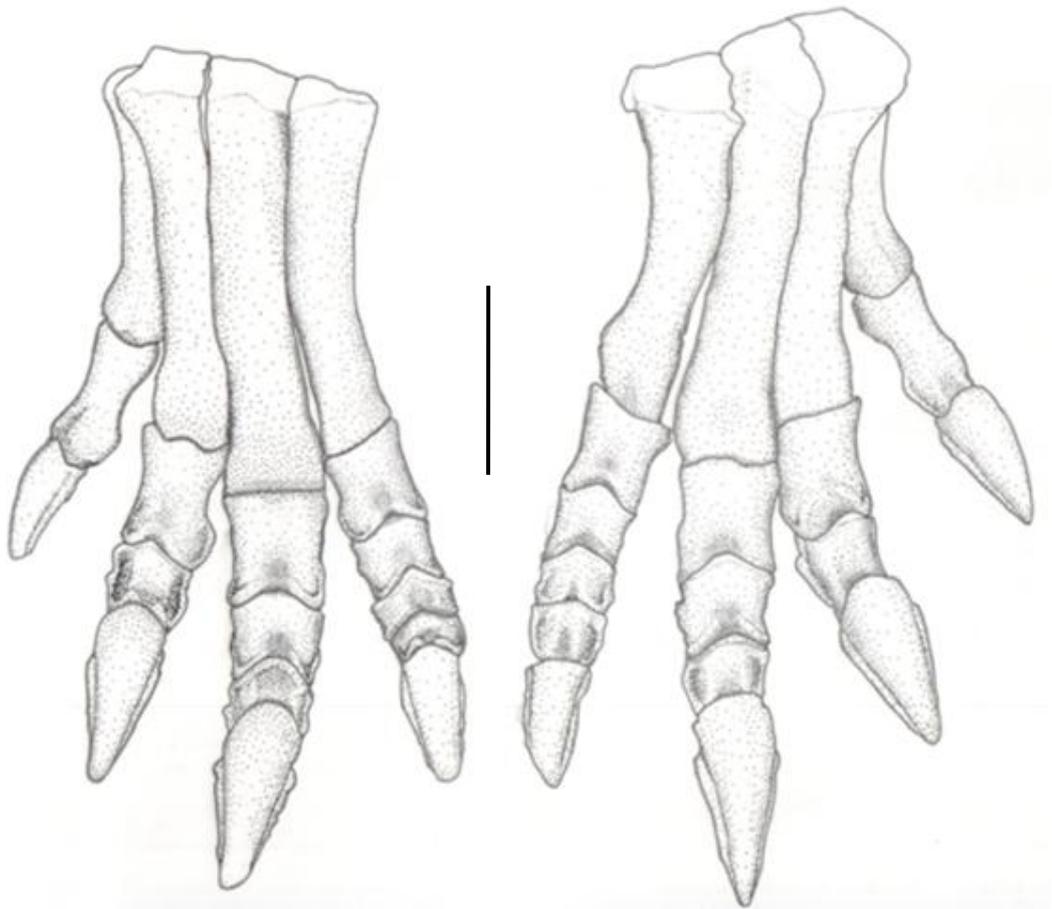

Figure 31 – sinistral and dextral pedal elements artificially assembled. Scale = 10cm. Note claw-shaped unguals in comparison to high iguanodontians, and longer, gracile metatarsals.

# 5.0 Discussion

Ontogeny is essential in understanding fossil taxa and their systematic placement in cladistic studies, and the ontogenetic fit of specimens should always be considered, especially when defining primary homologies in data matrices. Bennet (2008) emphasises this in a study on the ontogeny of *Archaeopteryx lithographica*, where previous attempts to classify specimens into several different taxa are refuted with a robust ontogenetic analysis. A similar study regarding the Pachycephalosauridae by Horner and Goodwin (2009) signifies how crucial prior understanding of ontogenic variation within taxa is; the authors eliminate two taxa of North American pachycephalosaurids, demonstrating using cranial ontogeny that the taxa represent a continuous growth pattern within a single species. However, this is based purely on cranial analysis; a more robust conclusion would be if the authors had implemented long bone histological analyses to confirm their findings.

The Ornithopoda have been subject to such a controlled analysis (Horner *et al*., 2009). Allometric growth patterns are examined using long bone histology in three "hypsilophodontid" genera (including *T. tilletti*); a somewhat generalised approach to a more complex and disparate phylogenetic situation. The study conclusively determines that *Dryosaurus altus* histologically resembles hadrosaur taxa, including *Maisaura* and *Hypacrosaurus* than phylogenetically closer iguanodontians (e.g. *T. tilletti*). Significantly, *D. altus* is also stated to be represented by no completely mature specimens (although determining absolute maturity is problematic), thus partially occluding its true identity and phylogenetic position; it may be that Dryosauridae requires repositioning closer to the Hadrosauridae, perhaps with inclusion in Ankylopollexia. This study develops beyond Chinsamy (1995), who after a similar ontogenetic analysis of *Dryosaurus lettowvorbecki* concluded that the genus should be either placed within the Hypsilophodontidae with *Valdosaurus* or as the sister clade to the Iguanodontidae.

With regards to *Tenontosaurus*, specimens require thorough ontogenetic analyses to determine if the species separation is an ontogenetic rift, determined primarily through careful histological studies on a wide sample range. Relying on the state of cranial fusion to determine age may be problematic, in that it does not account for allometric growth; thus future studies must combine long bone and cranial histology with other ontogenetic factors to determine absolute (or relative) age of specimens. Only then can phylogenetic sampling and analysis become based on the robust framework necessary for authentic taxonomic identity.

The current state of phylogenetic affinities within the Ornithopoda is in disorder, with multiple analyses each concluding a range of results based on an individual authors approach in defining either phylogenetic definitions or homology classification. The breakdown of Butler *et al*. (2008)'s data matrix into higher resolution groups (e.g. basal Ornithopoda, similar to Norman (2004)) is critical, rather than the use of a large data set for a large range of species or groups (or the generation of new multiple individual analyses based on interlinked phylogenetic characters). This, and the refinement of characters to become more clade-specific should increase finer-scale phylogenetic resolution, providing outgroup taxa are carefully selected as standards. For example, many of the homologies applied to the Ornithischia by Butler *et al*. (2008) are potentially variably redundant for many taxa and clades due to the substantial variety of ornithischian species, despite the initial universal and simple bauplan. Thus, as one increases towards the genus level of the taxonomic hierarchy, fewer homologies become applicable or relevant (e.g. those synapomorphies that aid definition of the Ankylosauria will have little to no relation to the synapomorphies of the Hadrosauridae within a total analysis of Ornithischia). One must however stress again the importance of truly understanding the affinities of the studied specimens prior to cladistic analysis; construction of synapomorphies without determining absolute generic or specific classification will ultimately lead to the breakdown of the cladogram. Sexual dimorphism is also a factor that must be observed in future analyses, if possible. Currently, the majority of studies either do not mention or do not include or understand that intraspecific gender variation can often be misinterpreted as interspecific character variation. Erection of similar contemporaneous species or genera may simply be the artefact of lack of understanding of sexual dimorphism. The same may be true of constructing data matrices, which rely heavily on comparing subtle synapomorphic variations – gender-specific variations (e.g. length of post-pubic rod) may appear to manifest as interspecific variations, and again this will lead to incorrect homology assumption, and false phylogenetic relationship establishment.

Norman (2004) undertook a phylogenetic analysis of the basal Iguanodontia, including *Tenontosaurus tilletti*. Due to the completeness of LL.12275, an attempt to recode the data matrix has been undertaken. 12 of an initial 67 these have now been recoded (18%) (with 2 remaining uncoded) (see appendix 6 for results).

Unfortunately, a phylogenetic analysis of basal euornithopods is beyond the scope of this study, but this degree of differentiation between two apparently identical species emphasizes that either, the matrix requires re-definition of synapomorphies to a greater detail, that ontogeny is occluding phylogeny to some degree, or simply that significant intraspecific variation occurs within tenontosaurs; it is perhaps a little implausible that this specimen represents a new species based on the general resemblance to previously described specimens.

Future work must employ the use of stratigraphy to delineate subtle variations within *Tenontosaurus'* evolution. Placement of individual specimens into a temporal stratigraphic context will potentially allow morphological patterns to be observed and demarcated on a fine-scale, given the wealth of specimens available over a broad span. This does not necessarily require the exact dating of individual temporal fossiliferous strata, merely the relative dating of one to another within a well-defined boundary (e.g. utilising volcanic horizons). One would expect to see a systematic variation through time, constrained presumably to genus level. Intraformational spatial variations are postulated to have little to no effect on morphology, considering the extensive coverage of palaeoenvironmental conditions. Closer comparison with extraformational specimens (i.e. from coeval deposits external to the Cloverly Formation) is similarly required, as it is likely that if the genus *Tenontosaurus* became isolated as Weishampel *et al*. (2010) suggests, then separate endemic pockets would establish (allopatric speciation) as features such as mountain ranges and intercontinental seas developed. This requires the careful integration of robust stratigraphic and tectonic data with available palaeontological records, providing an exact location for specimens can be provided.

With regards to the incongruity between Morrison and Cloverly faunas, external fossil-bearing units must be identified and ornithopod specimens placed into a local (North American) stratophylogenetic context. This could solve the apparent disparity between faunas, and help resolve the current problematic status of basal euornithopod relationships. Galton (2009) mentions there are specimens that require description or revision that could prove decisive for such a task.

This study does not conclusively determine where *Tenontosaurus* should be placed taxonomically. Placement within Iguanodontia however is dismissed, based on the simple fact that *Tenontosaurus* does not fulfil even a majority of the synapomorphic requirements (maximum 8-9 out of 20) for the clade (Sereno, 1986). The lack of prominence of the humeral head also distinguishes *T. Tilletti* from the Iguanodontia, based on the addition of this synapomorphy in Norman (1998). Given that Sereno (1986) provides the most recently known full definition of Iguanodontia, it seems conclusive that tenontosaurs are primitive to, but not inclusive within the current status of Iguanodontia. The phylogenetic characters may also require revision (e.g. "leaf-shaped" denticles are highly ambiguous). The suite of taxa currently occupying basal euornithopod positions is in critical need of a robust phylogenetic analysis, where emphasis is also placed on generating a tectonostratigraphically stable cladogram. The position of the sub-clade Camptosauridae also requires revision, as do the positions of South American ornithopod taxa and 'rhabdodontids'. The clade "Hypsilophodontidae" requires either dissolution (with taxa being assigned to either successive sister groups to Euornithopoda or erection as several smaller sister groups placed in various positions relative to Iguanodontia), or rigorous re-analysis, accounting for the apparent 'derived' state of several members relative to basal members of Iguanodontia. Synapomorphies distinguishing these two clades require similar modification. Prior to any future analysis however, ontogenetic studies (following Horner *et al.* (2009) for example) must be undertaken to define the precise growth stage of any specimens undergoing study.

The configuration of cranial elements of LL. 12275 remains problematic; many contacts are variably occluded by restorative materials. However, several features appear to differ unequivocally from the paratype skull described in Ostrom (1970). Given that Forster (1990) removed all previous defining cranial characters (they appear to be too universal or synapomorphic amongst ornithopods), it is impossible for the features observed here different to Ostrom (1970) to be considered diagnostic, as no robust underlying set is currently in place. One can postulate that the majority of these features relate to ontogeny, in that they vary in degrees of fusion and suture strength and placement. However, others, such as posteroventral extent of the external nares appear to be genuine morphological variations, possibly attributed to intraspecific variation. Given the tentative nature of these variations, they cannot be conclusively assigned to a separate species other than *T. tilletti* for the present, or assigned autapomorphic status, until comparative morphological differences can be elucidated with increased confidence (i.e.

personal observation of the paratype skull, and others exhibiting similar degrees of completion). Revision of cranial material is required; the numerous depictions of *T. Tilletti* skulls in the literature convey that no unequivocal consensus on the cranial morphology currently exists, and it will take more than one semi-complete skull to resolve this.

*Tenontosaurus tilletti* appears to have been a facultative biped based on strong hindlimb characters, but possibly utilised a quadrupedal stance for more rapid locomotion based on manual and humeral features. A biomechanical analysis of the forelimb is essential in constraining the ability of the forelimb a weight-bearing role.

## 6.0 Conclusions

The diagnosis of *Tenontosaurus tilletti* has been revised; instead of designated autapomorphies being based on multiple non-contemporaneous, partially incomplete, and possibly multi-growth stage specimens, one mostly complete, well-preserved specimen has been utilised to create a more robust definition. The diagnostic characters have been revised as follows:

1. Vertebral count 12-15-6-61(+)
2. Tail comprising at least 57% total animal length
3. Scapula with straight but with slight proximally and distally expanded caudal margin to approximately 110-120% depth
4. Coracoid with strong, subtriangular sternal process, and coracoid foramen completely closed off from articulation (40% coracoidal width)
5. Forelimb 68% hindlimb length, with equally robust humerus and ulna
6. Humerus dominated by strong, rounded, proximally confined deltopectoral crest; expansion approximately equal to humeral shaft thickness
7. Carpus comprising unfused intermedium, radiale and ulnare
8. Manus short and broad, with phalangeal formula 2-3-3-2?-2?
9. Ilium with gently ventrally dipping preacetabular process; approximately 60% iliac body length
10. Ilium with dorsoventrally expanded postacetabular blade and rugose caudal margin, and concave ventral border
11. Ilium brevis shelf less than 1.5 times mediolateral width of postacetabular ventral margin
12. Pubis with straight postpubic rod comprising 1.5x pubic body length, and projecting at 100° to prepubic blade
13. Obturator foramen closed off from articulation
14. Prepubic blade laterally compressed, depth non-linear and approximately 75% maximum ischium depth, with distal tip variably expanded
15. Shaft of ischium non-linear, proximal half dominated by lateral ridge, with crescentic obturator process one-third length of shaft
16. Femur with hemicylindrical lesser trochanter extending one-fifth femoral length longitudinally, one-third greater trochanter width and separated by thin fissure
17. Pes with phalangeal formula 2-3-4-5-(0)

This suite of characters is by no means irrefutable, and subject to variation as complete adult specimens are utilised to refine the values presented here. A single specimen exhibiting the observed degree of completion (relative to the holotype) is however deemed qualified in re-assessing the character state. The quantifiable manner in which each character is expressed greatly aids interpretation of features, in that they no longer source an independent bias based on an individual's interpretation of a statistically unbound term. Autapomorphy '3' designated by Forster (1990) (see section 4.0) is removed, as it is widely homologous (symplesiomorphic) throughout ornithischians. Autapomorphy '18' is also removed as it is deemed synapomorphic amongst Iguanodontia. No further complete additions or subtractions were made. The variations outlined in section 4.2 cannot be designated as autapomorphic, considering the overall general resemblance to previously described specimens. Many of the variations are likely attributed to ontogeny, and until a thorough analysis of other *Tenontosaurus* material is undertaken, in conjunction with information obtained from this specimen, one cannot add characters that may not be present within contemporaneous or otherwise related specimens.

No reason is observed why *Tenontosaurus* should be internally associated with the clade Iguanodontia, or within Hypsilophodontidae given its phylogenetically unstable position. Possible erection of a new clade, Tenontosauridae, is suggested to include taxa endemic to North America in the Early Cretaceous, upon future stringent revision of material, with placement as the sister clade to Iguanodontia. Several basal euornithopodan synapomorphies are obsolete and in need of more quantifiable phylogenetic definitions. Increasing clarity of this clade (and sub-clades) is critical if the current state of relationships is to be resolved.

Phylogenetic analyses require an approach from a multidisciplinary viewpoint; inclusion of tectonostratigraphical, ontogenetic, palaeoecological, and biomechanical data and sets of well-defined principal homologies are essential in elevating phylogenetic resolution and generating stratigraphically feasible ancestor-descendant relationships. Determination of homology weighting is also crucial, as this can greatly affect the systematic relative placement of an organism, or serve to occlude phylogenetic resolution.

Specimens assigned to *Tenontosaurus tilletti* are in critical need of re-analysis; a significant quantity of material is attributed to this species, and is potentially the result of inconclusive stratigraphic constraints and the vast temporal and spatial span occupied within the Cloverly Formation and coeval units. Future revision of this material is expected to reveal temporal variations on the species-level intrinsically linked to environmental evolution, as well as possibly provide clues to sexual dimorphism in contemporaneous, yet morphologically distinct tenontosaurs. Interspecific variation has been recorded previously, but not placed into any well-defined stratigraphic context, therefore patterns have been difficult to elucidate. Cranial material is in most vital need of revision based on the observed variations of this study, and those previously identified in passing in the literature. Significant diagnostic autapomorphies are expected to be revealed, considering the wealth of ornithopodan cranial material uncovered since the last description in 1970.

# 7.0 Acknowledgements

Initial thanks go to D. Gelsthorpe (UoMM) and M. Edwards (UoM) for entrusting the specimen into my care and providing storage, and to the former for generously providing photographs of when the specimen was assembled. Many thanks to P. Falkingham, J. Jepson, and K. Bates (UoM) for their ever-present support and direction. The expert guidance of D. Norman (UoC) directed the study and provided invaluable comments on specific topics to focus on, and is much appreciated. Gratification also is extended to L. Steel, S. Maidment, S. Chapman, and P. Barrett (NHM) for allowing access to specimens in their care, and especially to the former two for helpful discussion and assistance. A special thanks to J. Nudds (UoM) for his patient, friendly, and first-class supervision, for without which this project would not have ensued. Thanks John.

# 8.0 References

Bakker, R. T. and Galton, P. M. (1974) Dinosaur monophyly and a new class of vertebrates, *Nature*, **248**, 168-172

Barrett, P. M. and Han, F-L. (2009) Cranial anatomy of *Jeholosaurus shangyuanensis* (Dinosauria: Ornithischia) from the Early Cretaceous of China, *Zootaxa*, **2072**, 31-55

Bennett, S. C. (2008) Ontogeny and *Archaeopteryx*, *Journal of Vertebrate Palaeontology*, **28(2)**, 535-542

Boyd, C. A., Brown, C. M., Scheetz, R. D. and Clarke, J. A. (2009) Taxonomic revision of the  basal neornithischian taxa *Thescelosaurus* and *Bugenasaura*, *Journal of Vertebrate  Palaeontology*, **29(3)**, 758-770

Brill, K. and Carpenter, K. (2007) A description of a new ornithopod from the Lytle Member of the Purgatoire Formation (Lower Cretaceous) and a reassessment of the skull of *Camptosaurus*, In: Carpenter, K. (ed.), *Horns and Beaks, Ceratopsian and Ornithopod Dinosaurs*, 49-67, Indiana University Press

Burton,  D., Greenhalgh, B. W., Britt, B. B., Kowallis, B. J. Elliott, W. S. Jr. and Barrick, R. (2006) New radiometric ages from the Cedar Mountain Formation, Utah and the Cloverly Formation, Wyoming: implications for contained dinosaur faunas, *Geological Society of America Abstracts with Programs*, **38(7)**, 52a

Butler, R. J., Smith, R. M. H. and Norman, D. B. (2007) A primitive ornithischian from the Late Triassic of South Africa, and the early evolution and diversity of the Ornithischia, *Proceedings of the Royal Society*, **274**, 2041-2046

Butler, R. J. and Barrett, P. M. (2008) Palaeoenvironmental controls on the distribution of Cretaceous herbivorous dinosaurs, *Naturwissenschaften*, **95**, 1027-1032

Butler, R. J. and Galton, P.M. (2008) The 'dermal armour' of the ornithopod dinosaur *Hypsilophodon* from the Wealden (Early Cretaceous: Barremian) of the Isle of Wight: a reappraisal, *Cretaceous Research*, **29**, 636-642

Butler, R. J., Upchurch, P. and Norman, D. B. (2008) The phylogeny of the Ornithischian dinosaurs, *Journal of Systematic Palaeontology*, **6(1)**, 1-40

Butler, R. J., Barrett, P. M., Kenrick, P. and Penn, G. P. (2009) Testing co-evolutionary hypotheses over geological timescales: interactions between Mesozoic non-avian dinosaurs and cycads, *Biological Reviews*, **84**, 73-89

Carpenter, K. and Wilson, Y. (2008) A new species of *Camptosaurus* (Ornithopoda: Dinosauria) from the Morrison Formation (Upper Jurassic) of Dinosaur National Monument, Utah, and a biomechanical analysis of its forelimb, *Annals of Carnegie Museum*, **76(4)**, 227-263

Chinsamy, A. (1995) Ontogenetic changes in the bone histology of the Late Jurassic ornithopod *Dryosaurus lettowvorbecki*, *Journal of Vertebrate Palaeontology*, **15(1)**, 96-104

Cifelli, R. L., Gardner, J. D., Nydam, R. L. and Brinkman, D. L. (1997) Additions to the vertebrate fauna of the Antlers Formation (Lower Cretaceous), south-eastern Oklahoma, *Oklahoma Geology Notes*, **57(4)**, 124-131

Cifelli, R. L., Wible, J. R. and Jenkins Jr., F. A. (1998) Triconodont mammals from the Cloverly Formation (Lower Cretaceous), Montana and Wyoming, *Journal of Vertebrate Palaeontology*, **18**, 237-241

Coria, R. A. and Calvo, J.O. (2002) A new iguanodontian Ornithopod from Neuquen Basin, Patagonia, Argentina, *Journal of Vertebrate Palaeontology*, **22(3)**, 503-509

Coria, R. A., Cambiaso, A.V. and Salgado, L. (2007) New records of basal ornithopod dinosaurs in the Cretaceous of North Patagonia, *Ameghiniana*, **44(2)**, 473-477

Coria, R. A. and Salgado, L. (1996) A basal Iguanodontian (Ornithischia: Ornithopoda) from the Late Cretaceous of South America, *Journal of Vertebrate Palaeontology*, **16(3)**, 445-457

DeCelles, P. G. and Burden, E. T. (1992) Non-marine sedimentation in the overfilled part of the Jurassic-Cretaceous Cordilleran foreland basin: Morrison and Cloverly Formations, Wyoming, USA, *Basin Research*, **4**, 291-313

Dodson, P. (1980) Comparative osteology of the American Ornithopods *Camptosaurus* and *Tenontosaurus, Mémoires de la Société Géologique de France, **139**, 81-85*

Dodson, P. and Madsen Jr., J. H. (1981) On the sternum of *Camptosaurus*, *Journal of Palaeontology*, **55(1)**, 109-112

Elliott Jr., W. S., Suttner, L. J. and Pratt, L. M. (2007) Tectonically induced climate and its control on the distribution of depositional systems in a continental foreland basin, Cloverly and Lakota Formations (Lower Cretaceous) of Wyoming, U.S.A., *Sedimentary Geology*, **202**, 730-753

Forster, C. A. (1984) The palaeoecology of the ornithopod dinosaur *Tenontosaurus tilletti* from the Cloverly Formation, Bighorn Basin of Wyoming and Montana, *The Mosasaur*, **2**, 151-163

Forster, C. A. (1990) The postcranial skeleton of the ornithopod dinosaur *Tenontosaurus tilletti*, *Journal of Vertebrate Palaeontology*, **10(3)**, 273-294

Forster, C. A. (1990b) Evidence for juvenile groups in the ornithopod dinosaur *Tenontosaurus tilletti* Ostrom, *Journal of Palaeontology*, **64**, 164-165

Galton, P. M. (1971a) *Hypsilophodon*, the cursorial non-arboreal dinosaur, *Nature*, **231**, 159-161

Galton, P. M. (1971b) The mode of life of *Hypsilophodon*, the supposedly arboreal Ornithopod dinosaur, *Lethaia*, **4**, 453-465

Galton, P. M. (1973) The cheeks of Ornithischian dinosaurs, *Lethaia*, **6**, 67-89

Galton P.M. (1974a) The Ornithischian dinosaur *Hypsilophodon* from the Wealden of the Isle of Wight, *Bulletin of the British Museum (Natural History) Geology*, **25**, 1–152c

Galton, P.M. (1974b) Notes on *Thescelosaurus*, a conservative ornithopod dinosaur from the Upper Cretaceous of North America, with comments on ornithopod classification, *Journal of Palaeontology*, **48(5)**, 1048-1067

Galton, P. M. (1976) The dinosaur *Vectisaurus valdensis* (Ornithischia: Iguanodontidae) from the lower Cretaceous of England, *Journal of Palaeontology*, **50(5)**, 976-984

Galton, P. M. (1981) *Dryosaurus*, a hypsilophodontid dinosaur from the Upper Jurassic of North America and Africa postcranial skeleton, *Paläontologische Zeitschrift*, **55**, 271-312

Galton, P. M. (2007) Teeth of ornithischian dinosaurs (mostly Ornithopoda) from the Morrison Formation (Upper Jurassic) of the western United States, In: Carpenter, K. (ed.), *Horns and Beaks, Ceratopsian and Ornithopod Dinosaurs*, 17-47, Indiana University Press

Galton, P. M. (2009) Notes on Neocomian (Lower Cretaceous) ornithopod dinosaurs from England – *Hypsilophodon, Valdosaurus, "Camptosaurus", "Iguanodon"* – and referred specimens from Romania and elsewhere, *Revue de Paléobiologie*, **28(1)**, 211-273

Galton, P. M. and Jensen, J. A. (1975) *Hypsilophodon* and *Iguanodon* from the lower Cretaceous of North America, *Nature*, **257**, 668-669

Galton, P.M. and Powell, P. H. (1980) The Ornithischian dinosaur *Camptosaurus prestwichii* from the Upper Jurassic of England, *Journal of Palaeontology*, **23(2)**, 411-443

Galton, P. M. and Taquet, P. (1982) *Valdosaurus*, a hypsilophodontid dinosaur from the Lower Cretaceous of Europe and Africa, *Geobios*, **15**, 147-159

Godefroit, P., Codrea, V. and Weishampel, D. B. (2009) Osteology of *Zalmoxes shqiperorum* (Dinosauria, Ornithopoda) based on new specimens from the Upper Cretaceous of Nălaţ-Vad (Romania), *Geodiversitas*, **31(3)**, 525-553

Grigorescu, D. (2010) The Latest Cretaceous fauna with dinosaurs and mammals from the Haţeg Basin - A historical overview, *Palaeogeography, Palaeoclimatology, Palaeoecology*, doi: 10.1016/j.palaeo.2010.01.030

Head, J. J. (2001) A reanalysis of the phylogenetic position of *Eolambia Caroljonesa* (Dinosauria, Iguanodontia), *Journal of Vertebrate Palaeontology*, **21(2)**, 392-396

Hilton, R. P., Decourten, F. L., Murphy, M. A., Rodda, P. U. and Embree, P. G. (1997) An early Cretaceous ornithopod dinosaur from California, *Journal of Vertebrate Palaeontology*, **17(3)**, 557-560

Horne, G. S. (1994) A mid-Cretaceous ornithopod from Central Honduras, *Journal of Vertebrate Palaeontology*, **14**, 147-150

Horner, J. R., De Ricqles, A., Padian, K. and Scheetz, R. D. (2009) Comparative long bone histology and growth of the "hypsilophodontid" dinosaurs *Orodromeus makelai*, *Dryosaurus altus*, and *Tenontosaurus tilletti* (Ornithischia: Euornithopoda), *Journal of Vertebrate Palaeontology*, **29(3)**, 734-747

Huxley, T. H. (1869) On Hypsilophodon foxii, a new dinosaurian from the Wealden of the Isle of Wight, *Quarterly Journal of the Geological Society*, **16(1)**, 3-12

Jun, C., Butler, R. and Liyong, J. (2008) New material of large-bodied ornithischian dinosaurs, including an iguanodontian ornithopod, from the Quanto Formation (middle Cretaceous: Aptian-Cenomanian) of Jilin Province, northeastern China, *Neues Jahrbuch für Geologie und Paläontologie, Abhandlungen*, **248(3)**, 309-314

Kim, J. Y., Lockley, M. G., Kim. H. M., Lim, J-D. and Kim, K. S. (2009) New dinosaur tracks from Korea, *Ornithopodichnus masanensis* ichnogen. et ichnosp. nov. (Jindong Formation, Lower Cretaceous): implications for polarity in ornithopod foot morphology, *Cretaceous Research*, **30**, 1387-1397

Knoll, F. (2009) A large iguanodont from the Upper Barremian of the Paris Basin, *Geobios*, **42(6)**, 755-764

Kobayashi, Y. and Azuma, Y. (2003) A new Iguanodontian (Dinosauria: Ornithopoda) from the lower Cretaceous Kitadani Formation of Fukui Prefecture, Japan, *Journal of Vertebrate Palaeontology*, **23(1)**, 166-175

Lee, A. H. and Werning, S. (2008) Sexual maturity in growing dinosaurs does not fit reptilian growth models, *Proceedings of the National Academy of Sciences,* **105(2)**, 582-587

Li, R., Lockley, M. G., Makovicky, P. J., Matsukawa, M., Norell, M. A., Harris, J. D. and Liu, M. (2008) Behavioural and faunal implications of Early Cretaceous deinonychosaur trackways from China, *Naturwissenschaften*, **95**, 185-191

Lockley, M. G. and Matsukawa, M. (1999) Some observations on trackway evidence for gregarious behaviour among small bipedal dinosaurs, *Palaeogeography, Palaeoclimatology, Palaeoecology*, **150**, 25-31

Maidment, S. C. R. and Porro, L. B. (2010) Homology of the palpebral and origin of supraorbital ossifications in ornithischian dinosaurs, *Lethaia*, **43**, 95-111

Manning, P. L., Payne, D., Pennicott, J., Barrett, P. M. and Ennos, R. A. (2006) Dinosaur killer claws or climbing crampons?, *Biology Letters*, **2**, 110-112

Mateus, O. and Antunes, M. T. (2001) *Draconyx loureiroi*, a new Camptosauridae (Dinosauria, Ornithopoda) from the Late Jurassic of Lourinhã, Portugal, *Annales de Paleontologie*, **87**, 61-73

Maxwell, D. and Horner, J. R. (1994) Neonate dinosaur remains and dinosaurian eggshell from the Cloverly Formation, Montana, *Journal of Vertebrate Palaeontology*, **14(1)**, 143-146

Maxwell, W. D. and Ostrom, J. H. (1995) Taphonomy and palaeobiological implications of *Tenontosaurus-Deinonychus* associations, *Journal of Vertebrate Palaeontology*, **15(4)**, 707-712

Meyers, J. H., Suttner, L. J., Furer, L. C., May, M. T. and Soreghan, M. J. (1992) Intrabasinal tectonic control on fluvial sandstone bodies in the Cloverly

Formation (early Cretaceous), west-central Wyoming, USA, *Basin Research*, **4**, 315-333

Molnar, R. E. and Galton, P. M. (1986) Hypsilophodontid dinosaurs from Lightning Ridge, New South Wales, Australia, *Geobios*, **19(2)**, 231-239

Moreno, K., Carrano, M. T. and Synder, R. (2007) Morphological changes in pedal phalanges through ornithopod dinosaur evolution: a biomechanical approach, *Journal of Morphology*, **268**, 50-63

Norman, D. B. (1990) A review of *Vectisaurus valdensis,* with comments on the family *Iguanodontidae*. In Carpenter, K. and Currie, P. J. (eds.) *Dinosaur Systematics*, 147-161, Cambridge University Press

Norman, D. B. (1998) On Asian ornithopods (Dinosauria: Ornithischia). 3. A new species of iguanodontid dinosaur, *Zoological Journal of the Linnean Society*, **122**, 291-348

Norman, D. B. (2004) Basal Iguanodontia. In Weishampel, D. B., Dodson, P. and Osmólska, H. (eds.) *The Dinosauria*, 413-437, University of California Press

Norman, D. B. and Weishampel, D. B. (1985) Ornithopod feeding mechanisms: their bearing on the evolution of herbivory, *The American Naturalist*, **126(2)**, 151-164

Nydam, R. L. and Cifelli, R. L. (2002) Lizards from the Lower Cretaceous (Aptian-Albian) Antlers and Cloverly Formations, *Journal of Vertebrate Palaeontology*, **22(2)**, 286-298

Oreska, M. P., Carrano, M. T. and Lockwood, R. (2007) Palaeoecology of the Cloverly Formation (Lower Cretaceous) vertebrate fauna from microvertebrate sites in the Bighorn Basin, Wyoming, *Geological Society of America Abstracts with Programs*, **39(5)**, 10a

Organ, C. L. (2006) Biomechanics of ossified tendons in ornithopod dinosaurs, *Palaeobiology*, **32(4)**, 652-665

Organ, C. L. and Adams, J. (2005) The histology of ossified tendons in dinosaurs, *Journal of Vertebrate Palaeontology*, **25(3)**, 602-613

Ostrom, J.H. (1969) Osteology of *Deinonychus antirrhopus*, an unusual theropod from the lower Cretaceous of Montana, *Bulletin of the Peabody Museum of Natural History*, **30**, 165pp

Ostrom, J. H. (1970) Stratigraphy and palaeontology of the Cloverly Formation (lower Cretaceous) of the Bighorn Basin area, Wyoming and Montana, *Bulletin of the Peabody Museum of Natural History*, **35**, 234pp

Paul, G. S. (2007) Turning the old into the new: a separate genus for the gracile iguanodont from the Wealden of England, In: Carpenter, K. (ed.), *Horns and Beaks, Ceratopsian and Ornithopod Dinosaurs*, 69-77, Indiana University Press

Paul, G. S. (2008) A revised taxonomy of the iguanodont dinosaur genera and species, *Cretaceous Research*, **29**, 192-216

Pisani, D., Yates, M. A., Langer, M. C. and Benton, M. J. (2002) A genus-level supertree of the Dinosauria, *Proceedings of the Royal Society of London*, **269**, 915-921

Roach, B. T. and Brinkman, D. L. (2007) A re-evaluation of cooperative pack hunting and gregariousness in *Deinonychus antirrhopus* and other non-avian theropod dinosaurs, *Bulletin of the Peabody Museum of Natural History*, **48(1)**, 103-138

Ruiz-Omeñaca, J. I., Superbiola, X. P. and Galton, P. M. (2007) *Callovosaurus leedsi*, the earliest dryosaurid dinosaur (Ornithischia: Euornithopoda) from the Middle Jurassic of England, In: Carpenter, K. (ed.), *Horns and Beaks, Ceratopsian and Ornithopod Dinosaurs*, 3-16, Indiana University Press

Sachs, S. and Hornung, J. J. (2006) Juvenile ornithopod (Dinosauria: Rhabdodontidae) remains from the Upper Cretaceous (Lower Campanian, Gosau Group) of Muthmannsdorf (Lower Austria), *Geobios*, **39**, 415-425

Sanz, J.L., Santafe, J.V. and Casanovas, L. (1983) Wealden Ornithopod dinosaur *Hypsilophodon* from the Capas Rojas Formation (lower Aptian, lower Cretaceous) of Morella, Castellon, Spain, *Journal of Vertebrate Palaeontology*, **31(1)**, 39-42

Schachner, E. R. And Manning, P. L. (2008) 3D trauma analysis using X-ray microtomography in *Tenontosaurus tilletti* (Montana, USA), *Journal of Vertebrate Palaeontology*, **28(3)**, 136a

Sereno, P. C. (1986) Phylogeny of the bird-hipped dinosaurs (order: Ornithischia), *National Geographic Research*, **2**, 234-256

Sereno, P. (1999a) The evolution of dinosaurs, *Science*, **284**, 2137-2147

Sereno, P. (1999b) Definitions in phylogenetic taxonomy: critique and rationale, *Systematic Biology*, **48(2)**, 329-351

Sereno, P. (1999c) A rationale for dinosaurian taxonomy, *Journal of Vertebrate Palaeontology*, **19(4)**, 788-790

Sereno, P. (2005) The logical basis of phylogenetic taxonomy, *Systematic Biology*, **54(4)**, 565-619

Stokes, W. M. L. (1987) Dinosaur gastroliths revisited, *Journal of Palaeontology*, **61(6)**, 1242-1246

Tang, F., Luo, Z.-X., Zhou, Z.-H., You, H.-L., Georgi, J.A., Tang, Z.-L. and Wang, X.-Z. (2001) Biostratigraphy and palaeoenvironment of the dinosaur-bearing sediments in Lower Cretaceous of Mazongshan area, Gansu Province, China, *Cretaceous Research,* **22**, 115-129

Taquet, P. and Russell, D. A. (1999) A massively-constructed iguanodont from Gadoufaoua, Lower Cretaceous of Niger, *Annales de Paleontologie*, **85**, 85-96

Varricchio, D. J., Martin, A. J. and Katsura, Y. (2007) First trace and body fossil evidence of a burrowing, denning dinosaur, *Proceedings of the Royal Society*, **274,** 1361-1368

Weishampel, D. B. and Bjork, P. R. (1989) The first indisputable remains of *Iguanodon* (Ornithischia: Ornithopoda) from North America: *Iguanodon lakotaensis* n. sp., *Journal of Vertebrate Palaeontology*, **9**, 56-66

Weishampel, D. B. and Heinrich, R. E. (1992) Systematics of Hypsilophodontidae and basal Iguanodontia (Dinosauria: Ornithopoda), *Historical Biology*, **6**, 159-184

Weishampel, D. B. and Jianu, C-M. (2000) Plant-eaters and ghost lineages: dinosaurian herbivory revisited. In: H-D. Sues, eds. *Evolution of Herbivory in Terrestrial Vertebrates: Perspectives from the Fossil Record*, 123–143, Cambridge University Press

Weishampel, D. B., Jianu, C-M., Csiki, Z. and Norman, D.B. (2003) Osteology and phylogeny of *Zalmoxes* (N. G.), an unusual euornithopod dinosaur from the latest Cretaceous of Romania, *Journal of Systematic Palaeontology*, **1(2)**, 65-123

Weishampel, D. B., Csiki, Z., Benton, M. J., Grigorescu, D. and Codrea, V. (2010) Palaeobiogeographic relationships of the Haţeg biota — between isolation and innovation, *Palaeogeography, Palaeoclimatology, Palaeoecology*, doi:10.1016/j.palaeo.2010.03.024

Werning, S. A. (2005) Ontogeny and osteohistology of the ornithopod dinosaur *Tenontosaurus tilletti*, *Journal of Vertebrate Palaeontology*, **25(3)**, 128-129a

Wiffen, J. and Molnar, R. E. (1989) An Upper Cretaceous ornithopod from New Zealand, *Geobios*, **22(4)**, 531-536

Winkler, D. A., Murry, P. A. and Jacobs, L. L. (1997) A new species of *Tenontosaurus* (Dinosauria: Ornithopoda) from the Early Cretaceous of Texas, *Journal of Vertebrate Palaeontology*, **17(2)**, 330-348

You, H-L., Luo, Z-X., Shubin, N. H., Witmer, L. M., Tang, Z-L. and Tang, F. (2003) The earliest-known duck-billed dinosaur from deposits of late Early Cretaceous age in northwest China and hadrosaur evolution, *Cretaceous Research*, **24**, 347-355

# 9.0 Appendices

## 9.1 Appendix 1: Morphological Description of LL.12275 (*Tenontosaurus tilletti*)

### 9.11 Skull

*Premaxilla*: 'snout' or 'beak' comprised entirely of edentulous premaxilla - major contrast from *Heterodontosaurus* and *Hypsilophodon* which exhibit premaxillary teeth; premaxilla surrounds and envelops rostral sector of narial opening; curves sharply ventrally and flares out into transversely expanded 'beak'; groove runs down midline increasing slightly in depth anteroventrally; posterodorsal projections extend at 45° to horizontal with thin rod-like form; sinistral projection mostly absent, and dextral one missing dorsal extremity thus occluding assumed upper contact with absent nares.

Anterior margins vertical before becoming dorsoventrally canted into listric profile forming ventral border of narial opening; convexity extends as lateral groove along upper projections, gradually thinning and terminating one-third of length - feature absent in partially restored sinistral side; tip of 'beak' cratered with small circular to subcircular pits on anteroventral border; distinct symmetry about midline in cranial view - only heterogeneity being variably sized, rounded ventral projections adjacent to midline (synapomorphic of higher iguanodontians).

Contact between premaxilla and maxilla largely obscured - appears somewhat linear where visible; mediolaterally expanded ventral border ceases abruptly contacting more slender and planar maxilla immediately rostrodorsally to first tooth; paired premaxillae unfused in ventral aspect - each border slopes ventrolaterally from medial edge contacting outermost extremity of flared 'beak'.

*Maxilla*: mostly absent, although structure tentatively inferred based on bounding elements form; sinistral side comprises fragmentary shards with poorly preserved uninformative teeth; dextral side preserves ventral border with many well-preserved teeth - extends for 15mm dorsally before plaster covering; anterior contact with premaxilla visible up to one-third of distance up external nares; perceived that maxilla completely encloses antorbital fenestra, although potential role of absent lachrymal cannot be discarded. Contact with ventral half of jugal visible occurring as suture slightly dorsal to final tooth position; ventral border of orbit possibly formed by part of maxilla. Ostrom (1970) observed that the maxilla is "long and very high", and although the structure of the maxilla here is indeterminable, the dimensions are easier to estimate; this observation is deemed correct. Despite this, configuration of cheekbones largely speculatative due to prevalent absence of elements.

Maxillary tooth row extends from posterior extremity of premaxilla contacting between maxilla and jugal; alveoli dorsally concave - single tooth inhabits individual socket; teeth overlap each other slightly posteriorly; Dorsally from row, maxilla thickens laterally before sloping back ventromedially - truncated by lower premaxilla projection anteriorly and expansion of jugal posteriorly; slope contains linear row of anteroposteriorly arranged, evenly spaced sub-spherical pockmarks, located 10mm dorsal to tooth row.

*Jugal*: distinct sub-triangular 'arrow-head' shape; maximum height approximately maximum width; individual apices point rostrally and caudally, with third forming tapering process creating ventral half of lateral temporal fenestra. Unknown whether dorsal border of jugal or posterior process of maxilla forms ventral border of orbit. Contact with lachrymal also unknown.

Ventral border anteroventrally canted, continuing as ventral border of quadrate; thickened rostrolateral projection located approximately three-quarters of way towards maxilla. Posterior border jagged; does not appear sutured with quadratojugal - instead appears content to rest anteriorly to quadratojugal with slight overlap (characteristic or function of preservation or restoration unknown); border also curves inwards rostrally ascending into posterodorsal edge of "superior process" of Ostrom, (1970) forming lower quadrant of lateral temporal fenestra; border thinner than remaining jugal.

Dorsal margin forms somewhat 'exponential' curved contact with maxilla and postorbital; partially restored but exhibits symmetry on either side of skull so assumed true attitude of contact; contact with postorbital also unsutured; postorbital overlaps jugal laterally in thickened posterior border of orbit, forming robust 'column' between orbit and lateral temporal fenestra; rest of orbit is formed from frontal and presumably restored lachrymal.

*Nares*: mostly absent; dorsal aspect consists of disassociated fragments within plaster matrix adjacent to narial-frontal suture; prefrontals also both absent; form however is depicted by preserved contact with jugals; presumed suture with planar anterior edge directed mediolaterally before bending strongly posterolaterally; strongly symmetrical about longitudinal midline; border forms gently curved posterior three-quarters of dorsal margin of orbit.

*Postorbital*: ventral surface rests gently on sloping dorsal margin of jugal, curving and extending posteriorly forming upper half of rostral and dorsal margins of lateral temporal fenestra; anterior margin forms entire posterior edge of orbital before contacting frontal; border strongly deflects half-way up before thickening mediolaterally expanding into brain chamber forming base of frontal and parietal; dorsal contact fully sutured forming robust arch between lateral temporal fenestra, orbital and supraorbital fenestra.

Dorsal surface sub-horizontal with modest dorsal expansion two-thirds length posterior to postorbital-frontal suture; general appearance subrectangular, medially flattened process creating dorsal border of lateral temporal fenestra resting vertically against parietal.

Ventral border smooth continuation of ventral margins of both quadrate and jugal; jugal-quadratojugal contact represents transition from blade-like jugal to massive medially thickened quadratojugal, forming medially-projecting stout process contacting complex internal walls of braincase.

*Quadratojugal*: moderately thick dorsoventrally elongated element comprising anteroventral portion of lateral temporal fenestra - curves anterolaterally forming slender dagger-like process resting and abruptly ceasing against quadrate - partially obscured by 'random' restorative material projecting into lateral temporal fenestra. Rest of quadrate-quadratojugal contact vertical suture, terminating at horizontal process of frontal with no suture evident; lower half of quadratojugal greatly depressed in lateral aspect.

*Quadrate*: greatly thickened ventrally into posterodorsal-anterolateral expansion forming transversely robust articular head with mandible; general form dorsoventrally elongate sub-rectangle, dominantly narrow except anterior thickening conforming to quadratojugal; posterior border mostly vertical - expands 80% of way up curving anteriorly contacting paraoccipital process (non-sutured juxtaposition); narrowest point occurs immediately before curve; generates mildly convex form with concave extremities; dorsal extremity sub-rounded fitting somewhat weakly between frontal, paraoccipital process and squamosal, forming dominantly orthogonal quad-convergence.

*Frontal*: flattened transversely laying uniformly horizontal in dorsal aspect dominating posterior half of cranium; mid-section thickened ventrally; dorsal surface striated - striae run laterally from midline in symmetrical 'blossoming' pattern; midline suture between frontals strong and well-defined continuing posteriorly into parietals fusing bones into broad, robust structure covering brain chamber; frontals terminate posteriorly at parietal suture, therefore do not contribute to supratemporal fenestrae; contact jagged and difficult to trace continuously - defined by well-developed mediolateral ridges; expansion mirrored partially ventrally where parietals expand ventrally forming posterior wall of brain chamber before diverging and flaring laterally. Exact internal morphology largely obscured by copious restorative material; however general shape and structure somewhat discernible.

*Parietal*: thickened and massive forming paired fused tetra-ridged cross with exceptional symmetry; greatest width in dorsal aspect occurs at postorbital-parietal contact - forms massively thickened anterior wall descending into supratemporal fenestra forming anterior and medial walls. Two ridges occur pre-supratemporal fenestra (rostrally) curving medially meeting at longitudinal medial line before deviating posteriorly passing between supratemporal fenestrae, before diverging again into two fin-like ridges deflecting acutely at 45° from midline; inner curve of structure surrounds anterior half of squamosal and supratemporal fenestra with squamosal resting unfused ventrally; posterior walls of two posterolaterally-directed ridges strongly fused to supraoccipital.

*Squamosal*: sit unsutured or partially sutured between parietals, postorbitals, paraoccipital processes and proximal tip of quadrate - fairly massive, medially thickened and with sub-polygonal form; rest entirely posterior to supratemporal fenestra forming posterior wall with very thin lamina of bone; anterior border expands medially into thin, poorly developed process laying adjacent against ridge-intersection of parietals (i.e. point of greatest curvature); medial edge conforms to more robust parietal with strongly convex form; does appear to have been process running parallel to paraoccipital process - partially absent; process laid dorsal to quadrate or interposed between quadrate and paraoccipital process; lateral edge of squamosal horizontal and unsutured to adjacent postorbital.

Occipital surface comprises supraoccipital and paraocciptal; former massively constructed with subtriangular profile conforming to parietals; inclined posteroventrally contacting paraoccipital processes laterally. Occipital condyle formed mostly of basioccipital, although relationship difficult to discern as partially obscured by restorative material; forms from ventral deflection of paraoccipitals and supraoccipitals expanding and thickening posteriorly into dorsally-hooked sub-crescentic condyle. Paraoccipitals sub-vertical lateral expansions or processes, of the supraoccipital; sinistral process missing distal tip, dextral element mostly restored; anterior edge of processes lays flat against posterolateral-facing edge of squamosal, slightly contacting parietal; may also contribute partially to occipital condyle; Morphologically, processes are narrow, hooked blades curving ventrally resting against dorsal curve of quadrate.

*Lower* jaw: fairly robust, similar maximum and minimum length and width to cranium; parallel dorsal and ventral margins, straight in lateral aspect; depth consistent along length of dentary and predentary, until end of tooth row anterior to coronoid process; jaw angle approximately 25°, flaring slightly posteriorly.

Predentary caps symphysis; pincer-shaped (dorsal aspect); congruent depth with dentary; two lateral rami lie in dorsally-facing anterior dentary border – flare slightly at dorsal margin, thicken ventrally conforming to dentary thickness; terminates with associated border slightly before primary tooth, where curved predentary deflects laterally into linear tooth row. Anterior margin strongly curves ventrally, thinning and contacting dentary junction, cupping the symphysis suture (linear, sub-vertical); forms broad sheath (anterior aspect) expanding dorsally into lateral rami. Dorsal margin moderately straight with sharp crest between lateral and shallower-sloping medial borders. Irregular, small, semi-rounded projections from dorsal margin anteriorly; also marked with small circular holes - terminate anteriorly; larger, conical holes variably clustered. Inner border slopes posteriorly and laterally connecting with anteroventral dentary border.

Dentary with parallel dorsal and ventral borders; 12-13 teeth occupy linear row; 75% total lower jaw length between predentary and surangular; laterally slopes away from tooth row into robust, broadly curving outer margin, sharply terminating ventrally contacting the sub-vertical medial border – extends in linear fashion to inner tooth margin. Small pock marks (maximum 3mm diameter) form a band approximately 40% of dentary depth – comet-like voids with 'tail' pointing anteriorly.

Posteriorly (adjacent to penultimate tooth), surface smoothly curves ventrally into a gracile, somewhat rectangular "coronoid process" (lateral aspect). Ventral border deflects and expands ventrally (two-thirds length anteriorly from predentary contact) into thin sheet, expanding anteriorly and ventrolaterally into angular – forms 'flattened-S' shape in dextral ventral aspect.

Horizontal dorsal margin deflects orthogonally (ventrally) into straight posterior margin, curving posteriorly into small notch (directed dorsolaterally) before tapering into "coronoid process" of surangular (Ostrom, 1970). Sinistral surangular partially obscured and restored (mostly ventral portion); suture between dentary and surangular runs anteroventrally in smooth curve prior to small medial deflection, becoming more complex and ultimately tapering to terminate approximately half lower jaw length – at point of deflection, angular suture appears, running sub-parallel to lower jaw ventral border, terminating adjacent to point of beginning of coronoid process.

Dorsal border of surangular broadly expands ventromedially before twisting slightly into gracile posterior process – posterior-most tip absent, therefore extent unknown; process deflects dorsally and thins posteriorly.

### 9.12 Pectoral Girdle

*Sternum*: very thin, brittle elements, and well-ossified; form elongated 'kidneys'; moderately crescentic with rounded points; lay symmetrically about median plane immediately caudal to coracoids; degree of concavity or convexity unknown as cross-sectional profile disrupted by bending or warping (presumably taphonomic aspect). Smooth outer surface with rougher inner surface; caudal edge of ventral surface heavily stippled; apparent thickness uniform except at caudal and cranial borders where slightly thickened (approximately twice average thickness); both plates somewhat broader cranially than caudally with cranial, medial and caudal borders forming continuously curved margin; greatest curvature at extremities – intersects considerably less concave lateral border.

*Scapula*: long gracile component, approximately three times maximum width at glenoid, with variable thickness longitudinally; generally forms gently curved blade with thickened cranial third adjacent to coracoid.

Dorsal expansion represents flaring of blade to approximately 120% of width to equal thickness at glenoid extension; form mildly concave in dorsal aspect - presumably where medial edge conforms to structure of ribcage.

Thickness inconsistent - dorsal border approximately half thickness at glenoid expansion; also varies caudocranially - thickest in centre giving elongated ellipsoid profile; medial surface pre-expansion distally contains prominent bulge ("thickened ridge, or buttress" of Forster (1990)) central to blade - acts to almost double breadth; decreases in thickness immediately pre-expansion of glenoid, approximately two-thirds of distance from cranial border; expansion 'reflected' and continued ventrally along coracoid for three-quarters of width from midpoint of coracoid articulation.

Distal two-thirds of lateral surface moderately rugose with last 40mm becoming less striated - stippled instead; caudal border more concave than cranial border - gentler at dorsal edge where caudal divergence greatest; dorsal edge becomes slightly convex reflecting expansion; caudal border deflects proximally at 70° developing glenoid.

Above morphology of dextral scapula not perfectly mirrored in sinistral scapula: degree of divergence concordant, but dorsal border strongly convex and crescentic, generally symmetrical about longitudinal midline but still quite irregular; also partially thickened although highly variable along width - reason for feature unknown, although considering its inimitability it may be pathological resulting from lateral trauma on sinistral flank. Despite variation, cranial expansion at distal end just as slight and equal to sinistral scapula.

Cranioventral border obliquely slanted, obtuse, and moderately-rounded oval form; inclined lengthways caudoventrally; does not form part of scapula-coracoid suture; no visible acromion process as in *Hypsilophodon*, *Camptosaurus*, *Iguanodon* and *Ouranosaurus* (Dodson, 1980) – also absent in hadrosaurs, therefore possibly a derived feature. No clavicular facet as *Camptosaurus* (Dodson, 1980).

Where expansion of glenoid initiated, large depression with broad concavity begins in lateral surface - reflected on medial border by medially-directed expansion;

depression uniform except for lateral deflection and expansion of glenoid and extension of cranioventral border; flattens off immediately adjacent to scapula-coracoid suture with slight invariable thickening of ventral border; caudoventral border dictated primarily by form of glenoid; caudoventral edge forms from an abrupt deflection of dorsal border - initially flat before curving slightly caudally intersecting coracoid articular surface at most caudal point.

Glenoid fossa cranially depressed, caudal-facing surface orientated obliquely to coracoid articulation; develops acute angle with deflected caudal border of scapular blade; dorsal edge comprises thickest part of scapula giving distinct asymmetry compared to moderate expansion of cranioventral surface - role currently unknown. In ventral aspect glenoid laterally curved reflecting inward curving of medial border; plays no part in scapula-coracoid articulation; form reflected in opposite curvature of caudolateral border of coracoid.

Differences between scapulae distinct: primary bulge on medial surface much less prominent on sinistral member, forming no continuous trend towards articular surface - instead develops mild crease; immediately ventral small subcircular foramen located - genuine structure or result of taphonomy or preservation impossible to determine, thus considered ambiguous presently in context of morphological characteristics; absence in sinistral scapula may be result of infilling - moderately crystalline mass occludes this area of scapula. Another potentially significant feature is presence of 2-3cm long teardrop-shaped growth located on cranial margin approximately 80% of length from dorsal extremity; feature appears authentic and not resultant from improper restoration, thus presence proves somewhat puzzling.

*Coracoid*: broad, moderately thick with equidimensional width and height in lateral aspect; sinistral and dextral members virtually identical; sternal process develops as rounded tip with acute angle where ventral and caudal borders intersect; process fairly prominent and slightly hooked distally; dextral sternal process slightly more curved and sickle-shaped than respective sinistral development.

Articular surface with scapula strongly fused, thicker caudally, progressively tapers cranially conforming strongly to respective dorsal scapular border; continues developing glenoid fossa with crescentic form; lower two-thirds of articular surface greatly medially thickened before thinning ventrally into uniformly thin body; decrease less prominent along caudal border and glenoid surface; mid-way down articular surface rounded-ridge structure develops extending for approximately half coracoid width occupying one quarter of length - propagates into slight surficial depression.

Dorsal articular border vertical in lateral aspect; ventral and cranial surfaces orthogonal forming curved intersection; lateral surface smooth and lightly striated; caudal border obtuse to caudoventral-facing glenoid fossa surface curving gently into slightly thickened sternal process; mid-way down cranial border becomes mediolaterally expanded and laterally canted, which Forster (1990) defines as a "rugose muscle scar"; caudal border deeply but non-uniformly concave - more linear adjacent to intersection with ventral border at sternal process.

Cranioventral curvature deflects gently medially in medial aspect; slight depression in cranial third of body reflected in lateral expansion - maintains uniform thickness; only heterogeneity being slight cranial thickening.

Coracoid foramen circular with inwardly-sloping walls; centre is 25mm from articular surface with 12mm diameter concordant with Forster (1990) and dissimilar to *Camptosaurus* where it is more proximally placed; develops most prominently, but still shallow, on lateral surface centred relative to articulation as in Forster (1990); on medial surface (both coracoids) foramen completely obscured - only poorly preserved vestige remains as channel directed ventrally and orthogonal to articular surface.

### 9.13 Forelimb and Manus

*Humerus*: comprises transversely constricted shaft, massively expanded proximal head, robust pair of distal condyles, and well-developed strong deltopectoral crest, compared to the weaker form of *Camptosaurus*, *Hypsilophodon, Iguanodon* and *Ouranosaurus* (Dodson, 1980). Morphology strongly supports a quadrupedal mode of life.

Proximal caudal surface expands into moderately rugose humeral head (articular condyle) with semi-circular profile and minimal distal extent (approximately 10% of humeral length), tapering into caudal surface pre-constriction of shaft; apex slightly laterally offset from midline; proximal end transversely expanded into thick, caudally convex sheet progressively thinning distally; head forms caudocranially thickest proximal point; proximal surface bears  strongly convex profile; medial edge thicker than lateral edge - both intersected by thickened, discontinuous ridge expanding from proximal head.

Caudolateral surface exhibits shallow longitudinal channel running from articular surface to point on shaft adjacent to long axis of deltopectoral crest; on left humerus channel recurves medially wrapping obliquely around caudal surface of shaft and terminates at mid-point; caudal surface twists projecting anteriorly wrapping around lateral surface of deltopectoral crest inclined orthogonally to shaft; cranial surface mediolaterally concave with gentle longitudinal convexity; concavity converges distally into deep groove coursing adjacent and medial to deltopectoral crest - develops with strong curvature on thin and planar medial surface; groove terminates on shaft at point medial to deltopectoral crest apex.

Deltopectoral crest progressively thickens distally towards apex with maximum thickness achieved approximately half thickness of humeral head; cranial surface rugose and concave in medial view longitudinally and caudocranially; forms sharp medial border and somewhat recumbent lateral border; well-rounded, sub-cylindrical crest develops from cranial expansion of craniolateral surface creating gently concave surface ascending to apex which lays at 45% longitudinal humeral axis; distal surface recedes more rapidly than ascending proximal surface; lateral surface of deltopectoral crest mildly depressed distal to channel (approximately 45% of humeral length). Forster (1990) states this feature is between 40% for juveniles and 54% for adults confirming the ontogenetic position of this specimen.

Proximal medial surface well-rounded at articular surface, converging distally into uniformly thin and rounded surface; commences expansion and thickening at point approximately medial to refraction of apex of deltopectoral crest, merging into caudal surface of shaft - mergence complete at distal termination of deltopectoral crest.

Cranial surface post-deltopectoral crest distally becomes gently and broadly concave longitudinally as intercondylar groove develops; intersects medially skewed intercondylar groove on distal articular plane; caudal surface comprises gently concave shaft running distally into more proximally extensive, transversely narrower, but distally deeper caudal intercondylar groove. Shaft bowed slightly cranially in lateral aspect (more so distally); compensated by increased expansion of radial and ulnar condyles caudally which envelop deeper and well-confined groove; intercondylar grooves termed "coronoid fossa" and "olecranon fossa" by Forster (1990).

Distal tip of ulnar condyle with greater cranial expansion than caudally, forming part of broken ridge on anterior surface at heavily rugose distal articulate with slight thickening of radial condylar rim; mediolateral thickness approximately two times maximum caudocranial width; articular surface slightly expanded and rounded at each condyle, separated by flattened continuation of intercondylar grooves.

*Ulna*: moderately robust elements; rugose proximal articular surface dominated by massive and rounded olecranon process, cranial and lateral coronoid processes (CCP and LCP respectively) - latter two form an orthogonal association.

Olecranon process pervades onto articular surface of LCP with slight circumferential ridge tapering caudomedially into thin and shallow furrow; olecranon process flattened into cranially-directed planar slope extending onto laterally canted articular surface of CCP.

LCP has sub-triangular profile with well-rounded laterally projecting apex tapering rapidly into lateral surface; separated from CCP on surface by well-developed depression pervading no further than LCP; function of depression to accept associated radial head - also acts to inhibit distal continuation of CCP by creating mediolateral constriction. LCP expands proximally into ulnar head and distally into cranial surface of transversely narrow shaft.

CCP extends from cranial base of olecranon process into transversely narrow ridge; merges into shaft developing continuous concavity distally as it twists and expands in an equal and opposite manner to distal craniolateral surface to form distally expanded cranial surface.

Medial surface has craniocaudally expanded distal end, leading to concave median axis and slight longitudinal concavity; latter converges distally into well-developed depression at approximately one-third length of shaft on median axis. Cranial and caudal borders have well-rounded margins and are slightly flattened. Medial surface accommodates twisting of shaft by narrowing distally and rotating slightly caudally; this develops craniocaudally constricted, heavily rugose distal end of shaft.

Proximal end of caudal surface longitudinally convex accepting expansion of LCP - extends approximately two-thirds of length before twisting, forming expanded caudal and caudolateral borders that form gently arcuate, convex and rugose surface.

Twisting creates obliquely orientated long-axis of distal articulate to coronoid processes - surface rugose and sub-oval, canted craniolaterally where develops into thin cranially directed and rounded tab, separated from cranial border by gentle, proximally tapering expansion; circumference of distal head marginally thickened, and well-developed striae occupy distal sixth of surface adjacent to head; resultant cross-sectional profile of shaft is 'pinched-oval', developing distally into sub-rectangular form with marginally flattened surfaces.

Slight variation exists on left ulna: proximal depression absent on medial surface, instead forming slightly undulose plane; resultant groove well defined and longitudinally orientated; character distinct on both ulnae, but to varying degree; thus may be function of ontogeny as unlikely related to mortality or preservation/taphonomic processes.

*Radius*: simpler, slenderer and slightly shorter elements than associated ulnae; comprise well-developed head tapering into bowed shaft and distal end with equidimensional craniocaudal expansion to proximal head.

Head expanded craniocaudally, narrow transversely; rugose proximal articular surface; well-developed central depression surrounded by thickened, flattened ridge extending cranially developing rounded tab-like extension of proximally diverging shaft; expansion mirrored to lesser extent in caudal edge – thicker but not as extensive; leads to 'teardrop' proximal profile; head mediolaterally thickened to lesser degree immediately proximal to articular surface.

Proximal shaft well-rounded and oval cross-sectional profile; mediolaterally narrow due to flattened head; cranial surface pinches into thinner, rounded ridge extending to distal tip; caudal edge develops much gentler convexity retaining broad curvature – still thinner than proximal shaft transversely. Divergence of shaft gives conical form in lateral view; distal quarter of caudal and cranial borders converge slightly (cranial more so) becoming sub-parallel; dividing lateral surface striated and broadly concave caudocranially.

Long-axis of distal end oblique to proximal end – result of curvature and slight distal twisting. Cranial border slightly undulose; caudal border thicker and gently concave. Distal articular surface caudomedially canted, rugose, elliptical, occupying same area as associated ulnar surface. Craniomedial surface flattened proximal to articulate, tapers rapidly merging with medial surface. Small rugose area occupies surface on proximal cranial curvature – visible through distortion of otherwise linear striae.

*Manus*: metacarpal I (MCI): rhombohedral profile in dorsal view; lateral and medial borders concave; transversely expanded triangular rugose proximal articulate converges into stout shaft – consistently more rapid on ventral surface; well-rounded distal head; lateral surface dominated by shallow depression separating two rounded, rugose expansions projecting ventrolaterally and dorsolaterally; distal medial surface perforated by small circular fossa – may be of taphonomic origin; ventral and dorsal surfaces gently transversely convex; mid-section lateral and medial surfaces pinched into rounded narrow edge; distal head (articulate) well-worn – transverse flaring greater than dorsoventral thickening.

Metacarpal II (MC2): sub-rectangular proximal articulate – concave dorsoventrally, transversely convex with sharp edges; rapid convergence of surfaces into oval shaft with elongate mediolateral axis; shaft flexes slightly dorsally at midpoint giving gently sinuous surface in lateral aspect; ventral surface broadly concave longitudinally and flattened mediolaterally, with convex proximal end and concave distal end – separated by dorsoventrally constricted, semi-circular profiled shaft; distally lateral surface with rugose channel developing as surface diverges forming articular head – gently concave form separated by paired well-developed narrow ridges; distal medial surface rough and flattened with rounded dorsal and ventral borders; head flat mediolaterally, convex dorsoventrally – sub-rounded rectangular form with compressed lateral and ventral margins. Overall similar geometry to MCI but slightly broader and deeper.

Metacarpal III (MC3): generally similar geometry to MC2 but slightly less robust. Proximal articular surface elliptical, medial edge abruptly wraps perpendicularly around medial surface; surface concave mediolaterally and dorsoventrally, thickened circumferential border except at distal termination of deflected surface; dorsal surface proximally depressed – not reflected in the planar flattened ventral surface – laterally rugose 'bump' present, with rounded ridge terminating on dorsal surface orthogonal to lateral ridge (forms from lateral expansion of head approximately 2-3 times gentle medial expansion from shaft – leads to thinner and sharp ridge forming on proximal third); medial surface much wider and gently rounded – gives skewed oval profile of shaft; distal rugose lateral depression equal to that of MC2; dorsal and ventral surfaces gently concave in lateral aspect; distal articular head smooth, gently convex mediolaterally and dorsoventrally; margins well-rounded, thickened and moderately rugose.

Metacarpal IV (MC4): proximal head much more enlarged than distal head; separated by stout, dorsally convex and ventrally flattened shaft, with concave and medial and lateral surfaces; proximal head flat with mildly convex medial area, giving 'laterally-stretched' sub-oval form; surface strongly rugose adjacent to head; distal head flat transversely and strongly convex dorsoventrally; adjacent surfaces less-extensively stippled; lateral depression switched to medial surface (therefore correct metacarpal?); thickness uniform except for dorsal expansion of proximal head and sharp, sub-triangular lateral expansion.

Metacarpal V (MC5): more similar form to MCI than MCII-IV; longitudinally constricted into a much shorter, robust element; proximal head with sub-circular profile, flattened, medially canted and rugose, with slightly thickened dorsal margin; distal head similar but rounded and dorsoventrally compressed; shaft laterally and medially concave in dorsal aspect; medial edge very narrow before rapidly diverging into tips and shaft; dorsal surface swings around laterally in broad curve intersecting the flatter, dorsomedially sloping ventral surface; well-defined rugose area adjacent to distal head on medial surface; distal medial surface with small rounded expansion – develops ventrally as very narrow crest adjacent to rugose distal medial surface – joins at trinity with ventrolateral edge and ventral margin of distal head.

Medial phalanx of digit I small, robust element with deeply depressed proximal surface; proximal tip circumferentially thickened, sub-oval form; width approximately 1.5 times depth; dorsal surface well-rounded proximally; ventral surface flattened, gently convex exterior edges; distal end dominated by two well-developed longitudinal ridges pervading approximately half phalangeal length – diverge distally, separated by broad and shallow groove, expanding to form to rounded 'heads' on distal surface, wrapping around medial and lateral surfaces, contacting ventral border as rounded expansions; separated on the medial and lateral surfaces by shallow, broad depressions adjacent to associated metacarpal. Form mirrored to lesser magnitude in more dorsoventrally constricted phalanges (adjacent to metacarpals). Distal phalanx of digit III greatly compressed – transverse width approximately 2.5 times length. Proximal articular surfaces of metacarpals strongly united, except MC5, as noted by Forster (1990).

It is likely that this strengthened the carpus significantly, for the purpose of quadrupedal locomotion. This is somewhat dissimilar to the carpal and manual structure of *Camptosaurus aphanoecetes*, which retained a primarily quadrupedal stance aided by extensive manual modification with limited interphalangeal and carpal motion (Carpenter and Wilson, 2008).

Ungual phalanges: Longitudinally arcuate; variably rugose/striated surfaces; develop 'typical claw' profile with well-defined distal tips (perhaps sharp initially, but quite worn); distal surface becomes progressively convex and broader proximally – mirrored in rugose ventral surface; prominent medial and lateral ridges converge distally – narrow, rounded and moderately rugose; separated from phalangeal body by deep, well-defined groove extending for two-thirds length proximally (variable).

### 9.14 Pelvic Girdle

*Ischium*: dominated by long, moderately straight, gracile shaft; terminates distally in thickened, convex, rugose tip; proximally diverges into two heads orientated equally approximately 40° to longitudinal axis of shaft – dorsal attachment for ilium, cranioventral attachment for pubis; together encircle cranial acetabular portion; separated by smooth, concave, arcuate border with sharp caudoventral acetabular margin; develops progressive concavity proximally from shaft. Not 'footed' as higher iguanodontians.

Pubic process with gently caudocranially convex lateral surface; equally concave medial surface leading to gently curving pubic articulate - rugose, ventrally canted, thickened lateral rim, tapers anteriorly into thin point to retain form of acetabulum; articular surface twists slightly laterally; ventral edge thin and sharp, extending from much thicker shaft.

Iliac process with gently depressed, rugose articular surface and slightly thickened rim – expands dorsolaterally from acetabular intersection into sub-oval form; arcuate long axis reflecting greater flaring axially than transversely; lateral surface smooth, wrapping around head forming thickened extension of dorsal shaft border; medial surface gently twisted and convex. Long axes of heads oblique to each other; pubic process approximately vertical.

Shaft transversely narrow with proximally rounded quadrilateral form; ridge develops on lateral surface down midline intensifying convex lateral surface; medial surface flat and planar; sharp ventral and dorsal intersections; sub-parallel dorsal and ventral borders (excluding projection of obturator process).

Obturator process head expanded from ventrally projecting sheet; thickened apex directed ventrolaterally and slightly cranially; anteroventral and posteroventral surfaces approximately orthogonal and gently curved; tips of surfaces project beyond sheet, each separated from shaft by small cleft. Obturator process occurs approximately one-third ischial length similar to *Dryosaurus*, but approximately one-sixth the length of the ischial shaft (from the proximal end); separated from shaft by broad channel that accepts part of the postpubic rod. Not as proximally placed as higher iguanodontians.

Post-obturator process greatest mediolateral thickness occurs immediately prior to recession of lateral ridge at half ischial length; shaft broadens slightly and flattens into gently convex distal end similar to *Dryosaurus, Hypsilophodon, Thescelosaurus,* and *Parksosaurus* (Galton, 1981); post-lateral ridge, medial surface develops a median concavity broadening distally; well-rounded ventral edge and sharper dorsal edge; dorsal deflection of dorsal edge at point of broadest concavity –
projection of thin ridge developing convex border progressing into broader
proximal concavity; associated gentle distal concavity adjacent to tip; dorsal border
thickens slightly adjacent to lateral ridge development.

Distal end forms progressive ventrally thickening arc; ventromedial surface
develops distally expanding groove from approximately three-quarters ischial
length – gently concave, irregular channel form.

*Ilium*: longitudinal axis approximately two times height (excluding cranial process);
body dominated by robust postacetabular blade, massive ischiac peduncle and
stout pubic peduncle conjointly developing well-rounded dorsal acetabular margin.

Cranial process missing distal end; curves inwards ventrally in broad arc; medial
surface gently concave; lateral surface convex with mild concavity at midline;
process thickens dorsally; dorsal and ventral margins well rounded. Longitudinally
arrayed ossified tendons occupy small overhang on ventromedial border – possibly
extend past pubic peduncle caudally. Dorsal border intersected on medial surface
by well-developed, gently rounded ridge extending caudoventrally to terminate
approximately over acetabulum – enhances concavity of medial surface of cranial
process. No supracetabular crest (antitrochanter) as higher iguanodontids and
hadrosaurids (and *Lesothosaurus*).

Dorsal border oscillates in thickness and bearing; strongest over pubic peduncle,
thinnest over ischiac peduncle – associated with ventral dip in margin immediately prior to expansion of postacetabular blade approximately dorsal to mid-point of brevis shelf; refracts at approximately 60° into gently convex caudal margin. Caudal border strongly rugose, giving impression of intersecting and conflicting grooves often oriented at 45° to border; develops cranially into heavily striated blade.

Caudal ventral surface gently convex post-ischiac peduncle; slight depression on medial region leads up to massive ischiac head – occurs synonymously with mediolateral thickening correlated to expansion of peduncle; remainder of surface occupied by deep brevis shelf.

Postacetabular blade slightly constricted dorsoventrally with mirrored concavities; gently concave lateral surface, gently convex medial surface; expanded slightly on medial surface approximately one-third iliac height above cranial termination of brevis shelf; develops into main body cranially – reflected by maximum concavity on dorsal surface and slight lateral deflection of body above mid-point of acetabulum.

Ischiac peduncle massive, laterally rounded, tapering into point medially; heavily rugose articular surface with less-massive associated head on ischium; lateral expansion greater than greatly flattened medial expansion; rugose surface terminates at development of smooth, medially canted and broadly arcuate acetabular margin.

Pubic peduncle with damaged distal end, therefore structure indiscernible – clearly more slender than ischiac peduncle though; lateral surface smoothly curves inwards forming concave cranial border – extends caudally to form well-defined rounded ridge extending for length of peduncle; separates from flat caudoventral surface of acetabulum; medial surface dominated by broad askew channel separated from convergent caudoventral and craniodorsal borders; cross-sectional profile sub-triangular; initial 60° angle between pubic peduncle and cranial process.

Main body laterally flattened and medial surface dorsoventrally convex; medial body variably depressed for attachment of sacral ribs (Forster, 1990).

Right ilium displays several slight variations: slightly transversely wider brevis shelf; pubic peduncle diverges into two elements directed at 45° to body craniolaterally and craniomedially; postacetabular blade with flattened, straight and massively rugose posterior margin; ventral surface swings more strongly laterally to contact ischiac peduncle.

*Pubis*: *d*ominated by prepubic blade, moderately developed obturator foramen and long and slender postpubic rod; forms cranioventral margin of acetabulum; postpubic process disassembled and disassociated, but estimated to be approximately 1.5 times main pubic body length.

Prepubic blade extends cranially from body; dorsal margin broadly concave due to dorsoventral expansion of distal tip; progressively thins distally due to increasing transverse constriction –elongated ellipse profile distally; progressively arcuate profile dorsally in mid-section due to concavity on medial surface propagating down blade as shallow channel.

Acetabular articulate rugose, caudoventrally canted, medially depressed and dorsally thickened; deflects casually into puboiliac and puboischial surfaces; medially arcs ventrally in caudal aspect giving sub-crescentic profile. Puboiliac surface heavily rugose, flattened, lies cranioventral to acetabulum; abruptly terminates at thin ventral margin of blade; attaches to pubic peduncle of ilium; underlying lateral surface thickened immediately dorsal to gentle, broad concavity; puboischial surface develops from caudoventrally tapering acetabular articular surface – forms 'wedge' ventrally overlying obturator foramen; medial edge comprises very thin sheet projecting caudally; rugose ventral border encloses foramen.

Postpubic rod/process difficult to identify structurally due to degree of damage; projects from ventral surface below obturator foramen at approximately 100° to prepubic blade; lies in same dorsoventral plane as prepubic blade; tapers distally from initial oval form; lies against lateral surface of ischium (parallels it's longitudinal axis).

Obturator foramen dorsal to proximal genesis of postpubic rod; fully enclosed by postpubic rod, acetabular region and puboischial articulate caudally; obliquely inclined caudoventral long-axis –elliptical form; completely open/perforated; smooth borders; dorsal surface flares dorsomedially expanding into puboischial surface; ventral surface comprises well-rounded section of postpubic rod

**9.15 Hindlimb and Pes**

*Femur*: strongly robust element; proximal end dominated by massive, bulbous head and orthogonally broadened greater and lesser trochanters; distal end comprises two equally massive condyles and caudally projecting subsidiary condylid.

Femoral head projects medially and slightly cranially; caudal surface dominated by broad, shallow and distally tapering well-sculpted groove – merges into caudomedial surface of shaft (equivalent to proximal "fossa" of Forster (1990)); laterally develops into discontinuous rounded ridge – pervades as equidistant distally as associated groove; orientated orthogonal to long axis of femoral head; develops medially into strongly concave and proximally rugose shaft from distal tapering of head.

Femoral head separated from broad trochanters by distally flattened, caudocranially constricted neck; cranial surface develops much broader concavity than well-developed convergent groove caudally; degree of expansion greater cranially than caudally due to development of lesser trochanter. Lesser trochanter 'bullet-shaped' lateral profile; less prominent than *Camptosaurus* (Dodson, 1980); caudal edge defines concave cranial border of greater trochanter. Greater trochanter approximately three times width of lesser; lateral surface not flat - concave caudocranially; convex dorsal border – progressive caudally and cranially; trochanters separated by thin fissure – orientated mediolaterally; pervades further distally on lateral surface. Lateral surface of trochanteric region orthogonal to long-axis of femoral head; lesser trochanter somewhat oblique to greater trochanter; midline of greater trochanter extends cranially into rounded tip – may be incompletely fused to lesser trochanter.

Proximally, caudal surface of shaft dominated by greatly rounded and expanded distal pervasion of greater trochanter and flatter but equally wide extension of femoral head; asymmetrical interlaying groove flares proximally in accordance with femoral head becoming shallower; terminates distally approximately adjacent to mid-point of fourth trochanter. Lateral surface entirely uniplanar; deflects slightly caudally giving bowed appearance; reflected in medial surface – coincident with progressive thinning of shaft.

Cranial surface extends from broadly concave distal propagation of femoral head and rounded ridge-like lesser trochanter; medial surface of lesser trochanter deflects 90° into cranial surface, transmitting broad, somewhat deepened concavity; more medial surface progressively ascends cranially completely merging with lesser trochanter approximately mid-length of shaft – overall form of shaft massive, well-rounded and robust; flat lateral, gently sloping medial and flat caudal surfaces, similar to *Dryosaurus, Valdosaurus* and *Camptosaurus* (e.g. Galton, 1981). Unity terminated abruptly by distal cranial intercondylar groove; caudally shaft gently convex due to terminal expansions.

Fourth trochanter develops from distally thickened ridge beginning approximately one-quarter femoral length; twists to project caudoventrally into lobate form; axis curved, slightly thickened distally and continuously thins towards apex. Overall strong development to increase leverage for vertical backswing (Bakker and Galton, 1974).

Shaft twists slightly post-fourth trochanter distally so two condyles are set slightly obliquely to femoral head; shaft begins to expand mediolaterally approximately two-thirds length – coincident with caudocranial constriction. If interpretation of Forster (1990)'s "epicondyles" are correct, the medial is as equally gently convex as the lateral one, albeit narrower caudocranially.

Medial condyle with gently convex articulate; slightly thickened periphery; flat medial surface; develops from well-developed ridge extending from one-third shaft length, parallel to fourth trochanter; greatly expanded caudocranially into massive tip, slightly oblique to shaft. Lateral condyle more rounded and rugose - greatly mediolaterally expanded; craniocaudal expansion approximately two-thirds medial condylar.

On the cranial surface, distally expanding and deepening intercondylar groove develops, similar to *Dryosaurus*. On caudal surface, longer rapidly diminishing "condylid" separates deep, medially broad groove from more lateral shallow concavity – all three pervade same distance proximally. Deeper groove interpreted to be "popliteal surface" of Forster (1990). Medial condyle expands gently into this groove laterally. "Flexor groove" is thus interpreted as more distal expansion of "popliteal surface".

*Tibia*: lightly less robust than associated femur; similar morphology to *Camptosaurus* (Dodson, 1980); linear longitudinal axis (i.e. straight shaft). Proximal end flares craniocaudally – same degree as transversely expanded distal end to approximately twice shaft width; proximal end comprises enlarged cnemial crest (cranially projecting), orthogonal lateral condyle and massive caudal expansion of tibial head. Articular surface rugose – varies from caudally convex to gently medially canted cranially; lateral condyle inclined at 45° laterally; proximal radial head fits between cnemial crest and lateral condyle. Lateral condyle prominent (lesser than two associated proximal expansions still); curves slightly cranially; rugose periphery; separated from cranial projection by broad, rapidly tapering concavity and fibular head concavity – both converge into rounded, robust shaft (quarter tibial length). Fibula rests on craniolateral surface of cnemial crest - thin caudal edge articulates with cranial border of lateral condyle. Caudal projection thickness approximately 1.5 times cnemial crest – projects to same degree laterally as lateral condyle (more rounded tip).

Medial surface transversely expanded, converging distally into straight shaft, twisting laterally. Malleoli develop broad arc (distal aspect) into "inversed V-shape" of Forster (1990) (caudal aspect); well-developed, distally rugose ridge occupies caudal margin separating adjacent borders – tapers rapidly into proximally convergent shaft. Craniomedial border rounded, ascending into thin, sharp ridge pre-mergence with medial shaft surface – forms sharp deflection with cranial surface. Lateral malleolus exhibits rounded caudolateral apex – broader, rounder and larger than medial malleolus (develops from thinner expansion); lateral malleolus projects further caudally than craniomedial projection of medial malleolus from intersecting ridge. "Shallow flexor groove" of Forster (1990) on cranial surface intersecting malleoli – distal fibula fits here, articulating with cranial surface; articular surfaces rugose and slightly rounded to flattened (partially obscured by artificial attachment of tarsus). Overall very similar form to that described by Forster (1990).

*Fibula*: proximal end transversely flared 'goblet' form extending from constricted, slender neck (lateral aspect); similar morphology to *Camptosaurus* (Dodson, 1980), lateral surface marked by broad, rugose ridges and interposed grooves – extend for approximately 15mm before merging into converging shaft; only occur in medial section of lateral surface (gently convex caudocranially). Articulate ascends caudolaterally in gentle, linear incline; heavily rugose surface; lateral margin contacts proximal extensions of ridges (as small projections); elongated, slightly bowed oval form; rounded cranial and caudal ends; tip forced slightly laterally by expansion of more massive, robust cnemial crest.

Shaft develops from abrupt, rapid tapering of proximal head – gracile, twisting 90° at mid-point conforming to shape of associated cranial surface of tibia; flattened craniomedial surface forming sharp, prominent distal cranial ridge.

Distal end partially obscured by artificial attachment to calcaneum; appears sub-triangular, cranially thickened, and rounded (expansion terminates rapidly proximally); flat caudomedial surface with sharp margins. Shaft exhibits various cross-sectional profiles: hemispheric, crescentic, sub-rounded – flattening increases distally on medial surface. Shaft contacts tibia approximately two-thirds length (distally) – somewhat difficult to judge due to synthetic attachment. Caudolateral edge at least lies on cranial tibial surface. Generally exhibits minor overall variation to Forster (1990).

*Tarsus*: calcaneum and astragalus tightly interlock with respective fibular and tibial surfaces; attached to two distal tarsals as Forster (1990), forming continuous and strong articulation – attaches to metatarsal II, III and IV. All tarsal elements identical to Forster (1990); partial obscurity due to restorative assembly does not occlude any noteworthy features. Lateral distal tarsal absent; medial distal tarsal identical to Forster (1990).

*Pes*: twisted shaft of metatarsal I thickened, ventrally compressed; distal end set at 90° to dorsoventrally expanded proximal head – fits into concavity, articulating with lateral surface of metatarsal II. First phalanx (digit I) largest of all phalanges – distally depressed medially and laterally bordered by two crested ridges accepting ungual phalanx proximal articulate. Ungual phalanx similar to manual ungual phalanges; lateral and medial grooves well-defined, rugose dorsal surface, 'claw-shaped'; possibly capable of minor hyperextension. Digits II, III and IV: all intermediate phalanges are variably scaled versions of each other (as medial phalanx of digit I) – same morphology, variable dimensions.

Digit III largest; digit IV missing ungual phalanx (sinistral pes); digit II approximately identical length to digit IV. Metatarsal II second largest, proximally broadest accepting astragalus; cuboid distal end (slight ventral expansion); ungual phalanx equidimensional to digit III ungual; primary phalanx intermediate form between slender digit I element and robust digit III element. Phalangeal dorsal and ventral grooves equivalent to extensor and flexor grooves of Forster (1990); "large but shallow tendon insertions" correspond to medial and lateral depressions – on metatarsals II, III and IV, more prominent depressions extend further proximally on phalanx II. Unguals appear symmetrical conversely to Forster (1990) – not "asymmetrically twisted towards digit III".

Metatarsal III largest; slender, distally and proximally robust; straight shaft; flattened medial and lateral surfaces (even where metatarsals II and IV do not contact).

Metatarsal IV slightly shorter than metatarsal II; proximally mediolaterally expanded into triangular head; distally mediolaterally compressed, extending slightly ventrally; twists and diverges from metatarsal III distally (more so than metatarsal II).

No major differences to Forster (1990) except for the absence of the vestigial metatarsal V; Forster (1990) admits this is not always developed in *Tenontosaurus*, and the absence is not stratigraphically controlled. This may be a function of ontogeny, with the metatarsal reducing with age, or perhaps developing with age, although what function this would potentially provide is currently unknown.

## 9.2 Appendix 2: Comparative Morphology I – *Hypsilophodon foxii*

**9.21 Skull**

R2477 preserves a near-complete skull; predominantly disarticulated therefore sutural/contact relationships between elements difficult to discern (see Galton (1974a) for complete reconstruction).

*Premaxilla:* dextral premaxilla somewhat similar to *Tenontosaurus*; prominent difference in presence of premaxillary teeth projecting from ventrolateral surface - "thecodont" teeth (Galton, 1974a); paired foramina (variable size) on medial surface of premaxilla adjacent to each tooth.

Lateral surface not listric - instead ascends into ventral anterior process oriented at 45° encompassing external nares; process broad, transversely narrow and thickens slightly where contacts main premaxillary body; dorsal extent of process unknown as distal two-thirds absent; anterodorsal process has posterior surface continuing to form anteroventral margin of external nares before twisting laterally developing lateral surface of posteroventral process.

In lateral view anterior premaxillary border mimics posterior border in deflection angle and parallelism of associated process borders; ventral border thickened, expanding medially to form flat and horizontal ventral surface; anterior border sharply convex with anteroventral border curving ventrolaterally until intersection with posterior border; in rostral view premaxilla moderately concave laterally and slightly medially thickened; anterodorsal border develops arbitrary projections developing somewhat 'bumpy' appearance; sinistral premaxilla exhibits two small foramina ventrally - separated by two thin, parallel ridges.

Significant hiatus before first maxillary tooth; posterior border weakly contacts maxilla with large fenestra separating; rostrally-directed maxillary process contacts medial premaxillary border, terminating pre-symphysis - does not contact either process; maxillary process extends thin lamina ventrolaterally contacting premaxilla and forming posterior border of intervening fenestra; lamina posterodorsally canted from contact laying sub-parallel to ventral premaxillary process; result is 'reversed-tick' shaped fenestra broadening ventrally.

*Maxilla*:   ventrally convex post-tooth row anteriorly contacting premaxilla immediately pre-deflection of ventral process; tooth row arranged linearly with 10-11 teeth - each tooth occupies one dorsally concave socket.

Lateral surface bisected by horizontal semi-ridged surface ventral to tooth row; four more-or-less evenly spaced projections separated by 'u-shaped' basins positioned exactly laterally to anterior process on medial border - ventral surfaces form discontinuous linear plane; medial surfaces of projections contact dorsomedial maxillary lamina - slightly obliquely inclined with respect to lower surface; upper surface diverges anteriorly forming posterior margin of maxillary fenestra, which Galton (1974a) calls the "antorbital fossa"; anterior half of fossa develops from laterally overlapping and ventrally thickened, laterally convex dorsal extension.

Main body of maxilla subtriangular in section and dorsally thickened - extends anteriorly into well-rounded process fitting into premaxilla; process contains single dominant longitudinal groove with 2-3 parallel subsidiary flanking grooves, each divided by parallel ridge; anterior tip of process deflects dorsally; thin non-linear groove runs directly ventral to process within main maxillary body forming slight ventral concavity.

Main body with broad, slightly curved ventral surface - off this projects extremely brittle lamina that envelops part of antorbital fossa; also extends instantly ventral to fossa.

Lateral (ventral) sheet develops from vertical expansion of main body; thickens ventrally and very thin dorsally forming anteroventral margin of antorbital fossa; does not contact premaxilla with convex anterior border - forms posterior border of secondary slit-like antorbital fenestra; sheet has strongly concave profile, distinctly more so than dorsal sheet.

Medial (dorsal) sheet projects vertically, mildly concave laterally in section; stepped from ventral sheet; converges centrally anteroposteriorly projecting ventrally forming hook-like extension - fuses to lateral sheet at ventral apex of antorbital fossa; lateral sheet slightly overlaps most distal point immediately posterior to contact; also contacts lachrymal with upper half of oblique dorsal border.

Galton (1974a) states the "more dorsal part of the medial sheet is overlapped by the thin sheet of the lachrymal"; this is only apparent in medial aspect – if viewed from lateral aspect, the opposite is apparent, thus this comment is slightly ambiguous. The author also mentions that "there is a large fenestra anteriorly in the medial sheet of the maxilla"; this implies the fenestra is entirely within maxilla, which is misleading, as the maxilla only forms the ventral and posterior margins.

With the antorbital fenestra and fossa, a third opening develops posteriorly, formed by the dorsal border of maxilla and ventral edge of lachrymal; opening expands posteriorly.

*Lachrymal*: contacts medial maxillary lamina and jugal dorsal surface; ventral border slopes caudoventrally resting on maxilla and jugal; develops into posterodorsally orientated fold – lateral 'limb' vertical with gentle dorsal convexity tapering to point rostrally and posteriorly; ventral edge sharp, thin, slightly sinuous; internally hollow (between limbs); well-rounded 'hinge' – posterior and rostral borders perpendicular; 'axial trace' orientated caudorostrally; sloped fenestra penetrates posterior edge lengthways (reflected on anterior point); cylindrical fenestra punctuates dorsal border of lachrymal longitudinally – does not extend through internal hollow; continues posteriorly as open, posteromedially canted, flattened surface forming rostromedial margin of orbit. Complete connection of fenestrae unknown – designed for communication? Medial surface dominated by gentle convexity, separated from medially thickened dorsal border by shallow anterior groove and deeper, broader posterior groove – both ascend meeting at medially-deflected apex. Ventral border progressively rostrally convex – small fenestration approximately half of length. Medial surface forms dorsal border of lachrymal foramen. "Palatine bar" of Galton (1974a) equivalent to lateral process directed off palatine body – unsutured to lachrymal conversely to Galton (1974a) (may have been displaced); posterior groove is thin, possibly vestigial, ridge (intersects dorsal border) separating groove from parallel , slightly curved, deeper groove.

*Jugal*: lateral surface gently convex rostral profile; separated from maxilla by horizontal slit; rostral border partially absent – deflects ventrally into posteriorly convex border resting on lateral maxillary surface; ventral sheet deflects dorsomedially approximately 60° contacting lachrymal forming smoothly concave ventral margin of large orbit – deflection marked by discontinuous, longitudinal groove along hinge; oval foramen positioned ventrally orientated rostroventrally-posterodorsally; dorsal lateral surface extends further posteriorly diverging into two processes projecting posteriorly and posteromedially – latter thicker and more tabular than former (rod-shaped). Medial surface does not contact palatine; forms triple junction between grooved lachrymal surface, posteromedial process and third, sub-oval, thickened medial process (concave mediolaterally and anteroposteriorly) forming orbital base; lachrymal process strongly recurved ventrally into large hollow forming as passage between orbit and antorbital fenestra; thin ventrolateral wall.

*Quadrate*: rounded ventral edge, well developed "condylar region"; brief, obliquely inclined shaft projects, twisting slightly; condylar region anterolaterally-posteromedially orientated – forms two condyles (anterior marginally larger); rounded, rugose ventral surfaces; separated by very gentle concavity – ascends shaft terminating at onset of twist; posterior surface of distal condylar region rapidly converges into thin, rounded lamina extending to dorsal tip – horizontal, two-thirds of length traverses posteriorly, twists laterally intersecting triangular tip fitting into a socket in the squamosal (Galton, 1974a).

"Pterygoid flange" forms medially-directed, concave 'sail'; medially canted ventral border; two gently curved ridges connects surfaces posteriorly; main sheet deflects into progressively expanding arc; fenestra develops immediately post-expansion; posterior edge thin, sharp and slightly ventrally twisted; very thin ridge extends posterior to this border from condylar region, wrapping around anterolateral surface, extending dorsally – penetrated, but not halted by fenestra pre-culmination in anterior apex converging with lateral edge; continues as anterodorsal ridge extending to quadrate head. Pterygoid flange extends concavely from midline of shaft for two-thirds length, migrating and extending from anteromedial border of transversely flattened head. Anterior surface exhibits thickened articular border (distal periphery); flattened until pterygoid flange projects with concave ventral border with slight depression on anteromedial surface.

Dorsal aspect (R2477): nares absent, left prefrontal absent, posterolateral fragment of parietal absent. Frontals, postorbitals and parietals develop delicate 'casing' – strongly symmetrical, mostly fused (uncertain), rostrally frontals divergent (i.e. non-contacting).

*Parietal*: single element symmetrical about midline; dominated by anteroposteriorly concave and mediolaterally convex lateral surfaces extending posteriorly into two "posterolateral wings" twisting slightly in dorsal plane; converge in midline as thin, sharp ridge flaring laterally into two tapering, flattened hooks – rostral borders converge, curving inwards medially, in rectangular process fitting in between frontals with moderate overlap; progressively thickens anteriorly – whole structure longitudinally concave dorsally. Lateral borders flattened in ventral aspect – expand slightly upon lateral curving; slightly grooved along midlines (medially flanked by thin, sharp ridges pre-concave deflection); flanks converge posteromedially becoming steeper and vertical upon secondary lateral refraction.

Frontal contact heavily ridged on posteriorly canted surface; anterior border contacts postorbital – together project posteriorly overlapping margin of supraorbital.

*Frontals*: longitudinally elongate; strongly united pair divided by linear suture; dominates dorsal margin of orbit (with prefrontal and postorbital flanking); lateral border gently concave (dorsal aspect) – continue as posterior and anterior borders of prefrontal and postorbital respectively; deflects slightly dorsolaterally into flattened, ridged border ("insertion markings" of Galton (1974a)); slight transverse concavity; anterolaterally convex.

Smooth dorsal surface; anterolateral quadrant bears sheath-like groove (laterally canted, 'belemnite' shaped) accepting arcuate posterior process of prefrontal. Posterior border posteroventrally sloping; well-defined ridges; midline punctuated by triangular opening accepting anterior parietal projection. Orbit broadly laterally convex opening oblique to frontal axes. Ventral surface dominated by medially concave extension of lateral ridge (moderately rugose surface); develops mediolaterally into sharp, arcuate ridge – continues posterolaterally, deflecting sharply ventrally continuing on medial surface of postorbital in broad arc, remaining sub-parallel to orbital margin throughout length; separated from midline by 'bow-shaped' concavity; surface listric posteriorly, flaring laterally into prefrontals; "chokepoint" of surface slightly thickened into gently convex platform. Slight posterolateral rugose depression adjacent to ridge; continues more extensively and deeply depressed on postorbital; receives laterosphenoid head (Galton, 1974a); postorbital forms thickly fused plane with frontal.

*Postorbital*: sub-rectangular form, gently concave borders; forms broad posterior orbital margin and triple-junction with frontal and parietal (slightly anterior to supraorbital); sinusoidal suture with frontal – reflected ventrally; linear contact with parietal running posteroventrally – marks distinct twisting in internal supraorbital walls; postorbital develops thin, rounded margin (dorsal aspect) and deep, laterally canted wall (ventral aspect). Lateral surface flat anteroposteriorly; sharply laterally convex transversely upon orthogonal rotation - forms intersection between orbit and supraorbital.

Posterior and ventral processes variably slender, tapering distally; posterior element transversely thinner; both equally narrow; ventral process develops stout ridge on medial surface extending to diverge and enclose laterosphenoid socket - adjacent orbital surface strongly concave (compared to shallow posterior surface); ridge expands, thickens, flaring dorsally immediately pre-socket — very thin, subsidiary ridge extends, terminating at tip of posterior process; ventral process triangular cross-sectional profile, posterior rounded and sub-tabulate; medial process forms robust junction with parietal and frontal — reminiscent of a 'whale tail' in dorsal aspect.

*Prefrontal*: dextral element slots neatly into anterior frontal groove — extends here into two thin sheets: first tapers rostrolaterally, second thickens terminating ventrally; sharp, arcuate lateral border forms anterodorsal orbital margin — thickens into anterior junction between two sheets. Dorsal sheet gently inclined rostrally forming gentle concavity anteroposteriorly; thicker ventral sheet orthogonal to dorsal sheet forms anterior orbital wall with extended anteroposterior concavity. Anterolateral surface geometrically complex: small ridge extends from tip of dorsal sheet deflecting obtusely (more so than ventral sheet) intersecting orbital margin approximately half sheet length. Medial surface develops from expanded concavity of ventral surface of frontal — ventral ridge of frontal runs progressively concave along medial edge of ventral sheet.

*Teeth* (R2472 mainly): five premaxillary teeth (each element); 10/11 maxillary teeth; edentulous predentary; dentary >10 teeth.

Premaxillary teeth preserved *in situ* in specimen R197 – solitary significance is presence unobserved in *Tenontosaurus* and all higher ornithopods.

9 teeth preserved in R2477 (dextral maxilla): 2 oblique ridges culminate in longitudinal ventral ridge (varies from straight to moderate sinuosity); lateral surface medially inclined, medial surface laterally inclined. Anterior surface slightly depressed accepting posterior edge of adjacent tooth; lateral and medial surfaces enamelled; lateral surface exhibits single dominant central ridge with one to three subsidiary parallel ridges evenly spaced on either side – slightly denticulate margin (well-worn). Slight ridge separates crown from root ("cingulum" of Galton (1974a)) – ridges develop from this to crown tip; crown heavily compressed transversely; flattened surfaces, occasionally gently concave; medial surface equally ridged; ventral surface obliquely inclined to ridges (posteriorly canted).

### 9.22 Pectoral Girdle

*Scapula* (fig. 32): slightly shorter than humerus; blade slightly twisted with respect to articular head - caudal tip of blade oblique to articulate; deep, broadly convex medial surface where scapula conforms to ribcage. Triangular facet of Galton (1974a) observed forms prominent expansion caudal to articulation directly opposite glenoid fossa; long axis points down coracoid length; linear cranial border until caudal curvature as broad dorsal arc – 'scimitar' form, uniformly thin blade.

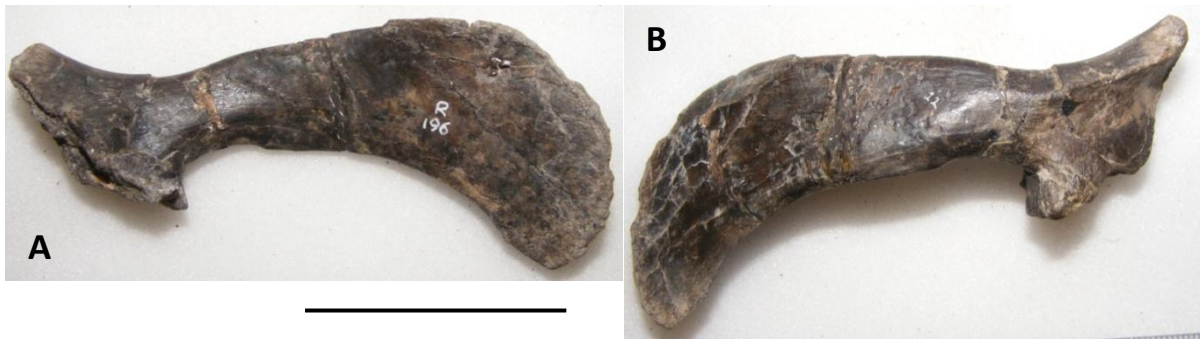

Figure 32 – Sinistral scapula in **A:** medial and **B:** lateral aspects. Scale = 5cm. Note degree of distal curvature and form of scapular blade compared to *Tenontosaurus*.

Sharp distal sector of caudal border – thickens proximally where blade transforms into broad, flattened sheet (reflected in anterior border); medial surface thickens developing into longitudinal ridge along surficial midline.

Dorsal surface arcuate, thickened along blade width, tapering cranially; lateral surface extends as broad concavity smoothly expanding dorsoventrally – cut by "clavicular facet" (Galton, 1974a), and ventrally by glenoid; intersected ventrally by lesser depression adjacent to glenoid.

Ventral surface deeply concave and rounded - does not contribute to glenoid fossa; proximal-most surficial curvature deflects reflexively cranially and medially expanding into scapular surface of fossa; surface medially canted, depressed with thickened periphery.

Proximal medial surface continues blade concavity; craniodorsally and cranioventrally divergent as caudally-directed ridge – pervades from coracoid articulation (and coracoid); reflected in lateral surface; medial surface convex dorsoventrally until proximal flaring.

Cranial border linear and vertical - except small kink ventral to mid-point corresponding to caudally-directed arcuate cleft from coracoid foramen dissecting articular surface running immediately ventral to rounded ridge; articulation thickness corresponds to scapular shaft greatest thickness.

*Coracoid*: Average thickness equivalent to scapular blade-shaft transition; medial surface concave craniocaudally and dorsoventrally; very mild dorsoventral convexity on lateral surface; medial surface contains well-developed depression in craniodorsal quadrant, thinning surface cranially; lateral surface with dorsal glenoid-facing rounded ridge orthogonal to articulate.

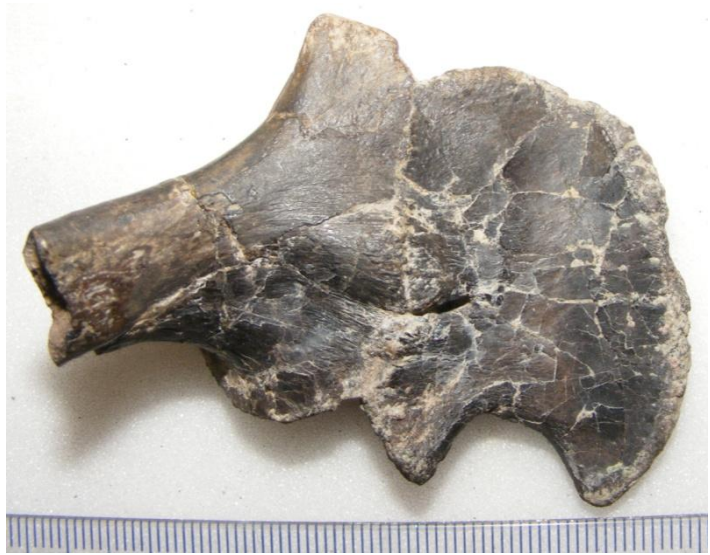

Figure 33 – Dextral coracoid (and associated proximal end of scapula) in medial aspect. Scale in mm. Note proximal placement of coracoid foramen, similar to *Camptosaurus*, but not *Tenontosaurus*.

Coracoid foramen circular, vertical walled (lateral surface), occurring mid-way down articulation approximately 40% coracoidal width; proximally placed foramen deeper on medial surface, extending obliquely into ridge – elongate longitudinally, inward-sloping walls.

Ventral border deeply concave – proximal end develops into massively thickened glenoid surface (less broad than associated scapular surface); distal end retains average coracoid thickness, curving cranioventrally into rounded, sub-triangular sternal process.

**9.23 Forelimb** (fig.34)
*Humerus* (R194): moderately robust; proximal expansion approximately 1.3 times distal width; prominent deltopectoral crest.

Proximal head heavily rugose with deep ridges; proximal humeral quarter exhibits rhombic form – proximal border caudally expanded; crescent form (dorsal aspect).

Medial and lateral borders converge into tightly constricted, sub-cylindrical shaft – thinnest immediately pre-deltopectoral crest; surface flat before anterolaterally directed crest projection (orthogonal elliptical ridge) – progressively thickens towards apex; concurrent with deltopectoral crest expansion, medial deflection into broadly concave margin – curves caudally; entire upper humerus trends posteriorly as deltopectoral crest diminishes before proximal transverse flaring; posterior edge broadly, longitudinally concave bordered by deltopectoral crest; slight medial depression immediately pre-deltopectoral crest – crest forms straight edge before deflecting obtusely into shaft; medial surface deeply concave (more linear distally). Deltopectoral crest much less extensive than *Tenontosaurus*; otherwise humeral form similar – *Hypsilophodon* represents a scaled-down form.

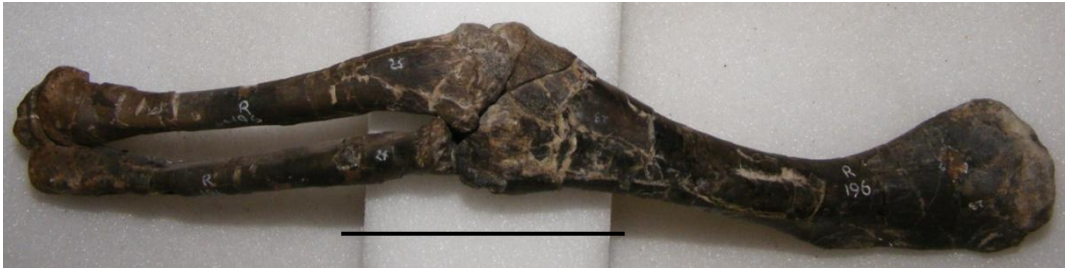

Figure 34 – Complete humerus, ulna, and radius in caudal aspect (specimen R196). Scale = 5cm.

Shaft fans proximally into twisted, expanded proximal head; distal intercondylar groove shallow and less continuous; ulnar condyle gently rounded (more so than radial condyle) – extends slightly less anteriorly than less robust radial condyle; non-symmetrical ulnar and radial condyles (anterior aspect) separated by smooth cleft developing bow-shaped distal border/ventral edge; cleft expands transversely into broad concavity ceasing abruptly approximately one-fifth length proximally against sharp lateral 'scar'; radial condyle develops well-defined ridge compared to rounded ulnar condyle – extends proximally curving into 'scar' forming entire anterior surface.

*Radius* (R194): distal head crushed; lateral surface wraps around cylindrical shaft (lateral aspect); forms slightly off-centre longitudinal ridge; flattens approximately two-thirds length immediately pre-proximal caudocranial flaring; cranial edge remains linear; caudal edge curves gently into moderately expanded proximal tip; dorsal rim set oblique to lateral surface. Proximal articulate depressed, circumferentially thickened; medial surface flattened – flares proximally and distally (approximately 1.7 times width); moderately concave medially with linear mid-section (craniocaudal aspect); transversely narrower pre-distal and proximal expansions.

Specimen R196 exhibits a well-preserved complete forelimb (minus manual elements) (fig. 34).

*Humerus*: rugose deltopectoral crest dominates humeral shaft; projects caudally; mediolaterally thickened; apex occurs 30% humeral length, apex progressively thinner; humeral head develops on caudolateral surface; separated from proximal caudomedial section by small depression. Cranial intercondylar groove broader, deeper (due to caudally expanding paired condyles), pervades further proximally into shaft (approximately quarter length); cranial and caudal intercondylar grooves connected by U-shaped cleft on distal articulate dividing two condyles. Condyles both exhibit identical caudocranial length; ulnar member more massive anteriorly; radial member develops thin cranially directed ridge; proximal aspect: ulnar condyle massive, sub-rounded intersections; radial condyle directed cranially, sub-triangular profile.

*Ulna*: olecranon process develops on proximal dorsal surface – bulbous, rugose, moderately developed; proximal surface transversely broad (continues into dorsoventrally canted ridge), converging into constricted shaft. Shaft medially concave, oval cross-section – concavity increases proximally becoming broader and deeper, coincident with distal craniocaudal expansion of shaft; develops as dorsal surface to ridge-like, thick expansion of shaft – sinuous medial border; lateral surface moderately concave proximally as ridge develops – concavity encloses rugose circular facet; lateral surface extends distally forming flattened surface on shaft – twists, merging with caudal surface forming gently convex distal end; surfaces separated by gently-rounded ridge extending from proximal caudolateral intersection; gentle distal twisting of shaft generates obliquely orientated distal end (to olecranon process and ridge axis). Caudal aspect, medial surface gently concave for length except for development of olecranon process; lateral surface exhibits gentle convexity (exactly opposite to respective concavity) with strongly concave proximal third – culminates in skewed development of olecranon process with gentler lateral growth into sub-triangular, dorsally rounded projection, extending orthogonal to larger cranial ridge. Distal "rugose markings" of Galton (1974a) may be muscle scars – quite obscure, extent difficult to determine.

*Radius*: proximal articulate – greatly sloping depression, smaller subsidiary caudal depression; shaft mid-section medially stretched cylindroid; medial ridge development rounded proximally and distally; equal proximodistal expansions; lateral surface flat distally and proximally, gently curved mid-section; caudal aspect, small well-rounded ridge proximally projecting caudomedially, moderately rugose surface, extending 20% radial length distally; distally (along same longitudinal axis), small rugose facet occurs immediately distal to flattened articular surface with slightly elevated proximal circumference. Distal end massive; no medial expansion (craniocaudal expansion evident) – begins immediately proximal to kink in shaft cranially; caudal shaft forms moderately convex border with concave extremities; anterior border completely gently concave.

### 9.24 Pelvic Girdle
Specimen R193 contains a near-complete disarticulated pelvic girdle.

*Ischium* (fig. 35): complete, well-preserved; in lateral view distal end forms greatly-expanded, transversely narrow blade (reminiscent of a 'sharks-head'); proximally blade thins dorsoventrally and thickens transversely; distal expansion manifesting as thin lamina of bone with slight variable thickness; cranioventral edge curves slightly ventrally at most distal point; caudal edge forms broadly convex profile intersecting ventral border with acute, rounded corner; ventral edge dominantly linear - slight deflection at dorsal intersection (also thins slightly); anteromedial and posterolateral surfaces smooth and undisturbed - except possibly with development of slight longitudinal groove where post-pubic rod rests on blade surface; axis of blade orientated at 45° to ischiac head.

At half ischial length blade transforms into progressively proximal thickening shaft with oval cross-section; becomes medially thickened in axis at 90° to length; shaft thins slightly caudodorsally and cranioventrally to greater extent; lateral surface twists post-obturator process forming ventral border; obturator process develops approximately half length of ischium - width equal to width of adjacent section of shaft; forms medially convex, thin lamina with somewhat quadrilateral form - this may be product of absent fragments; curved proximal border orthogonal to shaft. Distal border slightly damaged with shard missing, therefore exact form undeterminable.

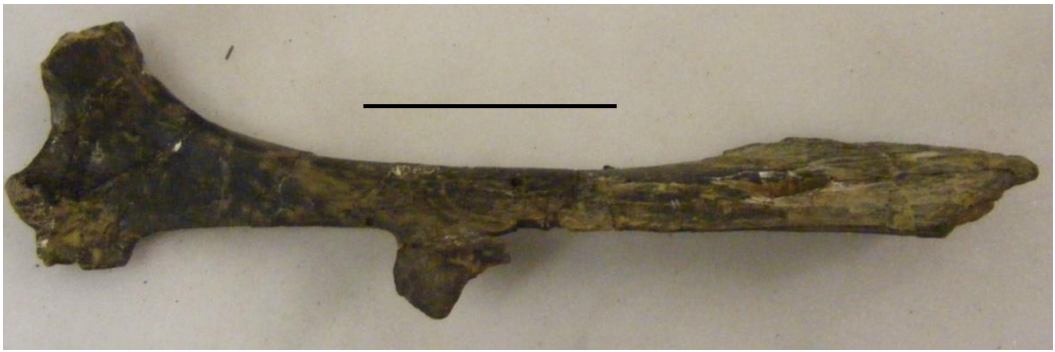

Figure 35 – Sinistral ischium, lateral aspect. Note placement of obturator foramen similar to *Tenontosaurus*, and complex twisting of shaft. Scale = 5cm.

Proximal 40% craniocaudally expanding concave surface resulting from adjacent shaft twisting; distal (upper) divergence develops into massive, rounded ischiac head; caudal edge forms from remnants of proximally tapering shaft – circular cross-section; dorsal border equates to smoothly concave acetabular border.

Proximal border of obturator process diverges into lateral surface – dorsal border creating the twisted shaft, propagating craniodorsally to intersect dorsal border immediately pre-expansion of ischiac head in tightly acute angle.

Width of ischiac head less than height of cranial expansion of lateral surface – develops cranial half of acetabular shelf; cranial expansion thins greatly ventrally as an extension of cranioventral continuation of shaft; cranial edge sharply deflects abruptly at midpoint with dorsal half becoming laterally convex and retaining thickness immediately post-deflection – marks change from isosceles profile to gently curved sub-rectangle. Dorsal view cross-section reveals highly constricted mid-section (at acetabulum); ischiac head approximately twice width of lower expansion; surfaces of two expansions orthogonal – both oblique to longitudinal ischial axis; proximal section of ischium displays less convex medial than lateral surface.

*Ilium* (fig. 36): caudodorsal region reconstructed (upper 40-60% of main body post-cranial process distally). Main body with broadly concave medial surface (caudal aspect); caudal extremity absent; uniform thickness post-ischiac peduncle distally; distinct ventral thickening dorsal to acetabulum; acetabulum transversely oblique connecting ischiac and pubic peduncles; ventral border medially canted (post-ischiac peduncle) – extends off dorsal surface of peduncle intersecting caudal edge sub-orthogonally. Post-peduncle process greatly thickened; connected to main body via small cleft (moderate depth); forms sinusoidal extension of pre- and post-peduncle borders. Ischiac peduncle massive (ventrolaterally expanded); blunted ventrolateral facing 'spear' (cranial aspect); sub-quadrilateral in lateral aspect; rounded dorsal borders expand into acetabulum and ventromedial brevis shelf.

Anterior process develops smooth arc (lateral aspect), with bladed 'scimitar' form; uniform thickness except for slight distal tapering; distal third bends gently ventrally; distally thins in dorsal aspect; deflects laterally from otherwise linear dorsal border; lateral surface revealed as process twists. Medial surface initially thickens medially (dorsal half) into elongated wedge – abruptly ceases sloping laterally into ventral half of process; remaining ventral surface gradually develops into medially expanded shelf/ledge – develops moderately convex form. Overall sinuous profile in cross-section (for proximal two-thirds); shelf gradually tapers to continue as flattened ventral border – where process begins to deflect ventrally, small subsidiary oblique ridges develop horizontally to cleave intersecting channel (between shelf and dorsal expansion) – had otherwise diverged forming flattened, vertical process tip.

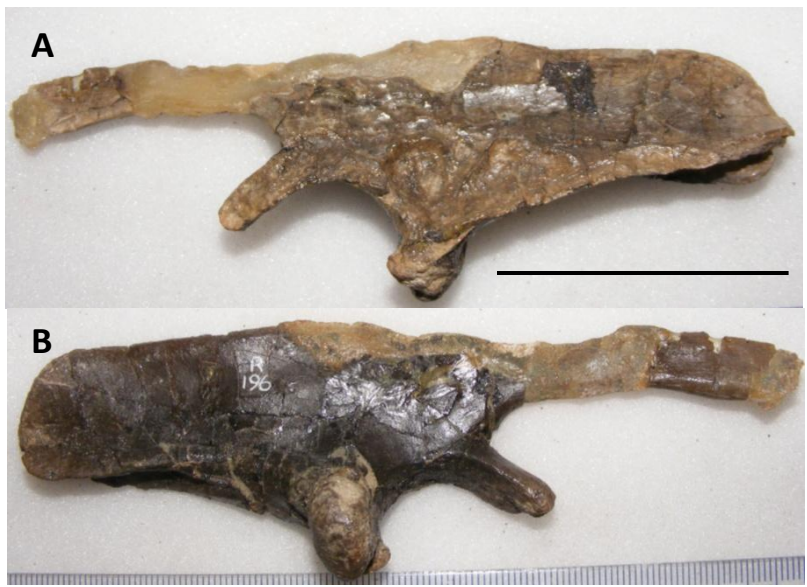

Figure 36 – Dextral ilium in **A:** medial and **B:** lateral aspects. Note relatively straight cranial process, and slightly wider ventromedial deflection (brevis shelf) compared to *Tenontosaurus*. Scale = 5cm.

Pubic peduncle set 30° to cranial process; much more robust structure, although partially deformed; ventral edge greatly flattened before strongly curving into ischiac peduncle developing dorsal acetabular margin; lateral edge forms sharp ridge; medial edge heavily damaged.

Caudal sector dominated by broad, convex brevis shelf (ventral aspect); develops cranially thickening ventrolateral projecting lamina approximately 90° to convex main body (medial aspect); lamina initiates development post-shelf cranially; brevis shelf expands marginally anteriorly contacting posterior process (blade); lateral and caudal borders orthogonal – exhibits spectacular symmetry in caudal aspect: ventral wall of main body divergent into two opposite lamina (dorsoventrally mirrored).

*Pubis*: comprises robust main body, rod-like prepubic process, slender post-pubic rod approximately twice length of prepubic process (most distal part absent). Dorsal surface comprises sharp border of postpubic rod; develops cranially into massively thickened, convex contact with ischium (puboischiac head) – deflects ventrally into broadened, grooved, slightly medially deflected prepubic process; groove ceases two-thirds process length.

Prepubic process smooth forming depressed oval cross-sectional profile; transversely broadened; ceases abruptly cranially; flattened medial surface – expands before immediately dipping caudally and flaring into broad depression dominating medial surface of body - ventral surface exhibits associated constriction; mimicked to lesser degree on lateral surface.

Smoother ventral border of main body much less concave than associated strong convexity of rough dorsal border – develops as gently obtuse refraction between two processes. Pubic/obturator foramen completely enclosed; lesser robust caudal wall; develops caudally to depressions; articular surface rough, thinning caudally where curvature strongest. Contact between postpubic process and body longitudinally orthogonal; slightly expanded before progressively thinning into medially concave blade; slight dorsal expansion immediately post-mediolateral thickening ("dorsal sheet" of Galton (1974a)).

Cross-sectional profile varies cranially: initially blade-like, medially transversely thickened, dorsoventrally alters into elliptical section, distally sub-triangular cylindroid (distal tip absent).

### 9.25 Hindlimb
*Femur*: best-preserved specimen R192a: near-complete left femur with distal end partially restored (fig. 37).

In medial aspect posterior border anteriorly concave; anterior border convex creating longitudinal gently curving arc; both proximal and distal ends expand caudocranially - proximal expansion of greater extent; shaft mostly uniform thickness (excluding fourth trochanter); lateral surface pinches into longitudinally-orientated ridge; medial surface fairly even, diverging at extremities covering lesser trochanter and femoral head, and outer and inner condyles distally; surface also punctuated by broad, shallow groove immediately anterior to fourth trochanter.

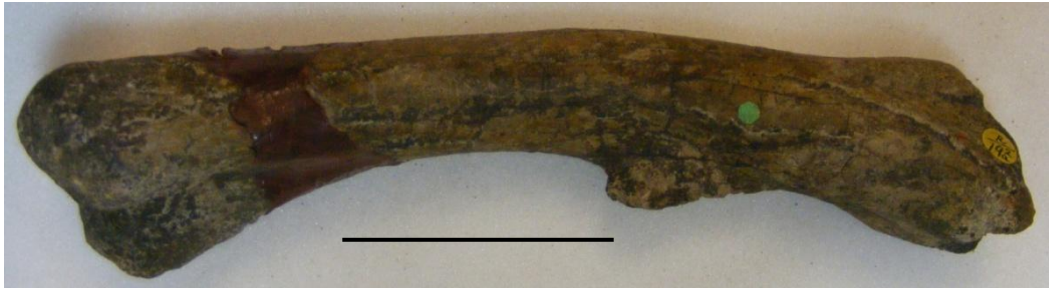

Figure 37 – Dextral femur in cranial aspect. Scale = 5cm. Note proximally placed fourth trochanter, and increased degree of curvature of shaft relative to *Tenontosaurus*.

Lesser trochanter canted cranially, developing as a cranial expansion orthogonal to femoral head; separated from head by shallow linear channel adjacent to lesser trochanter and part of lateral surface; significant fragment of femoral head is absent.

Greater trochanter separated from femoral head by depression for ischiac head of ilium (Galton, 1974a) forming an acute angle between both - depression forms rough convex strip terminating proximally in slight expansion separating femoral head and both proximal trochanters; lesser trochanter axis perfectly splits midline between femoral head and greater trochanter axes.

Fourth trochanter terminates 45% of femoral length distally - caudally directed and lobate form; distally thickens laterally; medial surface perpendicular to shaft axis; lateral surface forms from continuation of lateral surface of shaft, developing gentle concavity.

No development of cranial intercondylar groove due to gently convex and flattened cranial surface; inner condyle missing distal tip - appears same size as outer condyle, but with slightly different orientation and sectional profile; caudal intercondylar groove begins at 75% femoral length distally diverging mediolaterally to be cradled by two condyles; outer condyle provides more prominent bounding margin to groove than inner condyle; outer condyle with cranially canted proximal surface, oval cross-section; slightly more expanded caudocranially than inner condyle by approximately 20%; slightly convex in lateral aspect with borders gently converging into shaft proximally.

R195 femur distal quarter missing; displays several slight variations to R192a femur: fourth trochanter more 'fin-shaped' with sharper craniodorsal edge orientated 45° to longitudinal axis; femoral head more massive, smooth cranial surface - displays oblique concavity on medial surface, possible signs of strain directly distal to head.

*Tibia* (R5830 – distal end and articulation absent): proximal articulation flattened, caudomedially canted, rounded periphery; outer condyle with subsidiary 'parasitic' condyle (condylid – accepts fibula) – swiftly converge before merging into shaft. Inner condyle forms single element; slightly thicker ridge – separated by deep, longitudinal lateral groove from shaft (pre-convergence into shaft) – both condyles merge approximately quarter tibial length.

Oval profiled cnemial crest; obliquely inclined lateral border; merges with shaft approximately equidistant to distal condyles. Shaft develops prominent linear, longitudinal ridge on medial surface – sub-triangular profile with rounded apices.

*Fibula* (R5830): distal end absent; proximal end missing fragments. Forms gracile element; gently concave longitudinally (medial surface); expands slightly proximally – coincides with caudocranial concavity; shaft twisted slightly at mid-section – lateral surface rotates developing cranial surface.

## 9.3 Appendix 3: Comparative Morphology II - *Thescelosaurus neglectus*

*Femur* (fig. 38): head projects medially and slightly dorsally as massive, bulbous structure on dorsal articular surface; caudal surface cut by U-shaped ventrolateral groove beginning at proximal tip – merges with caudal surface immediately distal to head; head exhibits distinct signs of pervasive arthritis – extends entirely over articulate and down caudal surface of shaft ('gritty' appearance). Greater trochanter separated from head by shallow caudocranial depression forming from distally sloping lateral surface of head – sub-triangular profile (reminiscent of an ungual phalanx); posterior surface extends into lesser medial expansion with flattened surface – separated from shaft by deep, bowed groove extending midway distally (adjacent to fourth trochanter); caudal surface wraps caudolaterally around greater trochanter forming sharp, ridged proximal edge; intersected by prominent cranially-facing longitudinal groove – extends less distally than ridge, but extends proximally and expands slightly separating lesser trochanter; tapers approximately two-thirds length; semi-lunate form (lesser trochanter profile).

Lesser trochanter displays gently convex lateral surface; sharp ridge adjacent to cleft; thickens distally before converging into shaft; groove merges into surface and transversely broadened, linear lateral surface; proximal end rounded, deflecting into cranial border cranially. Cranial surface develops from abrupt cessation of trochanters and laterally broadening anterior surface of femoral head; broadly concave proximal end (caudocranially and longitudinally); thin medial and lateral borders. Concavity extends approximately one-third of shaft – terminates as fourth trochanter begins development; curves strongly medially into ventromedial concavity of femoral head; straightens immediately prior to fourth trochanter expansion.

Fourth trochanter projects caudoventrally; apex positioned approximately half femoral length; lobate form; sharp, thin medial edge; thickens internally; ventral surface deeply concave pre-intersection with linear shaft; contacts caudal surface with broad, gentle concavity – develops from constriction of shaft post-proximal caudal ridge; strongly curves into lateral surface (rounded edge); cut by more prominent caudal intercondylar groove and longitudinal ridge (extension of inner condyle); medially extends into outer condyle. Cranial surface separated by shallow, distally tapering groove from shaft; adjacent surface continues cranial surface of fourth trochanter as planar sheet – intersects main surface obtusely forming thin, non-extensive ridge; cranial surface extends from proximal concavity into flattened, medially twisted surface – becomes occupied by rotation and divergence of lateral surface extending from lesser trochanter. Cranial surface transforms into distal medial surface, flaring caudomedially conforming to distal expansion of outer condyle.

Lateral surface develops from cranially expanding distal continuation of lesser trochanter and lateral half of proximal caudal ridge – two-thirds shaft length rounded ridge dissects this (proximal extension of lateral inner condyle); alters surface from gently convex into caudally shifted and flattened, and orthogonal medially-directed surfaces (separated by rounded ridge); cranial surface develops into distally deepening sub-triangular depression (i.e. cranial intercondylar groove) – terminates abruptly against thickened distal margin; orthogonal to long-axis of lesser trochanter, oblique to proximal cranial concavity; remaining element of lateral surface rotates caudally into gentle edge (cranially concave); flares caudally distally into rounded, sub-polygonal caudal expansion of inner condyle.

Caudal surface gently arcuate longitudinally; intercondylar groove slightly more extensive proximally and deeper distally – not as symmetrical as cranial groove (result of more prominent expansion of inner condylid into sub-triangular to rounded ridge); intercondylar angle tighter than cranial surface; condylid separated by shallow convex (caudocranially) depression; outer condyle greater transverse expansion providing greatly concave form (reflected in femoral head); medial surface mostly absent; medial surface of outer condyle rugose, greatly convex mediolaterally and caudocranially with sharp , acute medial edge; 'hammer' form in medial aspect; caudal surface pinched into rounded tip, cranial half planar and polygonal; surface lies sub-parallel with longitudinal axis of craniomedial surface of fourth trochanter.

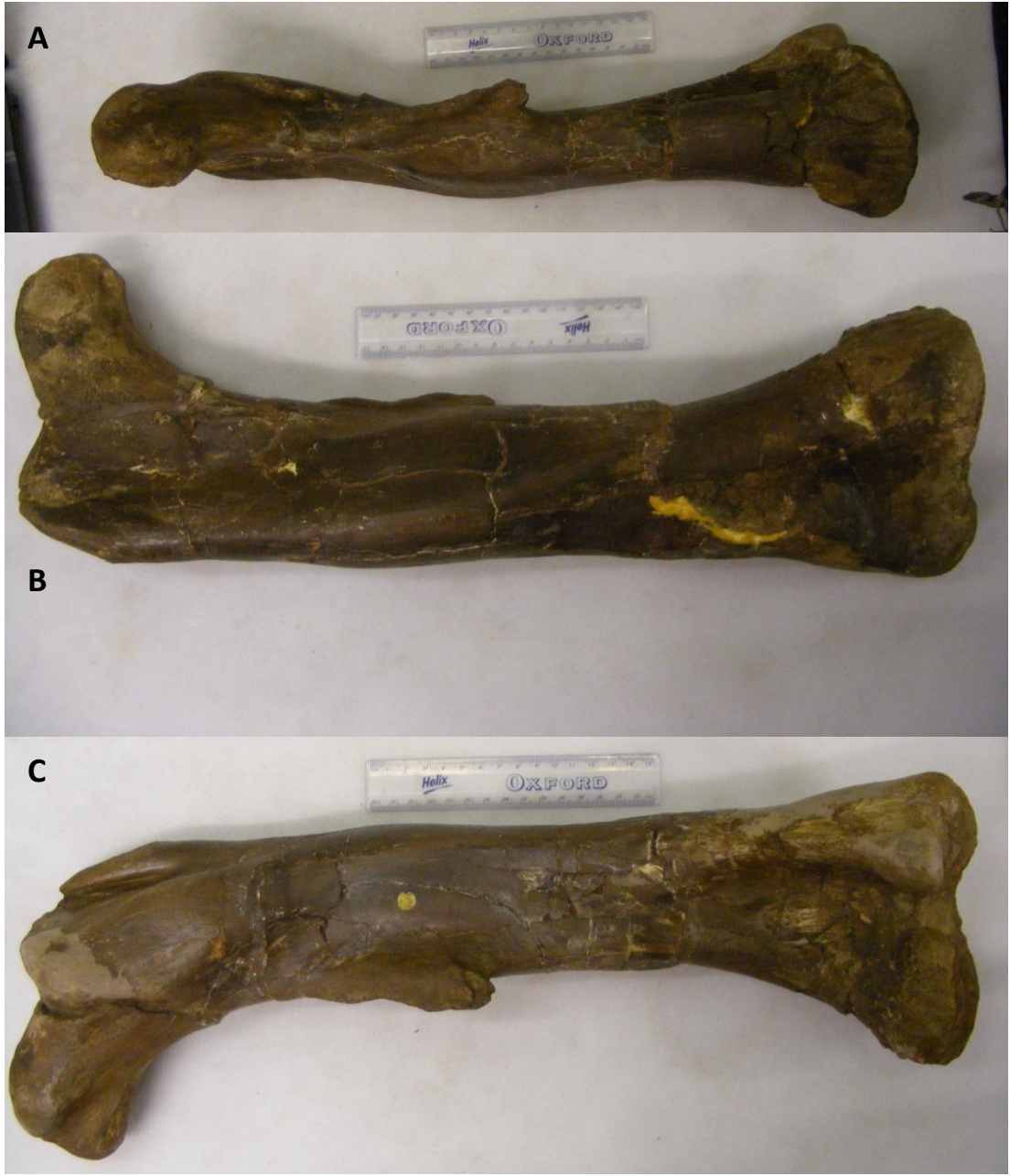

Figure 38 – Dextral femur in **A:** medial aspect **B:** cranial aspect **C:** caudal aspect. Scale = 15cm (ruler). Note cleft separating fourth and lesser trochanters from shaft, deep caudal intercondylar and shallow cranial intercondylar, reminiscent of iguanodontians.

## 9.4 Appendix 4:  Comparative Morphology III – *Valdosaurus canaliculatus*

R180 was previously regarded as a juvenile specimen of *Valdosaurus* from the Lower Cretaceous Wealden (comprising a partial dentary with one mature and two immature teeth), prior to reanalysis by Galton (2009) to Iguanodontoidea indet. The teeth are described here so that a comparison of iguanodontian, hypsilophodontian and tenontosaur teeth is possible.

*Dentary teeth* (fig. 39): form rounded sub-triangle (lateral view); gently convex cross-sectional profile due to medial thickening; lateral surface dorsoventrally convex; fits into deep, ventrally opening groove (alveolus); separated by thickened, rounded ridges converging to rounded point immediately ventral to lateral dentary surface. Individual teeth exhibit perfect symmetry about midline – round-crested ridge runs vertically and linearly through midline towards broadly rounded apex; flanked by convex-inwards subsidiary/secondary ridge (both sides) extending for tooth length; tertiary ridges occur between these and primary ridge parallel to secondary ridges – mimic gentle curvature pre-termination approximately one-third up tooth; tooth narrows ventrally to approximately half the distal/dorsal width. Ventral border slightly concave laterally, converging at terminus of primary ridge; margins strongly denticulate (approximately 16 denticles per half tooth) – become less prominent and more acicular medially; ventral denticles more tabulate (leaf-shaped?) – individual denticles very finely denticulated. Roots not visible; possibly >1 tooth per alveolus – perhaps indicates tooth replacement. Maxillary lateral surface fenestrated – not 1 fenestra per alveolus; lateral alveoli margins thickened into sinusoidal ridge.

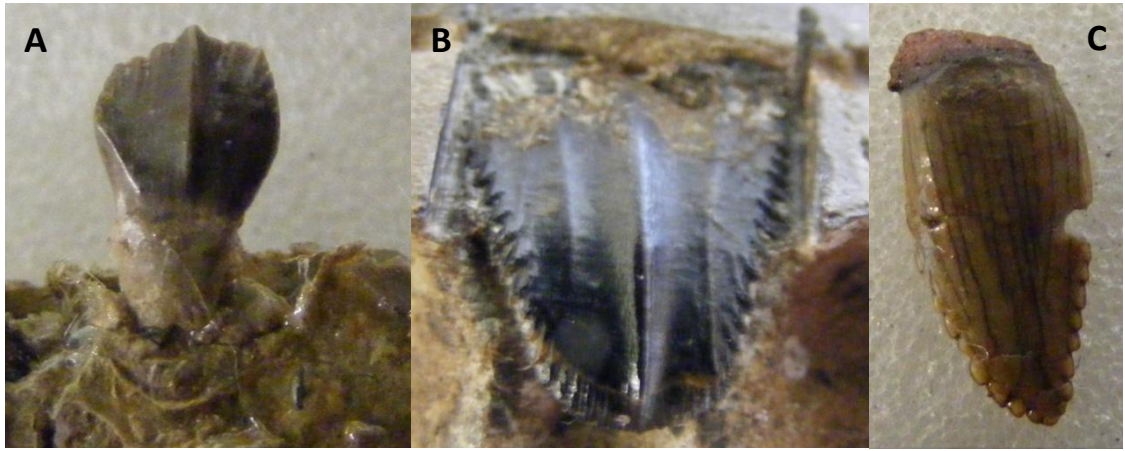

Figure 39 – **A:** dentary tooth of *Hypsilophodon foxii* (field of view approximately 5mm) **B:** dentary tooth of *Valdosaurus* (or Iguanodontidae. sp.) (field of view approximately 20mm) **C:** *Lesothosaurus diagnosticus* tooth (field of view 6mm) in lateral aspects Note lack of root in the iguanodontid form, and the addition of stronger secondary ridges, as well as the form of the terminus of the crown.

Specimen R185 is the holotype; material represented here are two paired femora and the dextral ilium. Given the size of the specimen and it's classification to the Dryosauridae (Butler *et al*., 2008), it provides a rational comparison to a sub-adult *Tenontosaurus*.

*Femur* (fig. 40): outer condyle massive, bulbous, well-rounded caudal projection; slightly flattened caudal surface; tapers proximally into thin rounded ridge terminating one-quarter shaft length; broadens slightly proximally. Inner condyle develops triangular sheet projecting caudally; gently compressed mediolaterally; tapers proximally equidistant to outer condyle – more sharply-crested ridge with flattened medial surface; slightly concave laterally. Intercondylar groove broad, shallowing rapidly proximally from condyles (caudal surface). Shallow depression separates inner condyle from distal lateral surface, merging with caudomedial border medial to condylar apex.

Both condyles gently striated on smooth cranial surface – distal end 'speckled'; cranial intercondylar groove similar length to caudal, forming much deeper u-shaped longitudinal cleft (progressively shallows proximally) – iguanodontian synapomorphy similar to *Tenontosaurus*. Outer condyle flattened; gently concave medially; medial edge becomes sharper and obtuse proximally; vertical contact with intercondylar groove; rounded sub-rectangular form (ventral aspect). Inner condyle gently arcuate, progressively curved laterally; twice broadness of outer condyle; thrice lateral extent of expansion than gentle medial outer condylar expansion; surface wraps around distal end until intersection with caudal projection; vertical walled contact with intercondylar groove; amorphous with tab-like extension (ventral aspect) – slopes into ventrally canted caudal projection,

Fourth trochanter projects as progressively caudomedially expanding ridge one-third femoral length (apparent apex); apex absent in both elements; appears to have sub-triangular form; separated by very broad depression (medial surface); shaft greatly more bowed than *Hypsilophodon foxii* and *Tenontosaurus*, with less twisting.

Femoral head massive, elongate medially; separated from greater trochanter by smooth, shallow depression on caudal surface; concave caudal surface relating to expansion of subsidiary, moderately rugose trochanter/process – sub-triangular profile, separated from femoral head tip by oblique depression (for ischiac head of ilium); caudal surface gently convex – develops thin ridge transversely on distal surface, extending fully around thickened ridge connecting head to shaft. Greater trochanter elongate perpendicular to femoral head medial-axis; forms abrupt, rounded edge with slightly inclined lateral surface; curves gently into caudal depression; constricts caudocranially into depression; oval cross-section; proximal surface broadly convex (caudocranially). Thin neck unites greater and lesser trochanters – orientated 45° to caudocranial axis on lateral border; two grooves facing craniomedially and caudolaterally also separate trochanters – former greater longitudinal extent but slightly shallower, latter deepens distally terminating abruptly with listric 'head'. Lesser trochanter semi-lunate (proximal aspect); forms thin rod constricting slightly before contacting convex shaft (craniolateral surface); moderately rugose proximal surface; proximal quarter of lateral surface gently striated longitudinally.

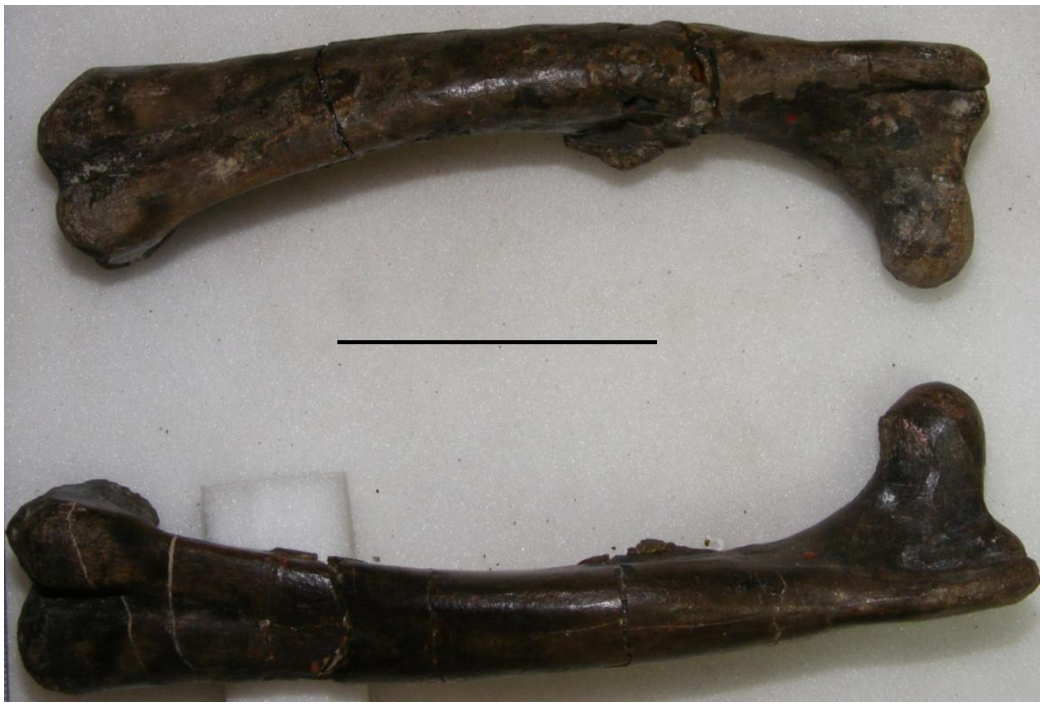

Figure 40 – paired femora of holotype, R185, cranial aspect. Note proximal placement of fourth trochanter, and shallow cranial intercondylar groove. Scale = 5cm.

*Ilium* (fig. 41): large, broad, shallow brevis shelf dominates – width approximately 1.5 times anterior process; posterior section (post-ischiac head) constricted transversely, triangular profile, rapidly tapering into thinner main body; flattened ventral border becomes gently rounded caudally. No supracetabular shelf (derived condition). Massive ischiac head; laterally expanded; ventrolaterally canted articulation. Acetabular margin rugose, equally distributed between ischiac and pubic peduncles. Pubic peduncle slender; abruptly terminates in expanded head; sub-crescentic profile.

Cranial process slender, blade-like; dorsoventral and caudocranially concave (medial aspect); thicker dorsal border overhangs concavity; thinner ventral border with sharpened medial edge; blade longitudinal axis linear – slight lateral deflection; lateral surface gently concave with flattened mid-section; ventral and dorsal borders descend approximately two-thirds length, 'tilting' process; ventral border convex, dorsal border equally concave; caudal edge vertical, flattened. Massive sub-triangular medial growth (rounded apex) on medial surface between ischiac peduncle and brevis shelf – rapidly tapers caudally into flattened medial edge of ventromedial surface (i.e. brevis shelf).

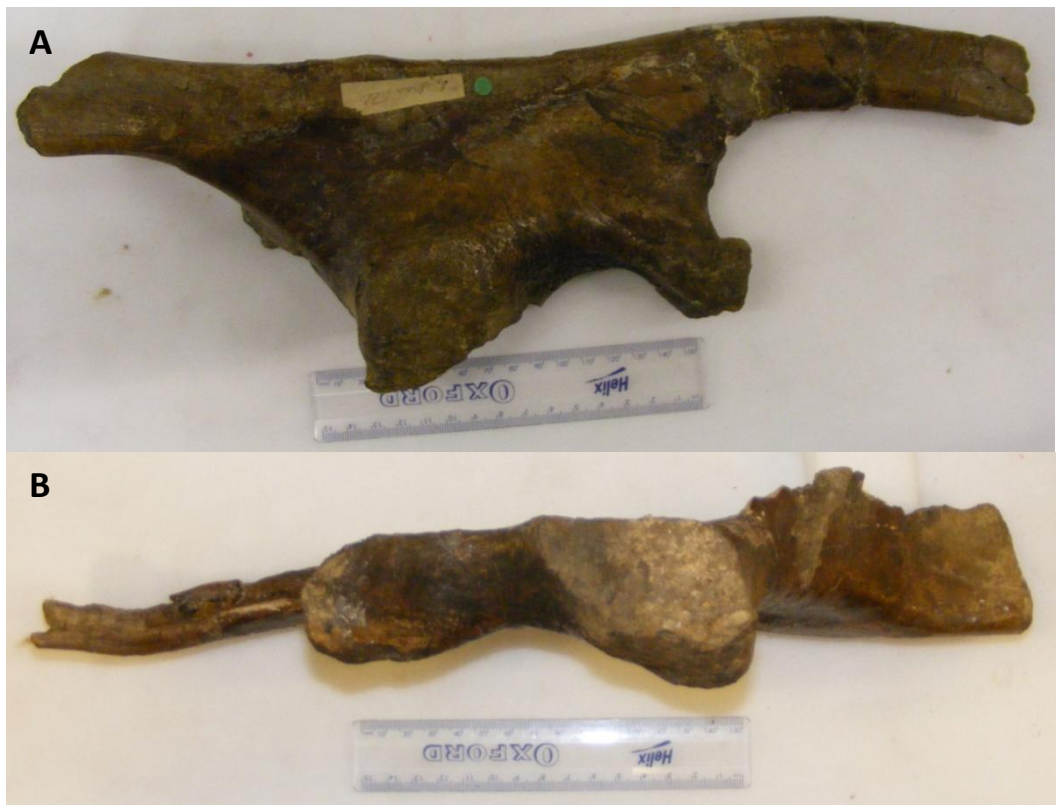

Figure 41 – Dextral ilium in **A:** lateral aspect **B:** ventral aspect. Scale = 15cm (ruler). Note medial extent of brevis shelf (strong bipedal adaptation) and relatively straight dorsal margin.

## 9.5 Appendix 5: Comparative Morphology IV – *Lesothosaurus diagnosticus*

*Lesothosaurus* is a basal ornithischian from the Early Jurassic of Stormberg, South Africa; its phylogenetic position is currently disparate, but its basal position with the Ornithischia is commonly acknowledged, grouped with other primitive ornithischians such as *Eocursor parvus* and *Pisanosaurus mertii* (e.g. Butler *et al.*, 2007). S. Maidment (pers. comm.) recommended a brief analysis of several elements of *Lesothosaurus* to provide a base reference for more derived forms, and that the pattern of morphological evolution would become easier to understand. A short description of post-cranial elements of *L. diagnosticus* is provided here for such a reason, and that it may prove useful in phylogenetic analysis as an outgroup to the Ornithopoda. RUB.17 is the adult syntype (presumed adult stage).

*Scapula* (fig. 42): quadrilateral distal end; sharp edges; dorsoventrally flared blade tapers proximally coincident with progressive thinning; concave form conforming to rib cage. Half scapular length, thickening resumes (greater rate than more gradual thinning) – well-rounded dorsal border. No twisting of shaft/blade. Curvature increases exponentially proximally (towards acetabulum). Thickness remains uniform for rest of proximal length post-point of maximum curvature. Lateral surface bears broad concavity dominating proximal end – connects with small depression separating thickened glenoid ridge and coracoid articulation - small rugose, cranially canted surface develops ventrally at abrupt termination of rounded ventral edge; continues as gently concave surface intersecting ventral tip of coracoid articulation.

Glenoid circumferentially thickened, flattened rugose internal surface; shallow depression separates from orthogonal articulation; ventromedial border wraps around to form medial articular margin; articulation caudodorsally canted and transversely concave. Small rugose 'bump' extends ventromedially on medial surface (at point of maximum shaft thickness).

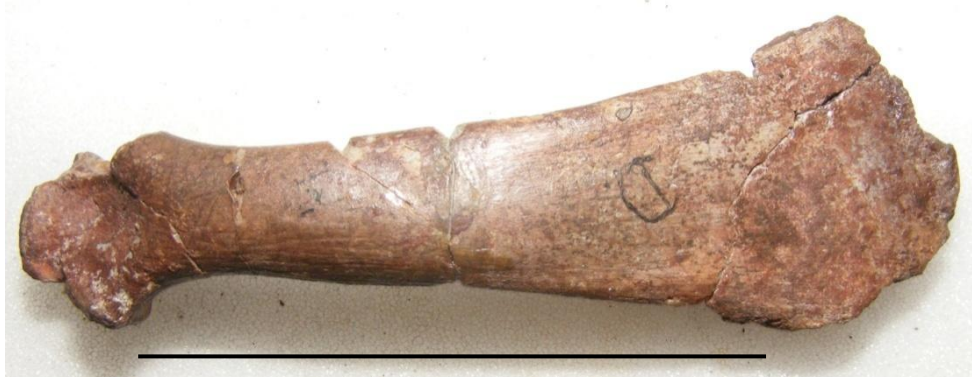

Figure 42 – Sinistral scapula, lateral aspect. Note relatively simple form compared to *Hypsilophodon* and *Tenontosaurus*. Scale = 5cm.

*Humerus* (fig. 43): poorly developed humeral head; proximal end transversely expanded (mediolaterally convex, dorsoventrally concave); deltopectoral crest develops from progressive distal expansion of craniolateral border – rounded apex approximately one-third length; shaft simple oval form – slightly cranially expanded immediately distal to deltopectoral crest; proximal border thickened cranially. Ulnar and radial condyles separated by listric depression (cranial surface) – gently tapers proximally; radial condyle gently rounded, extending into lateral border (caudal aspect), projects cranially into well-developed, rounded ridge with rugose, rounded tip. Ulnar condyle expands medially; broadly concave medial border; gently convex longitudinally and transversely; both cranial expansions gentle, converging one-third length of shaft (proximally); much gentler intercondylar groove (cranial aspect) – curves craniolaterally, less extensive than caudal intercondylar groove; rugose distal end, sub-rhombic profile. Radial condyle oblique oval (distal profile) – develops cranial ridge extending slightly further than ulna condylar expansion.

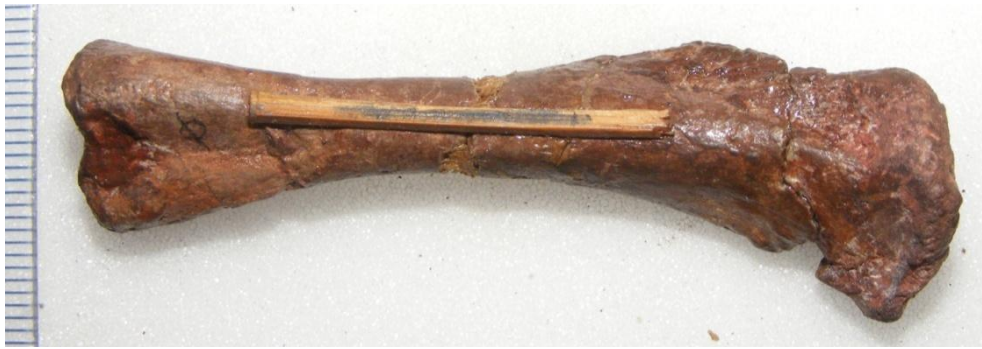

Figure 43 – Humerus, caudal aspect. Scale in mm. Note lack of prominence of both deltopectoral crest and humeral head, similar to 'hypsilophodonts'.

*Radius*: straight, slender element; caudocranially expanded proximally and distally; distal end rugose, more 'massive', proximal end gently striated; caudal border gently concave (lateral aspect) – cranial border slightly less concave; proximal medial border greatly exhibits well-developed depression tapering rapidly into shaft.

*Ilium* (fig. 44): ventral border flat, forming orthogonal association between lateral and medial surfaces – slightly kinked laterally at projection of anterior process. Anterior process flat, ventrally bent (distally); thins distally into flattened tip; medial surface develops slight longitudinal ridge dorsal to concavity between process and pubic peduncle – continues for entire process length, creating gentle dorsoventral concavity. Caudal end orthogonal, flat and vertical; broadly concave laterally. Lateral surface bears well-developed supracetabular shelf (primitive condition) - also found in Hadrosaurs implying independent acquisition (S. Maidment, pers. comm.). No brevis shelf development – ventral border twisted and crudely transversely ridged.

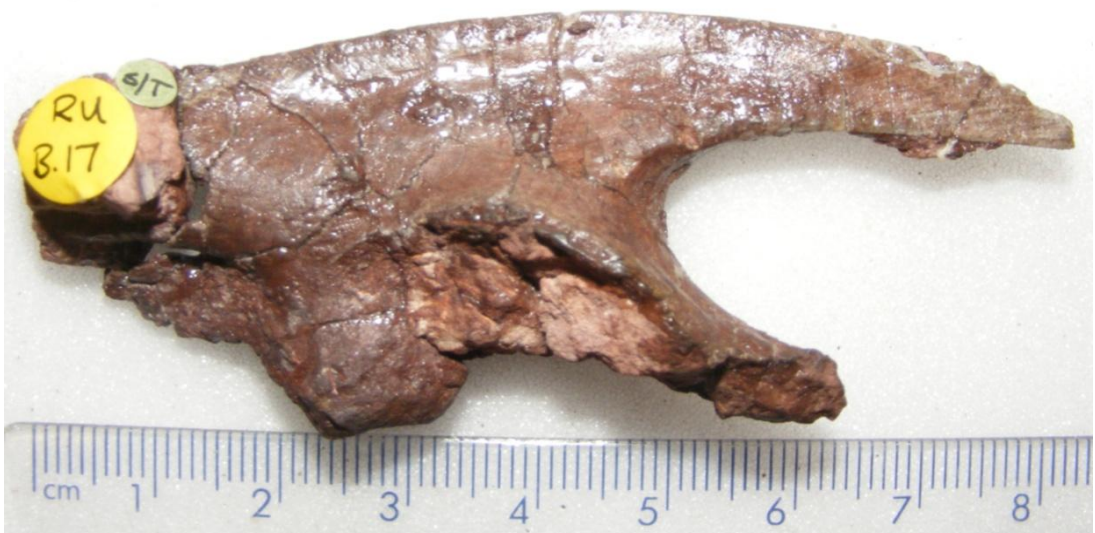

Figure 44 – Ilium, lateral aspect. Scale in cm. Note cranial extent of pubic peduncle, sheath-like cranial process, convex dorsal margin, strong supracetabular shelf, and lack of ventromedial expansion (primitive form).

*Femur*: variable form of femoral head: either thick, curved tab or sub-oval and massive; greater trochanter poorly developed – detached from lesser trochanter by distally converging cleft; fourth trochanter transversely broad, approximately one-third length of arcuate, transversely twisted shaft – projects caudomedially. Caudal intercondylar groove similar to *H. foxii* – more prominent than cranial surface (groove absent); cranial surface flattened with virtually no cranial expansion of rugose condyles; inner condyle projects further caudally than massive outer condyle – forms thick, rounded ridge extending half femoral length; long axis parallel to twisted axis of fourth trochanter.

## 9.6 Appendix 6: Character description and recoded data matrix

Synapomorphies from Norman (2004).

1. Dorsal aspect of the premaxillary rostrum: narrower than frontal width (0); wider (1).

2. Occlusal margin of the premaxilla not at all, or slightly, offset from the maxillary dentition (0); strongly offset (1).

3. Premaxillary teeth present (0); absent (1).

4. Opening of the external nares: confined to the area above the occlusal margin of the premaxilla (0); overlaps the rostral portion of the maxilla (1).

5. Lateral margin of the premaxilla slightly thickened (0); reflected dorsally to form a rim on the lower edge of the external nares (1).

6. Ventrolateral process of the premaxilla contacts nasal and maxilla (0); extends backward to contact the lachrymal/prefrontal and separates nasal and maxilla externally (1).

7. External opening of the antorbital fenestra shape: large and subtriangular (0); reduced and subcircular (1); not exposed externally (2).

8. Placement of antorbital fenestra: between lachrymal and maxilla (0); on rostrodorsal margin of the maxilla alone (1).

9. Lacrymal-maxilla contact: present (0); absent (1).

10. Palpebral: present (0); absent/fused to orbital margin (1).

11. Rostral end of jugal: tapers to a point (0); dorsoventrally expanded (1); expanded and bluntly truncated (2).

12. Jugal-maxilla suture: scarf junction (0); finger-in-recess articulation (1); large, corrugated vertical facet (2).

13. Jugal morphology: strap like with little undulation to lower border (0); marked ventral deflection (1).

14. Jugal-ectopterygoid articulation: present (0); absent (1).

15. Frontal shape: arched and narrow (0); flat in profile (1); rostrocaudally short and broad (2).

16. Frontal in orbital margin: present (0); excluded (1).

17. Paraquadrate foramen: present (0); absent (1).

18. Quadrate articular condyle: transversely expanded (0); narrow and subspherical (1).

19. Predentary ventral lobe: median process (0); strongly bifurcate distally (1).

20. Diastema: short (0); pronounced (1).

21. Dentary ramus: straight (0); deflected ventrally (1).

22. Dentary ramus: tapers rostrally (0); parallel dorsal and ventral borders (1); deepens rostrally (2).

23. Coronoid process of dentary: oblique (0); perpendicular and finger like (1); expanded apex (2).

24. Coronoid process position: laterally offset and dentition curves into the base (0); laterally offset and dentition separated from it by a shelf (1).

25. Surangular foramen: present (0); absent (1).

26. Angular visible on lateral surface of lower jaw (0); not visible (1).

27. Dentary crown shape: broad and shieldlike with more than one vertical ridge (0); narrow, approximately diamond-shaped, single median vertical ridge (1).

28. Dentary enamel: evenly distributed lingually and buccally (0); thin veneer bucally, thick lingually (1); exclusively found on lingual surface (2).

29. Marginal denticles: tongue shaped (0); curved and mammillate edge (1); much reduced to small irregular papillae or absent (2).

30. Cemented roots: not cemented (0); partially cemented (1); angular sided and rugose roots (2).

31. Alveolar trough: lateral wall contains a mixture of grooves and impressions of crowns (0); narrow, parallel sided grooves only (1).

32. Maxillary versus dentary crown width in lateral aspect: approximately equal (0); maxillary crowns narrower than dentary crowns (1).

33. Dentary crowns broad and shieldlike (0); mesiodistally compressed and lozenge like (1).

34. Maxillary crowns: equal to dentary crown width (0); narrower (1); lanceolate (2).

35. Maxillary crown ridges: no clear primary ridge (0); prominent primary ridge (1).

36. Occlusal surface on dentary: narrow—one tooth width (0); broad—two or more teeth in same alveolus form the occlusal surface (1).

37. Replacement crowns: one (0); two (1); three or more (2).

38. Dorsal neural spines: low and square (0); rectangular and height more than twice width (1); extremely elongate, height more than six times width (2).

39. Sacrum: seven or fewer vertebrae (0); more than seven (1).

40. Scapular blade: straight (0); curved (1); curved and flared distally (2).

41. Scapular acromion: prominent boss on the cranial margin of the scapula (0); the boss is reflected laterally (1).

42. Humerus: scapula length: approximately equal lengths (0); scapula longer than humerus (1).

43. Sternal shape: reniform (0); hatchet like (1).

44. Carpal structure: fully ossified and block like (0); reduced (1).

45. Metacarpal I shape: dumbbell-like (0); short and block like (1); absent (2).

46. Metacarpals II-IV: dumbbell-like and spreading (0); closely appressed (1); appressed, slender and elongate (2).

47. Manus digit I: present (0); absent (1).

48. Manus ungual I: claw like (0); conical (1); absent (2).

49. Manus unguals II and III: claw like (0); flattened, twisted and hoof like (1).

50. Manus digit III: four phalanges (0); three phalanges (1).

51. Preacetabular process of ilium long and laterally compressed (0); strongly downturned (1).

52. Dorsal margin of iliac blade: mostly smooth edged (0); strongly notched behind the ischial peduncle (1).

53. Dorsal edge of ilium above ischial peduncle: not thickened and bevelled (0); thickened (1); everted with pendent tip (2).

54. Ilium, postacetabular process: tapering caudally (0); low and rectangular (1).

55. Pubis, prepubic process: short and blunt (0); elongate (1).

56. Pubis, prepubic process: rod-shaped (0); laterally compressed, barnlike (1); short constriction and distal expansion (2); deep expansion (3).

57. Pubic shaft: ends adjacent to distal end of ischium (0); shorter than ischium, no pubic symphysis (1).

58. Ischium, shaft shape: straight (0); arched dorsally (1).

59. Ischium shaft: flattened in cross section (0); rounded in cross section (1).

60. Obturator process: absent (0); present near midshaft (1); present and close to pubic peduncle (2).

61. Tip of ischium: unexpanded (0); craniocaudally expansion to form a boot (1).

62. Femur: distal half of shaft curved caudally (0); straight (1).

63. Femoral fourth trochanter: pendent (0); triangular (1); crested eminence (2).

64. Femur extensor groove: open shallow trough (0); U-shaped groove (1); partially enclosed channel (2); fully enclosed tunnel (3).

65. Femur distal condyles: moderately expanded caudally (0); expanded caudally and cranially (1).

66. Metatarsal I: well developed and articulates with phalanges (0); slender and splint like (1); absent (2).

67. Pedal unguals: elongate and pointed claws (0); elongate but bluntly truncated (1); short, broad and crescentic with reduced or absent claw grooves (2).

*T. tilletti:*     0011001000 ??00100000 0100000100 0000000100 1?00000001 0000111001 0100000

LL.12275:     0011101000 0?1?1010000 0110100100 0000000101 1100000001 1000111111 0100000